\documentclass[11pt,a4paper]{article}
\pdfoutput=1
\usepackage{amsmath,amssymb,amsthm,mathtools,eucal,cite,color}
\usepackage[top=3cm,right=3cm,left=2.5cm,bottom=3cm]{geometry}
\usepackage{framed,graphicx,multirow}
\usepackage[T1]{fontenc} 

\providecommand{\U}[1]{\protect\rule{.1in}{.1in}}

\def\RR{\mathbb{R}}

\def\ZZ{\mathbb{Z}}
\def\HH{\mathcal{H}}
\def\KK{\mathcal{K}}

\def\inv{^{-1}}
\def\wh{\widehat}
\def\wt{\widetilde}
\def\ol{\overline}
\def\pa{\partial}
\def\lra{\leftrightarrow}

\def\Li{\mathrm{Li}}
\def\supp{\mathrm{supp}}
\def\bea#1{\begin{eqnarray}\label{#1}}
\def\eea{\end{eqnarray}}
\def\ba{\begin{array}}
\def\ea{\end{array}}
\def\bpm{\begin{pmatrix}} \def\epm{\end{pmatrix}}

\def\eref#1{(\ref{#1})}         
\def\sref#1{Sect.~\ref{#1}}
\def\aref#1{App.~\ref{#1}}
\def\erw#1{\langle #1\rangle}
\def\Erw#1{\big\langle #1\big\rangle}
\def\ERW#1{\Big\langle #1\Big\rangle}
\def\wick#1{\,\colon\!#1\!\colon}
\def\wickmv#1{\,\colon\!#1\!:_{m,v}}
\def\wickv#1{\,\colon\!#1\!:_{v}}
\def\ket#1{\vert #1\rangle}
\def\sla{\slash\!\!\!}
\def\sign{\mathrm{sign}}
\def\vv#1{\vert \vec #1\vert}
\def\skp#1#2{(\vec #1\cdot \vec #2)}
\def\Skp#1#2{(\vec #1\!\cdot\! \vec #2)}
\def\vnabla{\vec \nabla}

\def\ioi{\int_0^\infty}

\numberwithin{equation}{section}
\def\eps{\varepsilon}

\title{\vskip-10mm Infraparticle quantum fields \\ and the formation of photon
  clouds }

\author{{\sc Jens Mund$^a$, Karl-Henning Rehren$^b$, Bert
Schroer$^c$} \\ {\small
$^a$ Departamento de F\'isica, Universidade Federal de Juiz de Fora,} \\[-1mm]
{\small Juiz de Fora 36036-900, MG, Brasil, email: jens.mund@ufjf.edu.br} \\[1mm]
{\small $^b$ Institut f\"ur Theoretische Physik,
  Universit\"at G\"ottingen,} \\[-1mm] {\small 37077 G\"ottingen,
 Germany,  email: rehren@theorie.physik.uni-goettingen.de} \\[1mm]
{\small $^c$ Centro Brasileiro de Pesquisas F\'isicas, 22290-180 Rio de Janeiro, RJ, Brasil;} \\[-1mm] {\small Institut f\"ur Theoretische Physik der FU
  Berlin, 14195 Berlin, Germany,} \\[-1mm] {\small email:
  schroer@cbpf.br, schroer@zedat.fu-berlin.de}}

\date{September 2021}

\begin{document}
\maketitle

\begin{abstract}
A non-perturbative and exactly solvable quantum field theoretical
model for a ``dressed Dirac field'' is presented, that exhibits all
the kinematical features of QED: an appropriate delocalization of the charged field as a
prerequisite for the global Gauss Law, superselected photon clouds
(asymptotic expectation values of the Maxwell field),  
infraparticle nature of charged particles that cannot be separated from their photon
clouds, broken Lorentz symmetry. The model serves as an intermediate leg on a new
roadmap towards full QED, formulated as an off-shell theory, i.e.,
including a perturbative construction of its interacting charged fields. It also fills a
gap in recent discussions of the ``Infrared Triangle'', and points the 
way towards a new scattering theory for theories with massless
particles of helicity $\geq 1$, in which infraparticles can be
described at the level of charged fields, rather than just states.
\end{abstract}

\section{Introduction}
\label{s:intro}

The purpose of this paper is two-fold. On the one hand, we
  shall outline a new roadmap towards the perturbative off-shell
  construction of QED, including its charged fields,
that proceeds in a way that Hilbert space positivity is preserved at every
intermediate step. On the other hand, we present and discuss in detail an intermediate step
in this program, which can be performed in an exact non-perturbative
way. This ``intermediate model'' does not yet describe any interaction. It is rather a genuine off-shell construction
implementing a highly nontrivial change of the structure of charged free
fields. It thus manifestly captures salient infrared features of QED,
which are not visible (or not even addressed) in the standard
approach. It is of its own interest because it is the first of its
kind in four spacetime dimensions and may serve as an exactly solvable
model for developping infraparticle scattering theory.

In \sref{s:clouds} and \sref{s:infrap} of this introduction, we shall
discuss in some more detail the infrared features of QED to be addressed by our
exact model, which constitutes the main part of this paper.  Physically, those features are of course due to the long range of
the electromagnetic interaction, which is in turn due to the vanishing
photon mass. They will be implemented in terms of a ``dressed
Dirac field'' \eref{psiqc}, the main new object of our interest,
whose actual construction and analysis will be found
in \sref{s:4Dmodel}.

It is well-known that electrically charged particles are necessarily infraparticles,
  see \sref{s:infrap}. We want to convince the reader that a radical conceptual step of
using a new type of quantum fields for infraparticles
(``infrafields'') such as \eref{psiqc} is necessary to get to the roots of the (almost a
century old) infrared problems of QED. In a follow-up publication \cite{MRS4}, we
shall present a similar infrafield construction that is needed to
describe the formation of ``graviton clouds'' which determines the
gravitational properties of quantum matter.

The new conceptual mindset towards infrared quantum
field theory, and the ensuing need of new computational techniques
outside the routine
repertoire (some of which still remain to be developped) justify the length of
the present paper. We shall present many computations in more detail than usual,
but we strive not to overburden the main text, by deferring large
parts into the appendices.

The new approach to QED will be briefly outlined and commented in
\sref{s:roadmap} and \sref{s:comments}. It involves a perturbation theory of the 
charged infrafield of the intermediate model, that in this context is
regarded as a ``free infrafield'', see \sref{s:ptII}. There will appear a new type of
propagator called ``cloud propagator'' producing new types of IR-divergent loop
diagrams. Most noteworthy are indications of a new
  infrared  interference mechanism: The diagrams with cloud
  propagators describe an interference between the infrared divergences
responsible for the superselection structure of the dressed model
and the well-known infrared divergences of QED (that separately
cause velocity superselection and prevent scattering \cite{Wsoft}). The
interference produces a dynamical superselection rule: QED deforms the
rigid superselection rule of the dressed model, and the dressing relaxes velocity
superselection of QED so as to admit scattering.

We shall  put our approach in perspective with the
``Infrared Triangle'' \cite{Str} in \sref{s:triangle}, that links IR
features of QED scattering theory (Weinberg's soft photon theorems \cite{Wsoft})
with physics at the Penrose ``lightlike boundary'' of Minkowski
space. Our contribution to the Infrared Triangle is the computation and analysis of
expectation values of the quantum electromagnetic field in charged
quantum states in various asymptotic regimes, where the states are generated by the new infrafield. The results confirm (and shed some new light) on the views of
\cite{Str}, connecting them (through the spacelike 
asymptotics) with the infraparticle nature of charged particles.

Since the earliest days of Quantum Electrodynamics, one has
struggled with the redundancy of the description
\cite{Jo,Di,Ma,St} due to ``gauge degrees of freedom'', 
and with the lack of Hilbert space positivity in
covariant descriptions \cite{BLOT,Wein}. In our manifestly
positive reformulation of QED, the redundancy of the
description is directly related to the abundance of superselected
states of a given electric charge, which are commonly visualized as the
``shape of a photon cloud'' at spacelike infinity, and quantitatively
described by the asymptotic electric flux per solid angle.

Dirac \cite{Di} was one of the first to consider formulating QED
  with a formally gauge-invariant   electrically charged field of the form
\bea{dir} e^{iq\int_\Gamma dy^\mu A_\mu(y)}\psi(x)\eea
where $\psi(x)$ is the Dirac field and $\Gamma$ is a curve extending
from $x$ to infinity. Instead of the line-integral along a single
curve, smeared curves are also admissible. Similar
ideas were pursued by Mandelstam \cite{Ma} and Steinmann
\cite{St}. The question arises how such formal expressions can be defined. 

The formula \eref{dir} is also central to our approach. But we
  are led to it by a very different, dynamical motivation, that we
  shall expose in detail in \sref{s:roadmap}.
The guiding principle is Hilbert space positivity, rather than gauge
invariance which is no longer an a priori postulate. The latter will
  rather be an emerging feature following from the perturbative preservation of
  locality of observables, while the localization of charged fields may 
  (and must) be relaxed as suggested by \eref{dir}. See also the
  discussion in \sref{s:paradigm}.

For our purposes, it will be sufficient to be slightly more
restrictive than \eref{dir}: we choose straight curves (``strings'') of the form
$\Gamma_{x,e}=x+\RR_+e$ and average over the directions $e$ (in a way to be
specified). We shall write
\bea{phi}\phi(x,e):= \int_{\Gamma_{x,e}} dy^\mu A_\mu(y) = \ioi ds \, A_\mu(x+se)e^\mu
,\eea
and call it the ``escort field'' because it is evidently not an
independent field. We shall write $\phi(x,c)$ for its smearing with a real function $c(e)$ of total weight $1$. One may think
of the smearing as an ``average'', although $c(e)$ is not required to
be positive. Then we can rewrite the heuristic expression \eref{dir} in the form
\bea{psiqc}\psi_{qc}(x)=e^{iq\phi(x,c)}\psi(x) \qquad (\hbox{to be defined!})\eea
The point is that this formal expression is ill-defined because
\eref{phi} is infrared divergent. We shall provide in \sref{s:dressD} a non-perturbative
definition of the field \eref{psiqc} on a Hilbert space, which we
call ``dressed
Dirac field''\footnote{For a critical assessment of the terminology
  ``dressing'' see \sref{s:infrap}.}. It serves as an exactly solvable intermediate
model on the route to QED. It is a ``free infrafield'' that does not
have a nontrivial interaction, but that implements characteristic kinematical
features of QED that quantum fields in the usual axiomatic setting, such as
Wightman's \cite{Wight}, fall short of. The desired features of the
model are in fact {\em caused by} the cure of the infrared divergence,
while at the same time preserving Hilbert space positivity, see \sref{s:dressD}.

The most prominent of these features is the photon cloud superselection
structure. In this way, the undesirable 
gauge redundancy turns into a tool to ``manufacture'' charged states with
different photon clouds. Further consequences will be
discussed below.

We outline in \sref{s:roadmap} and in \sref{s:ptII} how full QED
should be constructed as a perturbation theory of the free infrafield model.

The proposed construction of QED is made possible with a new toolkit of quantum field theory
that allows to relax the localization properties of Wightman
fields. Notice that the escort field is localized along the string
$x+\RR_+e$, and the dressed Dirac field inherits its localization,
that extends to infinity. 
This departure from the standard axiomatics turns out to be not a
defect but a strength of our approach, taylored to implement the said
infrared features. 

{\em QED becomes the first -- and most prominent  -- instance of a
  new  setup for quantum field theory,} originating in various recent
conceptual insights gained by axiomatic approaches. A concluding \sref{s:paradigm}
  will place QED in a broader context of 
axiomatic QFT, and discuss some of the new emerging paradigms
concerning the nature of quantum fields.

Our construction has a long history. It grew out of a simple 2D model by one
of the present authors \cite{Sch1}, but was made possible in 4D only by
recent conceptual insights \cite{MSY} into the nature of quantum fields, that were
imported from the more abstract Haag-Kastler approach (``algebraic
QFT'' or ``Local Quantum Physics'' \cite{Ha}). The new element
is the possibility of quantum fields with a localization that extends
to spacelike infinity within (narrow) cones, called ``strings''. The
need for string-localization was also anticipated abstractly in \cite{BF} through
an algebraic analysis of localization properties of charged 
states in massive theories. It was not (or only secretly\footnote{by viewing the IR-divergent massless free field in 2D as a string-integral over its
  well-defined derivative $\pa_\mu\varphi$ (which is a pair of chiral currents).}) present in the old 2D model.

String-localization was introduced in \cite{MSY} because it can improve
ultraviolet divergences of perturbative interactions, and avoid the
indefinite metric state space of gauge theory (Krein space) at the same time. Without interaction, it is rather
an artificial possibility for all free fields 
creating particles of finite spin and helicity, with the benefit that 
it allows the construction of quantum stress-energy tensors for higher helicity
where a point-localized SET does not exist \cite{MRS1}. It is intrinsically
necessary only for the construction of free fields transforming in Wigner's infinite-spin 
representations for purely kinematical reasons: the increase of scaling dimension with
the spin can only be compensated at the expense of relaxing the
localization \cite{MSY,R1}.
It seems that string-localized fields are necessary for all
  interactions involving particles of spin or helicity $\geq
  1$, whereas interactions that can be described within the framework
of Wightman axiomatics are limited to low spin $\leq \frac 12$, and are mainly of
academic value.

The present paper emphasizes the mechanism how a suitable
string-localized free field in the interaction
density (or rather the ``dressing density''\footnote{\label{fn:dress}
  A note on terminology: We avoid the term 
  ``Lagrangian'' in order to avoid the impression of a Lagrangian
  quantization, which would require also a free Lagrangian. The latter
  plays no role here, because free quantum fields are constructed
  directly on Fock spaces. We avoid the term ``interaction'' because
  the dressing density at hand is a total derivative and does not lead
  to a genuine interaction. Accordingly, 
  we shall refer to the construction of the dressed Dirac field from
  the dressing density, which is only a kinematical part of the QED
  interaction density, as the ``dressing transformation''. })
passes its string-localization onto the resulting dressed Dirac field
in a new ``dynamical'' way, substantially different from the artificial construction of
string-localized Dirac fields in \cite{MSY}. In full QED as a perturbation
  of the dressed Dirac field, the interacting Dirac field remains
  string-localized. This not only resolves the conflict between locality and the 
global Gauss Law (\sref{s:clouds}), and restores the local Gauss
Law (\sref{s:pt-gauss}); it also makes explicit the  nature of the 
electron as an infraparticle, see \sref{s:infrap}.

\subsection{Global Gauss Law and photon clouds}
\label{s:clouds}
Quantum states of electromagnetically charged particles must yield
expectation values of the Maxwell field with a Coulomb-like fall-off
towards spacelike infinity like $r^{-2}$. Such states cannot be
generated from the vacuum by compactly localized charged field
operators $\psi$, due to
a standard argument \cite{FPS}: If Gauss' Law holds in QED, then the operator
of the electric flux through the infinite sphere
$$Q=\oint \vec E\cdot d\vec a = \lim_{r\to\infty}
\int_{S^2}r^2d\sigma(\vec e) \, \Skp e{E(0,r\vec e)}$$
must equal the charge operator that generates the $U(1)$
transformation of the charged field. But the integral commutes with the charged
field by locality. Contradiction.%
\footnote{\label{fn:fict}In standard perturbative QED, this
conflict appears in a different guise \cite[Chap.~10.B]{BLOT}: because
all fields are (anti-)local, the Gauss Law does not hold in the indefinite
metric state space. Instead, already the free Maxwell field satisfies a
modified Gauss Law $\pa_\mu F^{\mu\nu} = j^\nu_{\rm fict}$ with
the ``fictitious current'' (for simplicity in the Feynman gauge) $j^\nu_{\rm fict}=-\pa^\nu(\pa
A)$, see \eref{eq:fict}. Because the ``fictitious charge operator'' $\int d^3x\,
j^0_{\rm fict}(0,\vec x)$ cannot be written as a surface integral, it
has a non-vanishing commutator with the anti-local Dirac field with
QED interaction, and is the generator of the $U(1)$ symmetry. While the
fictitious current vanishes in the physical Hilbert space of the free
fields, it contributes a term to the expectation value of the charge
  operator in an electron state perturbed with the QED interaction, that cancels the total charge, cf.\  Table 1 in \sref{s:pt-gauss}.}

The conflict between locality and the long-range nature of the
  electromagnetic interaction, manifested in the Gauss Law, is tightly
  related to two other features: the existence of uncountably many
  superselection sectors in QED, distinguished by the ``profile'' of the
  asymptotic electric flux density at spacelike infinity as a function of
  the direction, that is often visualized as a ``photon cloud''
  accompanying charged matter sources; and the fact that charged
  particles cannot have a sharp mass (because they have to carry along their
  photon cloud): instead they are ``infraparticles'', see
  \sref{s:infrap}. Because the superselected photon clouds transform
  nontrivially under boosts, Lorentz invariance is broken in each
  irreducible sector.

  These long-established facts \cite{FPS,FMS,Bu2,Bu3} constitute more facets of
the infrared world of QED, that should be added to what was recently
called the ``Infrared Triangle'' \cite{Str},
whose corners are soft photon theorems, memory effects and new infinite
symmetries (large gauge transformations), see \sref{s:triangle}.

In our model, the dressed Dirac field is not a local field. The escort
field $\phi(x,e)$ is localized along the string $x+\RR_+e$, in the
sense that two such operators 
commute whenever their strings $x_i+\RR_+e_i$ are spacelike
separated. The dressed Dirac field inherits this localization (with a
smearing in $e$). In this way, the conflict of the global Gauss Law
with Locality is resolved, provided the commutator with the asymptotic
electric flux density has the correct $1/r^2$ fall-off. We shall verify this
behaviour in \sref{s:expFu} for the intermediate model (where the
total charge is zero) and for QED in first perturbative order in
\sref{s:pt-gauss} (where the total charge is $q$). The vanishing of
the total charge in the model is expected
because the model does not describe the QED interaction but is taylored to
exhibit the infraparticle nature of the electron, see \sref{s:infrap}.

The local effect of the long-range photon
cloud carried along by the dressed Dirac field should also guarantee the local
(differential) Gauss Law. We verify the local Gauss Law
for QED in first perturbative order,  
where the dressing cancels the charge density of fictitious
provenience in  the
Krein space approach (see Table 1 in \sref{s:pt-gauss}). This result highlights the power of the present approach, insisting on Hilbert
space positivity in every intermediate step without changing the
physical content of QED, while doing full justice to its infrared structure.

The dressing transformation (i.e., the non-perturbative construction
of the dressed Dirac field in \sref{s:dressD}) includes the construction of a
  Hilbert space with uncountably many superselection sectors (``photon
  clouds''). Also the subsequent step towards the perturbative
  implementation of full QED can be performed in a way that 
  manifestly preserves Hilbert space positivity, see \sref{s:ptII}. The combination of both steps defines QED directly in the
physical Hilbert space, which is a ``deformation'' of the Hilbert
space of the intermediate model. It is very different from the Gupta-Bleuler
or BRST Hilbert space, in which physical charged fields cannot be defined.

\subsection{Infraparticles}
\label{s:infrap}

The states generated by the dressed Dirac field are very similar to
the ``dressed states'' in the Chung and Faddeev-Kulish approaches to scattering
theory \cite{Chu,FK}, see \cite{Dy,Du3} for recent rigorous
treatments. These states include infinitely many soft photons in what
formally looks like a coherent state, but in fact belongs to a
superselection sector orthogonal to the
photon Fock state. In the FK states, the shape of the
photon cloud depends on the momentum of the charged particle, taylored
to cancel the vanishing of perturbative QED scattering amplitudes
due
to infrared divergences \cite{Wsoft, Wein}. In contrast, in the states of
the dressed model, the shape of the photon cloud can  be freely
chosen, but becomes coupled to the momentum when the QED interaction is turned on, see \sref{s:joint}.

The infrared superselection sectors are not separately Lorentz
covariant, because Lorentz transformations connect photon clouds of
different shape. Consequently, also the energy-momen\-tum spectrum of the sectors is not
Lorentz invariant. The spectrum of charged states in QED has a sharp lower bound (the mass of the free
electron) but is dissolved above the mass-shell, such that the mass
hyberboloid has zero weight \cite{Ha}. There are no states
which are eigenstates of the mass operator (as there are in the tensor
product of the Dirac Fock space with the photon Fock space), i.e., isolated electrons
separated from their photon cloud do not exist. Such states were
called ``infraparticle states'' \cite{Sch1,Bu2,Bu3}.
The states created from the vacuum by
the dressed Dirac field of our model enjoy these features.

The existence in QED of superselection sectors with
broken Lorentz invariance in each sector, and the
non-existence of eigenvectors of the mass operator in charged sectors, hence the
necessity for infraparticle states, are known for a long time as necessary
consequences of the nontriviality of the asymptotic flux density,
by abstract arguments \cite{FMS,Bu2,Bu3}. Although the global Gauss Law
  does not hold in our model, it provides the first explicit
  realization of these
features in the context of QED. It also allows to give an example for an infraparticle 
spectrum in \sref{s:dressD}. 

There arises a major challenge: scattering theory. The working horse
of scattering theory (LSZ theory), as well as the more conceptual
Haag-Ruelle theory (\cite[Chap.~12]{BLOT}) presuppose that
time-asymptotic particle states transform in one of the
irreducible positive-energy representations of the Poincar\'e
group (``Wigner particles''). In particular, they must have a sharp
mass-shell, and the asymptotic fields satisfy the Klein-Gordon equation. 
Those methods fail for infraparticles. Unlike with short-range
interactions, ``the photon cloud is not stripped off the electron'' at late times. Pretending that
electrons were Wigner particles, must lead to trouble, manifest in the
formal vanishing of the S-matrix in charged states computed with LSZ \cite{Chu,Wein}.
Practical ways out were presented in \cite{BN,YFS} and in \cite{Chu,FK,Dy,Du3}, all taking into
account photon clouds in one way or the other, related to Weinberg's
``soft photon theorem'' \cite{Wsoft}. These methods were
focussed on the S-matrix. In contrast, our model and the 
construction of QED based on it provide also the
charged quantum fields interpolating in- and outgoing states. They may
therefore serve to develop and test new variants of LSZ or
Haag-Ruelle scattering theory that should be applicable to fields
which create infraparticle states, see \sref{s:scatt}. As will become
clear, asymptotic two-infraparticle states are not expected to be (anti-)symmetrized
tensor products of one-infraparticle states, as it is the case for
Wigner particles. 

We want to stress an important point, just mentioned above: ``the photon cloud is not
stripped off the electron'' at late times. An undressing would occur when the
  photons were replaced by massive vector bosons. In the massive case, the
  motivation for the dressing transformation is that the coupling of
  massive vector bosons to charged matter with the Proca field is
  non-renormalizable, while the coupling with a massive gauge field, e.g., in
  the Feynman gauge, violates positivity. It thus allows a
  renormalizable coupling of massive vector bosons without positivity
  violation. However, the resulting dressing is an ``off-shell''
  (before the LSZ limit is taken) feature: clouds of massive
  vector particles disappear in the large-time LSZ 
  asymptotics thanks to the short range correlations of the massive
  vector boson field, and the asymptotic charged field describes free Wigner
  particles with a sharp mass-shell. Closely related is the fact
    that, in the massive case, a unitary S-matrix can as
    well be constructed from undressed charged fields
 with the BRST method. The dressing is only necessary to maintain
 Hilbert space positivity of off-shell correlation functions.

In contrast, this undressing scenario does not apply to electrically
charged particles. They remain infraparticles also in the asymptotic
regime, because their photon clouds lead to different sectors which no
asymptotic time  limit can undo. The long  range correlations of the photon clouds are manifest in the power law structure of correlation functions
of the dressed Dirac field, see \sref{s:dressD}, which in turn causes
the LSZ theory to fail.

We mention these facts in order to stress that the 
terminology ``dressing'' is not meant to suggest that an ``undressing''
were possible. It is possible in the massive case, but not in the massless case of QED.

Scattering theory is based on the subtle idea of ``separation of wave packets
at asymptotic times''. This gives us occasion to remember that Quantum Field Theory knows two
different kinds of localization. The first one, Born localization, is
a time-asymptotic feature of wave functions. It originates from
Quantum Mechanics
where it allowed Born (by his seminal ``scattering theory'' argument) to conclude
that the squared Fourier transform of the wave function represents the
momentum density. In contrast, Einstein Locality (vanishing
commutators at causal separation), also known as Causality, allows to
define localization of observables in algebraic terms -- one of the corner stones of
``Local Quantum Physics'' \cite{Ha}.  By this algebraic feature, QFT
goes far beyond ``QM with infinitely many degrees of freedom''. The
two concepts of localization are not directly compatible with each
other (e.g., in the sense that the smearing function of a quantum
field would have any direct relation to the localization of a particle
-- which is wrong; see the discussion in 
\cite{NW}.)

Born localization connects off-shell quantum fields to
on-shell particle states. It is an emergent asymptotic feature, that must be
controlled -- as in Haag-Ruelle theory -- with the help of Einstein
Locality and cluster properties. For this, it is indispensable to have
causal separability of the quantum fields that create particle
states. In the case of the string-localized dressed Dirac field in our model, it is
sufficient that the spacelike strings can be smeared in small solid
angles (localization in narrow cones \cite{BF}).

In our approach to QED, to be outlined in \sref{s:roadmap}, the interaction density has a string-dependent
part that  is necessary to formulate the theory without the use of
indefinite metric state spaces as is done in gauge
theory. It has no effect on the dynamics, but it is expected
to counteract the infrared divergences of the S-matrix, see \sref{s:joint}. The difference
to the Faddeev-Kulish approach is that the state-creating charged
infrafields inherently carry the compensating dressing factor with
them, and this factor has not to be attached to the states ``by
hand''. 

But instead of the option to pursue 
perturbation theory with the full QED interaction in one stroke (as advocated, e.g., in
\cite{Sch2}), our approach in this paper is to implement the non-dynamical part of the interaction beforehand.
This can be done non-perturbatively and leads to the model of the
dressed Dirac field, which is of its own interest not least as a
testing ground for new methods of scattering theory, e.g., \cite{Bu1,BPS,He2,DM}.

The model confirms the abstract criterium for the
infraparticle nature of charged states \cite{Bu2}, formulated in terms
of  asymptotic properties of expectation values of
the electromagnetic field at spacelike infinity. These expectation
values exhibt classical Coulomb-like behaviour. The asymptotic flux density operators commute with
  all local observables, so that their expectation values distinguish
  superselection sectors, mentioned above. Thus, the
classical behaviour of quantum expectation values emerges naturally in
the asymptotic regime, without a limit like $\hbar\to0$. The 
  long-range photons are responsible for
  the decoherence that is necessary, e.g., for ``classical
  measurements''. The decoherence is thus an intrinsic part of the theory, affiliated with
its infrared features, and needs not be accounted on environment
effects.\footnote{As it was emphasized, e.g., in \cite{La}: there is no
``classical world'' next to our quantum world.}

Also the lightlike asymptotics of expectation values of the electromagnetic
field in charged states can  be computed explicitly, \sref{s:ll}. Our model
therefore provides an underpinning to the Infrared Triangle (see \sref{s:triangle}). It
is a ``missing link'' in the form of an exactly solvable genuinely quantum field
theoretical model in which states with the abstract properties
discussed in \cite{Str} (some of which are imported from classical electrodynamics), 
are created from the vacuum by an infrafield. 
Our computations confirm
the classically expected $1/r$-behaviour of the transversal components
and the $1/r^2$-behaviour of the longitudinal components. By virtue of
the PCT theorem, it also 
satisfies the matching conditions of \cite{Str} that form the
cornerstone of the ``new symmetries'' in the Infrared Triangle, see \sref{s:triangle}.

The superselection of  photon clouds in QED has an interesting
``entropic'' side aspect. Formally, the states created by
exponentiated field $e^{iq\phi(x,c)}$ are coherent photon states in a positive-definite
subspace of Krein space. But because of the IR divergence, they are
not elements of that space, but lie in different superselection sectors. This
is also reflected in terms of entropy: In recent work \cite{CLR},
relative entropies between vacuum and coherent states have been
computed in QFT. When the states are restricted to the subalgebra of a
half space (a ``Rindler wedge'' in space and time), the relative entropy is
finite and can be computed as an integral over certain classical
Cauchy data associated with the coherent state. Applied to the
would-be coherent state generated by the exponentiated escort field,
the lack of decay (in the direction of the string) of the Cauchy data
makes the entropy diverge. This behaviour is in marked
  difference to relative entropy of states in DHR sectors \cite{Ha}, which is
  expected to be finite. The infrared superselection sectors are thus
  far ``more distant'' from the vacuum than sectors that arise due to
  a compact symmetry group.
The qualitative difference may also become visible  in thermal states.

In the case of helicity 2, entropic aspects play an important role in the
  understanding of black hole  thermodynamics, and more
generally for states of the gravitational field \cite{CLR}. The latter has by now
not been taken serious as an infrafield. It may well be that an
approach similar to the present approach to QED may provide new
impulses to the understanding of gravitational physics \cite{MRS4}.

\subsection{The roadmap towards QED}
\label{s:roadmap}

The new roadmap starts with a chain of equivalences, to be briefly
explained below. They are schematically
displayed as follows
\bea{eq:equ}
\mathrm{QED}\stackrel{1.}=
\{\psi_0,F\}\big\vert_{L(c)}\stackrel{2.}=\{\psi_0,F^K\}\big\vert_{L^K(c)}\stackrel{3.}=\{\psi_0,F^u\}\big\vert_{L^u(c)}
.
\eea
The various interaction densities appearing in \eref{eq:equ} will be
explained in Items 1--3 below. 
$\{\Phi_i\}\big\vert_L$ is understood as the algebra of the fields
$\Phi_i\big\vert_L$, constructed from the free fields $\Phi_i$ with the
interaction density $L$. We have in mind the framework of causal perturbation
theory \cite{EG}, which provides the interacting
fields $\Phi\big\vert_L$ as power series in
integrals over retarded multiple commutators of the free fields
$\Phi(x)$ with $L(y_i)$, see \cite{DF}. 
E.g., the first order perturbation of a field $\Phi(x)$ with an interaction
density $L$ is
\bea{eq:Phi1} \Phi^{(1)}(x)=i\int d^4y\, R[\Phi(x),L(y)]:=i\int d^4y\, \theta(x^0-y^0)[\Phi(x),L(y)].
\eea
By the Wick expansion, retarded commutators turn into retarded propagators times Wick
products of fields. The retarded propagators are products of
commutator functions with Heaviside functions. Such products are  
apriori not defined as distributions, and ``renormalization'' can be
understood as the process of defining them, possibly with some freedom
to be fixed by suitable renormalization conditions, see \cite{EG}.

Causal perturbation theory assumes the presence of a
spacetime cutoff of the coupling constant. The removal of this cutoff
is called the ``adiabatic limit''. Before the adiabatic limit is
taken, integration by parts will produce boundary terms whose
vanishing has to be controlled in the limit. The adiabatic limit is in
fact the most delicate issue, especially in massless theories, also in
local approaches \cite{Du2}. We shall, however, ignore these
difficulties throughout the present paper; for some case studies see \cite{MRS2}.

``Equivalence'' in the above means equality of all correlation
  functions. Correlation functions determine the fields as
  operators on a GNS-reconstructed Hilbert space \cite{Wight}. We are
  thus addressing the ``off-shell'' construction of the quantum fields
  themselves, rather than ``on-shell'' S-matrices.
  
We briefly explain the equivalences \eref{eq:equ}, referring to the main body of the
paper for more technical details.

{\bf 1. String-localized potentials.} It is well-known that a
local potential for the Maxwell field on a Hilbert space does not
exist. Let $F_{\mu\nu}(x)$
be the Maxwell field defined on the Wigner Hilbert space (the
Fock space over the unitary representations of helicity $\pm1$ of the
Poincar\'e group \cite{Wig}). For any $e\in \RR^4\setminus\{0\}$, the field\footnote{\label{fn:Ie}
The short-hand notation $I_e$ will be used
  throughout. It assumes that the integrand has sufficiently fast
  decay. In that case, one has $(e\pa)(I_e f)(x) = I_e((e\pa)f)(x) =
  -f(x)$. } 
\bea{AFe} A_\mu(x,e):=I_e(F_{\mu\nu}e^\nu)(x) \equiv\ioi ds\, F_{\mu\nu}(x+se) e^\nu
\eea
is a potential for the
Maxwell field: $\pa\wedge A(x,e)=F(x)$, as a consequence of the Bianchi identity
(homogeneous Maxwell equation) satisfied
by $F(x)$. It may be thought of as an axial gauge
potential, except that the string direction $e$ is not fixed but also transforms under
Lorentz transformations. By definition, $A(x,e)$ is localized
along the string $x+\RR_+ e$, in the sense that two such fields
commute whenever their strings $x_i+\RR_+e_i$ are spacelike
separated.

The study of in- and out-going multi-particle states 
  requires sufficient causal separability of the
  fields that create these states. This requirement
  excludes fields localized along timelike strings. In this paper, we shall assume $e^2<0$. Because
$A(x,e)$ is homogeneous in $e$, we are free to normalize
$e^2=-1$.

$A(x,e)$ is still a potential for $F(x)$, when it  is smeared
with a function $c(e)$ of total weight 1: $\pa\wedge A(x,c)=\pa\wedge A(x,e)=F(x)$. 
This observation allows a reformulation of QED in which the
interaction density is
\bea{Lc} L(c)=q\,A_\mu(c)j^\mu.\eea
In contrast to the standard gauge-theory approach, $A_\mu(c)$ is not
an autonomous field quantized on a Fock space with
indefinite metric, but  it is a functional of the free Maxwell field
on the Wigner Hilbert space. In particular, it creates from the vacuum
only physical states with 
the two transversal modes of helicity $\pm 1$.

Consequently, the interaction density \eref{Lc}
is defined on the tensor product of the Wigner Hilbert space and the
free Dirac Fock space. Formulating QED with $L(c)$ thus avoids from the outset
issues with unphysical states and indefinite metric.

The string direction is an auxiliary quantity, on which observables 
should not depend. Indeed, the variation of $A_\mu(e)$ with respect to
$e$ is a gradient
$\pa_{e^\kappa} A_\mu(e) = \pa_\mu \big(I_eA_\kappa(e))$, so that 
$\pa_{e^\kappa}L(e) = \pa_\mu\big(I_eA_\kappa(e)j^\mu\big)$ is a total derivative. Then the $e$-dependent part of the
classical action $\int d^4y\,L(y,e)$ vanishes by Stokes' Theorem. The
``lift'' of this property to the quantum theory, i.e., the
string-independence of $T e^{i\int d^4y\, g(y)L(y,c)}$ in the
adiabatic limit $g\to1$, is a
renormalization condition. When it is fulfilled, then the perturbation
of observables
with $L(c)$ does not depend on $c$. This approach (not just for QED) was advocated in \cite{Sch2}.

{\bf 2. Embedding into Krein space.} The second equivalence in
\eref{eq:equ} embeds the Maxwell field as $F^K=
\pa\wedge A^K$ into the Krein (= indefinite metric) Fock space of the  local potential
$A^K(x)$, for simplicity in the Feynman gauge. The equivalence holds,
simply because the embedded fields $F^K$ and $A^K(e):=I_e(F^Ke)$ have
the same correlation functions and propagators as $F$ and $A(e)=I_e(Fe)$.

When the Maxwell field is embedded into the Krein space
(indefinite-metric space) of the usual local description
with a Feynman gauge potential $A^K$ such that $F^K=\pa\wedge A^K$, one finds \cite{MRS2}
\bea{AKFe}
A^K(x,e):=I_e(F^K_{\mu\nu}e^\nu)(x)  = A^K(x)+
\pa_\mu\phi(x,e),\eea
where $\phi(x,e)$ is given by \eref{phi} (with the Feynman gauge
potential $A^K(x)$ in the place of $A$).  This is how the escort field enters the stage in
our approach.

The escort field makes the string-independence of the total action
manifest: When the
interaction density \eref{Lc} is embedded into the Krein space, it
splits accordingly into two pieces:
\bea{LLKe} L^K(c)= q\,A^K_\mu(c)j^\mu = q\,A^K_\mu j^\mu + q\,\pa_\mu \phi(c)j^\mu.\eea
The latter, string-dependent part is a total derivative. When
appropriate Ward identities can be fulfilled (confirmed in low
orders), this implies that the renormalized interacting observables
are string-independent, and the equivalence 2. in \eref{eq:equ} holds.

It must be stressed here that $\partial_\mu \phi(e)$ in \eref{AKFe} is well defined,
while $\phi(e)$ is a logarithmically divergent
field that requires an IR cutoff, see \sref{s:dressD}. This divergence
will become responsible for the superselection structure of QED, to be discussed in \sref{s:4Dmodel}.

{\bf 3. The physical Maxwell field in Krein space.} In the Krein
space, the free field strength $F^K=\pa\wedge A^K$ is not
source-free:
\bea{eq:fict}
\pa^\mu F^K_{\mu\nu} = -\pa_\nu(\pa A^K) = j^{\rm fict}_\nu
\eea
is the ``fictitious current''
mentioned in footnote \ref{fn:fict}. The null field $(\pa A^K)$ has
vanishing self-correlations and correlations with $F^K$. But its non-vanishing commutator with $A^K$ is responsible
  for the non-trivial commutator between the electric flux of $F^K$
  at infinity and the Dirac field with interaction $q\, A^K_\mu j^\mu$,
  responsible for the failure of the global Gauss Law in Krein
  space the standard approach.

  $F^K$ creates from the vacuum states with transversal
and longitudinal modes.
To cure this unphysical
feature of $F^K$, we introduce in \sref{s:dressMD} another potential
\bea{eq:Au} A^u_\mu(x) := A^K_\mu(x) + u_\mu I_u(\pa A^K)(x) \equiv A^K_\mu(x) + u_\mu
\ioi ds \, (\pa A^K)(x+su)
\eea
and its field $F^u=\pa\wedge A^u$. 
Here, $u$ is a future timelike unit vector, for definiteness $u=u_0\equiv\bpm 1\\[-1mm]\vec 0\epm$. 
The first important difference is that in contrast to \eref{eq:fict},
$F^u$ is the (embedded) {\em physical} free Maxwell field (isomorphic to $F$ on the Wigner Hilbert
space). It satisfies 
\bea{eq:nofict}
\pa^\mu F^u_{\mu\nu} = 0,
\eea
and does not create unphysical longitudinal null states. In fact (see \sref{s:dressMD}), its electric and magnetic components 
\bea{eq:EB} \vec E:=-\vec\nabla A^{u}_0 - \partial_0\vec A^u =  - \partial_0(\vec A^K - \vec
\nabla\Delta\inv(\vec \nabla \vec A^K)), \quad \vec B :=
\vec\nabla\times \vec A^u=\vec\nabla\times \vec A^K\eea
are the familiar {\em transversal} Maxwell fields. 

The second important feature is that $A^u$ and consequently
also $F^u$ live on the positive-definite subspace ({\em not} the
Gupta-Bleuler quotient space)
\bea{eq:Hu}\HH^u= \big\{\Phi\in \KK: (uA^K)^-\Phi=0\big\}
\eea
of the Krein Fock space. $(uA^K)^-$ stands for
the annihilation part. Thus $\HH^u$ is the subspace generated by the
three spacelike components
$a_i^*$ of the creation operators of $A^K_\mu$. 

$F^u(x)$ appears to be localized along the timelike string
$x+\RR_+u$. But its self-correlations are the same as those of $F^K$,
hence its commutation relations are local. The situation highlights the
  ``relative'' algebraic nature of the concept of localization deduced
  from commutation relations (Einstein Locality): Relative to itself, $F^u$
  is not ``more non-local'' than $F$. That $F^u$
  is non-local relative to $A^u$, $A^K$, or $\phi$, should rather be
  blamed on the latter
  un-observable fields which contain unphysical degrees of freedom.

We define the string-localized potential $A^u(e):=I_e(F^ue)$
associated with $F^u$. 
Then it holds
\bea{Auphi} A^u_\mu(x,e) = A_\mu^u(x) + \partial_\mu \phi(x,e).\eea
For $e\in u^\perp$, the escort field in \eref{Auphi}
is the same as in \eref{AKFe}, also living on $\HH^u$. By
\eref{Auphi}, the arguments as after \eref{LLKe} also apply for the
string-independence of the perturbation theory with the interaction
density
\bea{LLue} L^u(c)= q\,A^u_\mu(c)j^\mu = q\,A^u_\mu j^\mu + q\,\pa_\mu
\phi(c)j^\mu.\eea
The densities $L^K(c)$ and $L^u(c)$ are defined in terms of observables $A^K(e)=I_e(F^Ke)$ and
$A^u(e)=I_e(F^ue)$, resp., and differ from each other only by a
term involving the null field $(\pa A^K)$. Consequently, they give
rise to the same perturbative expansions for the interacting
observables, as claimed in \eref{eq:equ}.\footnote{The fields $A^u$ and
$A^K$, which would be sensitive to the difference, do not appear in \eref{eq:equ}!}

{\bf 4. The dressing transformation.} 
  While the free fields in the last three entries of \eref{eq:equ} are
point-localized, the interaction densities are string-localized relative to the free fields.
In causal perturbation theory, this feature a priori can jeopardize locality of the
interacting fields. However, by \eref{LLKe} and \eref{LLue}, the
dangerous string-localization of the interaction density resides entirely in
the escort term $q\,\pa_\mu \phi(e)j^\mu$.

Here, the crucial
observation (and a main topic of the present paper) comes to bear:
{\em Because the escort term is a total derivative,
  that part of the interaction can be
implemented by an exact
construction: the ``dressing transformation'', in which one can establish
locality directly without relying on perturbation theory.}

The
dressing transformation gives rise to a string-localized field $\psi_{qc}(x)$,
formally displayed as in \eref{psiqc}. This is how the ``dressed Dirac
field'' enters the stage.

$\psi_{qc}$ is non-perturbatively defined on a
non-separable GNS Hilbert space, that decomposes into uncountably many superselection
sectors, labelled by the smearing functions $c$. The non-perturbative construction in \sref{s:dressD}
is exactly solvable. It parallels a
two-dimensional model \cite{Sch1} that for the first time introduced infraparticle
fields.

The algebra of the dressed Dirac field can be extended to
include the potential $A^K_{\mu}$ or the potential $A^u_\mu$. The
former extension is
represented on an indefinite space containing the vacuum representation of
$A^K$. The subalgebra $\{\psi_{qc},A^u_\mu\}$ is defined on a
proper Hilbert subspace, see \sref{s:ext}.  

Our idea for QED amounts to subject the dressed Dirac field $\eref{psiqc}$ to the first part of the interaction in the split
\eref{LLKe} (the standard QED interaction $L^K=q\, A^K_\mu j^\mu$)
which poses no problems with locality, see Item 5 below:
$$\psi_0\big\vert_{L^K(c)}= \psi_{qc}\big\vert_{L^K}.$$
We call this the
``hybrid approach'' because of the detour through Krein space,
although the final theory lives on a Hilbert space. The intermediate
theory of the dressed Dirac field  without the QED interaction is
regarded as an autonomous model, that already captures
essential infrared features in a kinematic way, as outlined in \sref{s:clouds}
and \sref{s:infrap}.

{\bf 5. Towards QED.} After the dressing transformation, it remains to implement the remaining parts
$$L^K=q\,A_\mu^K j^\mu, \quad \hbox{resp.}\quad
L^u=q\,A_\mu^u j^\mu$$ of the interaction densities \eref{LLKe} resp.\
\eref{LLue} so that
\bea{eq:hyb}\{\psi_0,F^K\}\big\vert_{L^K(c)}
\stackrel{4.\mathrm{(a)}}= \{\psi_{qc},F^K\}\big\vert_{L^K}\quad \hbox{resp.}\quad\{\psi_0,F^u\}\big\vert_{L^u(c)}
\stackrel{4.\mathrm{(b)}}= \{\psi_{qc},F^u\}\big\vert_{L^u}
\eea
are all equivalent by \eref{eq:equ}. These equivalences have been
established in lowest orders; to be valid in all orders, they will require the fulfillment of
appropriate Ward identities and the control of the adiabatic limit, see \cite{MRS2}.
The right-hand sides are perturbative expansions around the
dressed Dirac field. The latter plays the role of a ``free'' field,
that we shall refer to in this context as the ``free infrafield''. 
Both right-hand sides of \eref{eq:hyb} stand for 
expansions of the same final theory: QED.
However, they
have complementary benefits and drawbacks. The former expansion is
only defined on the
indefinite non-separable Krein space extension of the model
$\{\psi_{qc},F^K\}$,
while the latter expansion is defined on the proper
non-separable Hilbert space of the model $\{\psi_{qc},F^u\}$. Another
charme of the latter expansion is that the unphysical longitudinal photon degree of freedom (which is responsible for the IR properties of
charged states, cf.\ the discussions in \cite{BCRV, MRS2}) appears only in the
dressed charged field, and neither in the Maxwell field nor in the
interaction density.

Conversely, the benefit of the former expansion is that its
interaction density is point-localized: the
string-localized perturbation of the local free Dirac field is
equivalent to a point-localized perturbation of the string-localized free infrafield.
In contrast, the interaction density of the latter expansion
(involving $A^u$) is non-local.
This latter feature again
jeopardizes the perturbative argument for locality of the
interacting fields.

{\em The synthesis is that locality can be controlled in the former, and
positivity can be controlled in the latter expansion.}  Because
both expansions yield the same correlation functions, QED is perturbatively constructed as a local {\em and}
positive QFT, with string-localized charged fields
$\psi_{qc}\big\vert_{L^K}$, whose perturbative expansion is equivalent
to that
of  $\psi_{qc}\big\vert_{L^u}$ which is defined on a Hilbert
space. For an explicit illustration of the equivalence invoked here, see \sref{s:pt-gauss}.

We stress once more that the program is a roadmap. Many
technical details remain to be filled in. In particular, the
equivalences \eref{eq:hyb} hold only in the adiabatic
limit, and upon performing the adiabatic limit, boundary terms have to be
controlled. Furthermore, it must be established that Ward  
identities for the conserved Dirac current can be satisfied in every
order of perturbation theory also in the present setting. Those issues
will not be analyzed in any detail in this paper. 

Our focus will be instead on the exact dressing transformation and the
analysis of the features of the resulting ``intermediate'' models
$\{\psi_{qc},F^K\}$ and $\{\psi_{qc},F^u\}$.

\subsection{Comments}
\label{s:comments}

The use of string-localized fields is essential in our
approach to QED. First
of all, it resolves the conflict between (assumed) anti-locality of the
Dirac field and the global Gauss Law (\sref{s:clouds}): the string-localized dressing
density turns the anti-local free Dirac field into the string-localized
dressed Dirac field, for which there is no such conflict for the
simple geometric reason that the string extends to infinity. Second, it
allows to formulate QED directly on a Hilbert space. 
The possibility of a manifest Hilbert space positive perturbative
expansion of QED stands behind the celebrated ``mystery'' why the
manifestly non-positive Krein space approach in the end of the day
produces a unitary S-matrix.
Third, it allows to isolate the infrared sector structure as an effect of the
logarithmically divergent escort field. 

It should be mentioned that the same strategy applies as well
  for other models with electrically charged fields, like scalar QED,
  cf.\ \sref{s:paradigm}.

It has been objected \cite{BCRV} that a formulation of QED based
  on (a functional of) the
Maxwell field such as \eref{AFe} rather than an independent potential cannot be capable of describing charged
states, due to the absence of longitudinal photon degrees of freedom in
the Maxwell field. This objection is true when QED is formulated 
exclusively in terms of observable fields. But our construction includes
unobservable charged fields. The model in fact describes a
``transfer'' of longitudinal photon degrees of freedom onto the Dirac
field, where they become responsible for the super\-selected photon
clouds of charged states and the loss of the sharp mass-shell of
charged particles. These features survive (with modifications) in the
full QED. They were long ago anticipated on the basis of
axiomatic considerations (see \sref{s:clouds}, \sref{s:infrap}), and
have led to a renormalized construction of non-local charged states
satisfying Gauss' Law \cite{MS1,MS2}.

Along with the 
longitudinal degrees of freedom, the escort field also transfers its string-localization onto the
charged field, thus solving the problem with the global Gauss Law by linking IR superselection charge with electric
charge, see \sref{s:clouds}. However, the resulting string-localization of the dressed charged
field is of a very
different kind than that of the mere string-integration $I_e$ in the
string-localized potential or the escort field itself: the latter can be
``undone'' by the directional derivative  $-(e\pa)$ as in footnote \ref{fn:Ie},
and has no effect
on the particle states created by the field. In contrast, the dressed
charged field creates infraparticle states that survive in the
asymptotic time limit.

Regarded as an autonomous subtheory, the dressed Dirac model with fields  $\psi_{qc}$
with different $c$, but without the inclusion of the Maxwell field
$F^u$, is a string-local theory with infraparticle states, that
is a valuable testing ground for infraparticle scattering theory. An interesting feature of this model is that the
string-localization is manifested by commutation relations
involving a complex phase depending on the string smearing, see \eref{any}.

The issue of the timelike string-integration $I_u$ in $L^u$ and $F^u$ is more
serious. It makes correlations of the Maxwell field $F^u$ with $A^u$
and with $\phi(e)$ nonlocal. But $F^u$ is a perfectly local field, 
only the unobservable dressed
Dirac field and the auxiliary potential and escort fields are non-local relative to $F^u$. In this way, the $\{\psi_{qc},F^u\}$ model becomes a
testing ground for the Infrared Triangle. In
the full QED this non-locality is absent because diagrams involving
$I_u$ inside $F^u$
cancel term by term against terms arising from the interaction $L^u$, as witnessed by the equivalence 3.\ in
\eref{eq:equ}, see also \sref{s:pt-gauss} and \sref{s:systcons}. 

As far as full QED is concerned, the original idea naturally arose from the definition of the potential \eref{AFe} for the
Maxwell field \cite{MSY,Sch2,MRS2} that has no unphysical degrees of
freedom because it is defined on the physical Wigner Hilbert space of helicity
$\pm 1$.
This suggested to reformulate QED with the interaction density
$L_{\rm int}(e)=q\, A_\mu(x,e)j^\mu(x)$ \cite{MSY,Sch2}, to be smeared in $e$
with a function $c(e)$ of total weight one.
Since $F_{\mu\nu}$ has scaling dimension 2, $A_\mu(x,e)$ has scaling
dimension 1, so that the new formulation of QED is power-counting
renormalizable and ghost-free at the same time.
The
improvement of the scaling degree is a general feature of
string-localized potentials. It also makes string-localized
``massive QED'' with a massive vector boson power-counting renormalizable on a
Hilbert space, whereas the coupling to the
Proca field would not be renormalizable, while the
coupling to a massive gauge field would not be ghost-free. See more on
this in \sref{s:paradigm}.

There is a price for the remarkable advantages of string-localized
QFT: one has to ensure that the interacting observables are 
local fields. This  requires that the renormalization can be done in
such a way that all observable fields and the S-matrix $Te^{i\int d^4x\,
  g(x)L_{\rm int}(x,c)}$ in the adiabatic limit $g\to1$ are independent of the auxiliary
variable $e$.\footnote{Although the interacting Maxwell field is
  string-independent, our construction provides string-dependent
  charged states in which its expectation value is
  string-dependent. Our formulation thus differs from the
  external field
  approach in \cite{DW} where the Maxwell field itself depends on $e$,
  cf.\ also the external-field example in \cite{MRS2}.} A necessary (and perhaps sufficient) condition is
that the dependence of the interaction on
the string is a total derivative ($d_e L(e)= \partial_\mu
Q^\mu(e)$ on the Hilbert space -- which is the case here and in many other models of
interest, including Yang-Mills and Abelian Higgs). At first
sight, the causal (i.e., respecting the localization) 
renormalization theory in position space involving string-localized
fields \cite{JJJ} is considerably more demanding than that of point-localized 
Wightman fields, as developped in \cite{EG}. Remarkably (at
  least for a large class of interactions including those of the Standard
  Model), loop diagrams involving string-smeared string-localized
  propagators need to be renormalized only at coinciding points
  (rather than at intersecting strings) \cite{G}, so that the benefit
  of the improved UV scaling dimension   \cite{MSY} fully comes to bear.

The hybrid approach (i.e., the detour through the Krein space by
embedding the Maxwell field into the Krein space where the
interaction can be split as $L(e)=L^K+\pa_\mu V^\mu(e)$ where $L^K$ is
string-independent and point-localized) gains some
flexibility for the construction. Unlike in the local BRST
formalism based on $L^K$ alone, taking 
all terms together ensures positivity at every step. Unfortunately,
such a split is not always possible, e.g., in the case of Yang-Mills
\cite{G2}.

In the present paper, a variant of the hybrid approach
isolates the escort contributions as a separate step. These contributions are responsible for the dressing of the Dirac field, and make
manifest the pertinent infrared features of QED. They also play the main role in creating a localization dichotomy
between point-localized interacting observables and string-localized ``charged''
fields, see \sref{s:paradigm}.

In this setting, the string-localizated renormalization
theory is expected to become more transparent\footnote{Emphasizing QED as a perturbation of the dressed
  model not only constitutes a most efficient reorganization of the
  perturbative expansion. The
  use of Weyl formulas rather than expansions of the exponential
  series also provides an inherent control of infrared structure, see
  \sref{s:systcons} and \sref{s:joint}.}. It keeps all local observables and the S-matrix
string-independent; the only place where the strings make themselves
felt is in the charged fields and hence in
the states in which the observables are evaluated: notably the
expectation values of the Maxwell field in spacelike and lightlike asymptotic directions.

\subsection{The Infrared Triangle}
\label{s:triangle}

The ``Infrared Triangle'' provides a unified view at several issues
related to massless particles of helicity 1 and 2 (photons in QED, gravitons in
General Relativity). The first corner of the triangle concerns soft
photons (resp.\ gravitons). For QED, the photon clouds accompanying charged particles
are responsible for the failure of the formal LSZ prescription as if
the electron were a Wigner particle with a sharp mass-shell. The
characteristic ``shape'' of these difficulties in momentum space were captured by soft photon theorems
\cite{Wsoft}. Prevailing prescriptions to deal with this issue are inclusive
cross-sections \cite{BN,YFS} or dressing factors on the S-matrix
\cite{Chu,FK,Dy}. They do not address the fact that the
charged field must be an infrafield.

The second corner of the triangle is related to lightlike infinity
($\frak I^\pm$ in Penrose terminology). The $r^{-2}$-decay of the
electric field in spacelike directions $\frak i^0$
is accompanied in lightlike directions by a $\lambda\inv$-decay of its radiative transversal
components and a $\lambda^{-2}$-decay of
its radial component. The radial and transversal components
are not independent from each other: they are related by an asymptotic
version of the Gauss Law at lightlike infinity \cite{BG,Str}, cf.\
also \aref{b:maxw}
\bea{dErE}\pa_U E_{r,\infty} (U,\vec n) = \big( \vec\nabla_n - \vec n
\Skp n{\nabla_n} \big)\cdot\vec E_\infty(U,\vec n) .\eea

The behaviour of the Maxwell field at lightlike infinity provides an analogue
to the ``memory effect'' of gravitational waves, by
producing an observable effect on test particles ``living on $\frak
I^\pm$''. In \cite{BG}, this effect is computed with the help of
  \eref{dErE} as a ``kick'' on the  test particle when a pulse of radiation passes by. 
In our model, the radiation originates from the
  string-localized dressed Dirac field, and neither massless charges
  \cite{HMPS} nor the dynamical coupling of massive matter to the
  electromagnetic field as in full QED must be invoked.

The asymptotic Maxwell field satisfies an autonomous quantum algebra at
lightlike infinity (\aref{b:ash}), that may be viewed as characteristic ``initial
conditions'' (analogous to time-zero canonical commutation relations)
for the commutation relations of the Maxwell field in the bulk
\cite{Ash}.
Like the latter, this algebra admits uncountably many inequivalent
Hilbert space representations, that in the case at hand correspond to
the photon cloud superselection sectors. In contrast to spacelike
infinity, where the Maxwell
field algebra becomes commutative, and the sectors signal a classical
behaviour (cf.\ \sref{s:infrap}), the algebra remains genuinely
quantum (noncommutative) at lightlike infinity.

The third corner of the triangle relates to symmetries that go well
beyond charge conservation. The radial component of the electric field
on $\frak I^+$ has a limiting value at $\frak I^+_-$, where $\frak
I^+$ touches on $\frak i^0$. Remarkably, consistency of scattering
amplitudes requires that this limiting value must
coincide with the limiting value on $\frak I^-_+$, where $\frak I^-$
touches on $\frak i^0$, i.e., the ``interpolation'' through $\frak
i^0$ corresponding to spacelike limits with increasing boost parameter
connects to identical values. This matching condition was known
earlier in a classical setting, see  \cite{He1,He3}. Because the limiting values are
functions of the direction $\vec n\in S^2$, this matching constitutes
an infinite number of conservation laws. There is a second set of
conservation laws which manifests itself in a change of sign of the
radial component of the magnetic field between $\frak I^+_-$ and
$\frak I^-_+$. 

The matching conditions between the electric and magnetic fields
  at $\frak I^+_-$ and  at $\frak I^-_+$ (which are spheres) are taken
  as the conservation laws associated with  a ``new symmetry'',
  (see \cite{HMPS,KPS}, and \cite{Str} for a textbook coverage). When
  smeared with test functions $\eps$ on the sphere, the electromagnetic field
  operators at $\frak I^\pm_\mp$ are interpreted 
  as generators $Q_\eps$ of ``large gauge transformations'' whose
  gauge parameters do not die off at lightlike infinity. Moreover,
  they are not globally defined due to a twist entailed by the
matching condition. They are considered as the electromagnetic counterparts of BMS
super-translations and   super-rotations in General Relativity.
See also \cite{He3} for a critical assessment of some of the
  interpretations forwarded in \cite{Str}.

Our work contributes to this scenario a genuine quantum model in the
bulk (\sref{s:ll}), with charged states created by charged
  fields. In this quantum setting, the matching condition turns out to be an
  instance of  the PCT theorem. The generators $Q_\eps$ nontrivially transform the charged matter field
(dressed Dirac field) in the bulk by a complex phase. This is
possible because the string-localization of the charged field extends to infinity.
The Maxwell field in the bulk is invariant (by locality). While
the generators $Q_\eps$ preserve the  superselection sectors, their
expectation values depend on the sector.

We have discussed how our model was motivated by 
Hilbert space positivity and locality, and how the superselection structure of
QED and the infraparticle nature of charged particles emerge.
The model thus integrates the Infrared Triangle into a much bigger ``simplex'' of
infrared aspects, connecting to most fundamental conceptual issues of Einstein
Causality, Hilbert space (i.e., the probability interpretation of
quantum theory), PCT, and the very notion of a
particle. The insights gained include the message that infraparticles
require thorough adjustments to
scattering theory as compared to LSZ theory (which assumes Wigner particles
and their free equations of motion), see \sref{s:scatt}. They also bring back to memory the 
No-Go theorems of the 1980's showing that algebraic principles 
of QFT do not allow the charged field of QED to be a point-localized
Wightman field.

\section{The old two-dimensional model}
\label{s:2D}

As a preparation, we begin with a look back at the construction of a kinematically much
simpler (string-less) two-dimensional model for infraparticle fields, due to
one of the authors \cite{Sch1}. Deviating from the original literature, we present the
model in a way that will easily generalize to 4D dressing
transformation for QED, cf.\ \sref{s:4Dmodel}. 

This exactly solvable 2D model is obtained by taking
the massless limit $m\rightarrow0$ of a scalar field $\varphi$ in $d=1+1$
dimensions whose derivative couples to a spinor current
\cite{Sch1}. 
The dressing density 
\bea{L-Sch}
L_{\rm dress}  =g\,j^{\mu} \pa_{\mu}\varphi = g\,\pa_{\mu}\big(\varphi j^\mu\big) 
\eea
gives rise to the equations of motion
\bea{eom-Sch}(i\gamma^{\mu}(\pa_{\mu
}-ig\pa_{\mu}\varphi)-M)\psi=0,\qquad(\square+m^{2})\varphi=0.
\eea
The exact solution is the dressed spinor field  
\bea{psi-Sch}\psi(x) =\wick{e^{ig\varphi(x)}}\psi_{0}(x),\eea
where $\psi_0$ is the free spinor field, and $g\in\RR$ the coupling constant. It lives for $m>0$ in a tensor product space of the
two particles of mass $M$ (spinor) and $m$ (scalar).

The dressed field belongs to the free Borchers class, i.e., it is
relatively local w.r.t.\ the free scalar and spinor fields
\cite{Bor}, cf.\ \sref{s:ptI}. This implies that the model leads to a trivial S-matrix $S=1$.
This is expected because $L_{\rm dress}$ is a total derivative. It can also be seen by applying the large-time LSZ scattering theory to $\psi$
in which case one finds $\psi_{\rm in}=\psi_{\rm out}=\psi_{0}$.

The non-perturbative construction of the massless limit proceeds as follows. 
One first defines normal-ordered Weyl operators of the massive field smeared with
real test functions $g(x)$
$$\wick{e^{i\varphi(g)}} =
\frac{e^{i\varphi(g)}}{\erw{e^{i\varphi(g)}}},$$
where $\erw{e^{i\varphi(g)}}=e^{-\frac12 \erw{\varphi(g)\varphi(g)}}$. By the
Weyl formula, their correlation functions are
\bea{weyl}
\erw{\wick{e^{i\varphi(g_1)}}\dots \wick{e^{i\varphi(g_n)}}} =
\prod_{i<j} e^{-\erw{\varphi(g_i)\varphi(g_j)}},
\eea
where (with $d\mu_m(k)$ the properly normalized Lorentz invariant measure on the mass-shell)
$$\erw{\varphi(x)\varphi(x')}\equiv w_m(x-x')=\int d\mu_m(k) \,
e^{-ik(x-x')}.$$
In the massless limit, $\lim_{m\to0}w_m$ diverges logarithmically at
$k=0$. The fact that $\lim_{m\to0}w_m(g)$ exists only for test functions with $\wh
g(0)=0$ means that the massless free field does not exist, but its
derivative $\pa_\mu\varphi$ is well-defined.

In order to extend the two-point function to unrestricted test
functions $g$, the Fourier transform has to be extended to $k=0$. This
is done by writing $e^{-ikx}=(e^{-ikx}-v(k)) + v(k)$ where $v$ is any
test function with $v(0)=1$, and accordingly splitting
$$w_m(x)=w_{m,v}(x)+d_{m,v}$$ into a renormalized part
$w_{m,v}(x)$ and an $x$-independent part $d_{m,v}$. The renormalized part has a finite limit
(as a distribution)
\bea{logx}w_v(x)=\lim_{m\to0}\int d\mu_m(k) \, \big(e^{-ikx}-v(k)\big)  =
-\frac1{4\pi}\log (-\mu_v^2x_-^2),
\eea
where $x_-^2 = x^2 -i\eps x^0$ in the distributional sense, and the
scale parameter $\mu_v$ depends on the regulator function $v(k)$. The
$x$-independent part diverges like $d_{m,v}\sim\log
\frac{\mu_v^2}{m^2}$. 
With test functions, 
$$w_m(g_1,g_2)=w_{m,v}(g_1,g_2)+\wh g_1(0)\wh g_2(0)\cdot d_{m,v},$$
where $\wh g_i(0)= \int d^4x \, g_i(x)$ will play the role of a ``charge'',
see below. The regularized two-point function $w_{m,v}$ defines a (non-positive)
linear functional $\omega_{m,v}(e^{i\varphi(g)})=e^{-\frac12 w_{m,v}(g,g)}$ on the Weyl algebra.
Then, we define the renormalized exponential field:
$$\wickmv{e^{i\varphi(g)}}\,:= \frac{e^{i\varphi(g)}}{\omega_{m,v}(e^{i\varphi(g)})}=e^{-\frac{\wh g(0)^2}2 d_{m,v}}\cdot
\wick{e^{i\varphi(g)}}.$$
When all commutator contributions from
the Weyl product formula and renormalization factors are collected, their
correlation functions exhibit an overall factor $e^{-(\sum_i \wh g_i(0))^2d_{m,v}}$.
{\em The crucial effect of the logarithmic divergence of $d_{m,v}$ is that it enforces a superselection
rule:} In the limit $m\to0$, $d_{m,v}$ diverges to $+\infty$, hence this prefactor becomes zero and the correlation function
vanishes unless $Q:=\sum_i\wh g_i(0)=0$:
$$e^{-(\sum_i \wh g_i(0))^2d_{m,v}}\to \delta_{Q ,0}= \left\{\ba{cl} 1
  & \hbox{if}\,\, Q=0, \\ 0 &
  \hbox{if}\,\, Q\neq0.\ea\right.$$
Thus, $q_i:=\wh g_i(0)$ is a conserved charge whose total $Q=\sum_iq_i$ is superselected.

The state defined by this limit satisfies the positivity requirement
because the vacuum state on the massive Weyl algebra underlying
\eref{weyl} is positive, and the limit preserves
positivity. The finite massless correlation functions of $\wickv{e^{i\varphi(g)}}=
\lim_{m\to0}\wickmv{e^{i\varphi(g)}}$ are
\bea{corr}\erw{\wickv{e^{i\varphi(g_1)}}\dots \wickv{e^{i\varphi(g_n)}}}
= \delta_{\sum_i \wh g_i(0),0}\cdot \prod_{i<j} e^{-w_{v}(g_i,g_j)}.
\eea
The GNS Hilbert space of the state defined by \eref{corr} is
\bea{Hbos}\HH^{\rm bos}=\bigoplus_{q\in\RR}\HH^{\rm
  bos}_q,\eea
where $\HH^{\rm bos}_q$ is densely spanned by
$\wickv{e^{i\varphi(g)}}\Omega$ with $\wh g(0)=q$. It holds
$$\wickv{e^{i\varphi(g')}}\HH^{\rm bos}_{q}\subset \HH^{\rm
  bos}_{q+\wh g'(0)}.$$
Each subspace $\HH^{\rm bos}_{q}$ carries a representation of the
IR-regular ``neutral'' Weyl subalgebra generated by $W(g_0)=e^{i\varphi(g_0)}$ with
$\wh g_0(0)=0$, from which the derivative field $\pa_\mu\varphi$ can be
obtained by variation w.r.t.\ $g_0=-\pa_\mu g_0^\mu$. This representation differs from the vacuum
representation by the automorphism $$W(g_0)\to
e^{[\varphi(g),\varphi(g_0)]}\cdot W(g_0),$$
where the (imaginary) commutator is IR-finite thanks to $\wh g_0(0)=0$.

One is tempted to think of 
$\wickv{e^{i\varphi(g)}}\Omega\in \HH^{\rm bos}_q$ as ``coherent states''
of the massless Bose particles of the model. But  if
$q=\wh g(0)\neq0$, these states do not belong to the boson 
Fock space. Namely, the scalar product with the Fock vacuum is the one-point function of the
exponential field, which is zero by the charge superselection;
and the same is true for the scalar
product with all vectors in the Fock space generated from the vacuum
by $\pa_\mu\varphi$.
The representation on $\HH^{\rm bos}_q$ with $q\neq 0$ is inequivalent
to the vacuum representation. Similarly, any two states
with different charge $q\in\RR$ define mutually inequivalent 
representations of the neutral Weyl subalgebra and the massless field
$\pa_\mu\varphi$.

Because the direct sum \eref{Hbos} is uncountable, $\HH^{\rm bos}$ is a non-separable Hilbert
space, while each $\HH^{\rm bos}_q$ is a separable representation space of the
massless field $\partial_\mu\varphi$.
This illustrates how the infrared divergence of the massless scalar
field  infers a very rich representation theory for its derivative.

In order to turn to the dressed spinor field \eref{psi-Sch}, we consider
the operator $\wickv{e^{i\varphi(g)}}$ when the test function
$g(x)$ is a multiple of a $\delta$ function. At the same time, we can
get rid of the $v$-dependent mass parameter $\mu_v$ in \eref{logx} by
a normalization factor without physical relevance. For
$q\in\RR$, it will be convenient to write 
 $$V_q(x):= \mu_v{}^{\frac{q^2}{4\pi}}\lim_{g(x)\to q\delta_x}
 \wickv{e^{i\varphi(g)}} \quad (\hbox{intuitively}\,=\mu_v{}^{\frac{q^2}{4\pi}}\wickv{e^{iq\varphi(x)}})$$
 and refer to these objects as  ``vertex operators''. Their
 correlation functions are 
\bea{vert}\erw{V_{q_1}(x_1)\dots V_{q_n}(x_n)}
=\delta_{\sum_iq_i,0}\cdot \prod_{i<j}(-(x_i-x_j)^2_-)^{\frac{q_iq_j}{4\pi}}.
\eea
Because $(-(x_i-x_j)^2_-)^\alpha$ are well-defined distributions, the
vertex operators are well-defined as (conformally
covariant with scale dimension $\frac{q^2}{4\pi}$) quantum fields.  
Then the dressed spinor field \eref{psi-Sch} is defined by fixing a
value $q>0$ and putting
\bea{psidress}\psi_q(x):= V_q(x)\otimes \psi_0(x), \quad \ol{\psi_q}(x) = V_{-q}(x)\otimes\ol\psi_0(x).\eea
It is defined on the Hilbert space
\bea{H0dress} \HH^{\rm dress} = \bigoplus_{n\in\ZZ} \HH_{nq}^{\rm bos}
\otimes \HH_n^{\rm fer},
\eea
where $\HH_{nq}^{\rm bos}\subset \HH^{\rm bos}$, and $\HH_n^{\rm fer}$
is the subspace of the Fermi Fock space $\HH^{\rm fer}$ of
Dirac charge $n$. (The cyclic subspace of $\psi_q$ may be a proper
subspace of \eref{H0dress} because of the diagonal restriction of the
test functions when \eref{psidress} is smeared in $x$.) In particular, $\HH^{\rm dress}\subset \HH^{\rm  bos}\otimes \HH^{\rm fer}$, but the inclusion is such that the
simple tensor product structure is lost. This accounts for the fact
that the tensor factors in \eref{psidress} are not independent fields
of the model: the boson clouds cannot be physically separated
from the charged particles.

The correlation functions of the dressed spinor field 
are just products of \eref{vert} with the free spinor correlation functions.

Because the state \eref{corr} is translation invariant, $\HH^{\rm bos}$ carries a unitary
representation of the translations. By the analytic structure of the
power law correlation functions \eref{vert}, it has positive energy
and the
spectral density in charged states is an inverse power of the mass
operator.

The spectrum of the translations of the dressed spinor field is the sum of
the spinor and boson spectra. Consequently, the states
created by the dressed spinor field have a sharp lower mass limit $M$ (the mass
of the free spinor field), and the spectral density of the mass
operator continuously reaches down to $M$. In fact, the mass operator
has no normalizable eigenstate with eigenvalue $M^2$, because there
are no states of energy zero in
$\HH_n^{\rm bos}$ ($n\neq 0$).

The model therefore describes infraparticles: massive particles for
which the mass operator has a sharp lower bound but no eigenstates: the
mass-shell has spectral weight zero. The infraparticle must not be viewed as an ordinary
particle plus a cloud. Instead, it forms an entity that cannot be
split up.

The analytic simplicity of the correlation functions permits an explicit
calculation of their Fourier transform in the form of convolution
integrals, to study the amount of dissolution of the mass-shell near the lower bound of
$p^{2}\geq M^{2}$. The large-time fall-off of the two-point function is increased from $t^{-1}$ for the free spinor
field to
$t^{-1-q^{2}/4\pi}$. This in turn implies that the LSZ scattering
theory cannot be applied to theories with infraparticles.

\section{Four dimensions: The dressed model}
\label{s:4Dmodel}

\subsection{The QED dressing transformation}
It remained unclear for a long time how infraparticle fields
can be realized in dynamical models in $3+1$ dimensions. The main
obstacle is the non-existence of local fields of scaling dimension $0$
in $3+1$ dimensions, to play the role of (the massive approximants of) $\varphi$
in the  $1+1$-dimensional model of \sref{s:2D}. An
interesting model was presented in \cite{BDMRS} with a ``Maxwell field of
helicity zero''. By defining the superselection sectors of this model as algebraic
automorphisms, the problem of their implementation by charged fields
was circumvented.

The new idea is to construct fields of scaling dimension $0$ in $3+1$ dimensions as string integrals over
free fields of dimension 1.  In the case of QED, the escort field \eref{phi} plays the role of the logarithmically divergent massless scalar field
of the two-dimensional model.\footnote{A slightly simpler model (without much physical motivation) would be the
string integral $(I_e\varphi)(x)$ over the massless scalar field
$\varphi$. Its treatment (and ensuing IR features) would be almost identical to the escort field,
except for the absence of Lorentz indices and contractions with the
string direction $e$. }

\subsubsection{Basic idea}
\label{s:idea}
The equation of motion of the Dirac field with interaction $L_{\rm
  dress}(e) = q\,
\pa_\mu \phi(e)j^\mu$,
$$(i\gamma^\mu(\pa_\mu -iq\pa_\mu\phi(e))\psi(x)=M\psi(x),$$
has the classical ``pure gauge'' solution
\bea{psiq-cl}\psi_{q,e}(x) = e^{iq\phi(x,e)}\psi_0(x).\eea
Notice that by definition, $A(e)=I_e(Fe)$ is gauge invariant. By
$A=A(e)-\pa \phi$, a gauge transformation 
amounts to $\phi\to \phi-\lambda$, and $e^{iq\phi}\psi$ is gauge
invariant. Dirac \cite{Di} (and others \cite{Ma,St}) suggested the same formula with non-local
expressions in $A$ in the exponent, subject to a source condition. Dirac characterized the difference
between general solutions to this condition and a special one
(corresponding to $\phi(x,e)$ smeared with $c_0(e)$, see below); he did not address 
the question how to define \eref{psiq-cl} as a quantum field, and in
particular not issues of localization, that are prominent in our approach.

To define \eref{psiq-cl} as a quantum field on a Hilbert
space, will require the choice of a Lorentz frame (a timelike unit
vector $u\in H_1^+$, where $H_1^+$ is the unit forward
hyperboloid). It also requires $\phi(e)$ to be smoothly smeared over string
directions $e\in H_1\cap u^\perp$, where $H_1$ is the hyperboloid of spacelike
unit vectors, with a real smearing function $c$. The latter is
required to have unit total weight $\int d\sigma(e)c(e)=1$, in order
that $A_\mu(x,c)$ still is a
potential for $F_{\mu\nu}$.

A special case may be illustrative.
We shall denote by $u_0\equiv\bpm 1\\[-1mm]\vec 0\epm$ the standard
time unit vector, and by $c_0(\vec e)=\frac1{4\pi}$ the constant
smearing function on the sphere $H_1\cap u_0^\perp=S^2$. Then, by \eref{phi0}, 
$\phi(c_0) =\Delta\inv \Skp\nabla {A^K}$. Its gradient is 
    the familiar longitudinal vector potential. Consequently, if
    $\rho(\vec x)$ is a (classical) static charge density distribution
    with Coulomb electric potential $\phi_{\rm Cb}(\vec x)=-\Delta\inv\rho(\vec
  x)$, one may rewrite the time-zero Weyl operator $e^{i\int d^3x\, \rho(\vec x)\phi(\vec
  x,c_0)}$ as $$e^{-i\int d^3x\, \phi_{\rm Cb}(\vec
  x)\,\Skp\nabla{A^K(\vec x)}}.$$ This operator 
creates a ``coherent purely longitudinal photon state'' with charge
distribution $\rho$, (cf.\ \cite{Kay}). It coincides with Dirac's
special solution \cite{Di}. As emphasized in \cite{BCRV},
the longitudinality is essential for this purpose\footnote{The
    fact that $A(e)=I_e(Fe)$ has only physical degrees of
    freedom,  was taken as an argument against the idea to couple the
    QED Dirac field via $A(e)$, see \sref{s:comments}. The present hybrid setting exhibits that the longitudinal degrees of
  freedom reside in the escort field that couples to the charged field. See also \cite{MRS2}.}. However, the richness of our model lies in its admitting
    arbitrary smearing functions $c(\vec e)$ of unit total weight,
    generating more general photon clouds without rotational symmetry,
    and hence with additional transverse degrees of freedom, 
    see \sref{s:expFu}, and allowing more flexibility of localization.

Here is what will be achieved:
Because the Dirac current is
conserved, $L_{\rm dress}(c)$ is a total derivative and not expected
to generate a nontrivial S-matrix. But the dressed 
Dirac field exhibits the desired kinematical features: It creates infraparticle
states with superselected photon clouds labelled by the smearing
function $c(e)$. It inherits the
string localization of the escort field, so as to resolve the conflict
with Gauss' Law discussed in the Introduction.
The photon clouds dissolve the mass-shell of the
electron with a sharp lower bound. They break the Lorentz invariance
of each sector, as expected \cite{FMS}, but Lorentz invariance of the
dressed Dirac field can be restored in a reducible Hilbert space
representation, \sref{s:Lor}. They 
cause expectation values of the Maxwell field to decay asymptotically in spacelike directions like $r^{-2}$,
so as to allow finite flux distributions over the asymptotic sphere,
see \sref{s:expFu}. The detailed features of asymptotic expectation
values in lightlike directions fulfill the assumptions expressed in
\cite{Str} (and the positive electron mass
poses no difficulties), \sref{s:ll}.

The actual dynamical interaction of QED comes with the second
  step, as outlined in Item 4 in \sref{s:roadmap}. This will be discussed
in \sref{s:ptII}.

\subsubsection{Non-perturbative construction}
\label{s:dressD}

We give a  non-perturbative construction of the ``dressing factor''
$e^{iq\phi(x,c)}$, where $c$ is a real smearing function for the
string direction. In \sref{s:expFu}, we shall see that the smearing function
$c$ determines the ``shape
of the photon cloud''. We
proceed as in \sref{s:2D} via a massless limit. 

The two-point function of the massive Feynman gauge potential is given
by \eref{eq:AKAK} with $\mu_0(k)$ replaced by $\mu_m(k)$. 
Viewed as an inner product, it is indefinite, but it is positive
definite when restricted to the spacelike components
$\mu,\nu=1,2,3$ in some Lorentz frame\footnote{\label{fn:lor}That this
  convention breaks Lorentz invariance, is unproblematic because
  photon clouds break Lorentz invariance anyway \cite{FMS}, see also
  \sref{s:expFu}.
  In \sref{s:Lor}, we shall restore Lorentz covariance in
  a reducible representation.}, which we choose for
definiteness  to be given by the standard future time unit
vector $u_0\in H_1^+$.

The massive escort field is defined by the same expression
\eref{phi} in terms of the massive Krein potential. It is positive
definite when restricted to string directions $e\in H_1\cap u_0^\perp$ of the form
$e=\bpm0\\[-1mm]\vec e\epm$ so that $-(ee')=\Skp e{e\,'}$ is Euclidean. We denote its two-point function by $w_m(x-x';e,e')$, given by
\eref{phiphi} with $\mu_0(k)$ replaced by $\mu_m(k)$. For $m>0$, this
is a well-defined distribution, but it diverges logarithmically at
$k=0$ in the limit $m\to0$.

As in \sref{s:2D}, we choose a real (and rotation symmetric) regulator test function $v(k)>0$ with $v(0)=1$. The split
$e^{-ikx}=(e^{-ikx}-v(k))+ v(k)$ splits
\bea{wm}
w_m(x;e,e')=w_{m,v}(x;e,e') +
d_{m,v}(e,e'),\eea
where by \eref{phiphi}
\bea{dmv} d_{m,v}(e,e')  = -(ee')\int \frac{d\mu_m(k) \,v(k)}{((ke)-i\eps)((ke')+i\eps)}.\eea
$w_{m,v}(x,e,e')$ has a massless
limit $w_v(x,e,e')$ because $e^{-ik(x-x')}-v(k)$ vanishes at $k=0$,
while $d_{m,v}(e,e')$ diverges as $m\to0$. An explicit expression for
$w_v(x,e,e')$ can be found in \eref{a:phiphi} and \eref{H}.

The subsequent analysis is largely parallel to
the two-dimensional case of \sref{s:2D}. Details can be found in
\aref{b:constr} and \aref{a:vertex}. The normal-ordering w.r.t.\ the non-positive functional defined by
$w_{m,v}(x,e,e')$ differs from the normal-ordering 
w.r.t.\ the massive vacuum state by a divergent factor. For $g\in
\mathcal{S}(\RR^4)$ and $c\in
C_\RR^\infty(S^2)$ we define
\bea{Wv} \wickv{e^{i\phi(g\otimes c)}}\, := \lim_{m\to0} \wickmv{e^{i\phi(g\otimes c)}}   = \lim_{m\to0}e^{-\frac12
  \wh g(0)^2d_{m,v}(c,c)}\cdot \wick{e^{i\phi(g\otimes c)}}.\eea
The correlations involve a factor $e^{-\frac 12 d_{m,v}(C,C)}$,
where $d_{m,v}(C,C)$ is \eref{dmv} smeared with $C(\vec
e)=\sum_i \wh g_i(0)c_i(\vec e)$. This factor converges to $\delta_{C,0}$ because $d_{m,v}(C,C)$ 
diverges to $+\infty$ unless $C=0$.
The main difference is that the resulting superselected quantity is not just the
total charge $g=\sum_i\wh g_i(0)\in\RR$ as in 2D, but the weighted sum over the
string smearing functions
$C\in C_\RR^\infty(S^2)$ :
\bea{corr4}\Erw{\wickv{e^{i\phi(g_1\otimes c_1)}}\dots
  \wickv{e^{i\phi(g_n\otimes c_n)}}}
= \delta_{C,0}\cdot \prod_{i<j} e^{-w_{v}(g_i\otimes c_i,g_j\otimes c_j)}.
\eea
Thus, the superselection
structure remains uncountable even when $\wh g_i(0)$ will later be restricted to discrete values of
electric charges $\pm q$.

Vertex operators $V_{qc}(x)$ are obtained by admitting
$g_x(\cdot)=q\delta_x(\cdot)=q\delta(\cdot -x)$ in \eref{corr4}, and
an irrelevant finite $c$-dependent rescaling \eref{a:V} for convenience. Their correlation functions are
\bea{VVV}
\Erw{V_{q_1c_1}(x_1)\dots V_{q_nc_n}(x_n)} = \delta_{\sum_i q_ic_i,0}\cdot
\prod_{i<j}
\Big(\frac{-1}{(x_i-x_j)^2_-}\Big)^{-\frac{q_iq_j}{8\pi^2}\erw{c_i,c_j}}
e^{-\frac{q_iq_j}{4\pi^2}\wt H(x_i-x_j;c_i,c_j)},\quad \eea
where
$\erw{c_i,c_j}$ is a quadratic form on real smearing functions
in $C_\RR^\infty(S^2)$, see \eref{cc}, 
and $\wt H(x;e,e')$ is a
Lorentz-invariant and in all three variables separately homogeneous 
distribution, see \eref{H}. This concludes the non-perturbative construction of 
vertex operators $V_{qc}(x)$ in four dimensions.

Vertex operators $V_{qc}(x)$ are string-localized in
the spacelike cones $x+\bigcup_{e\in \supp(c)}\RR_+e$. In general,
they commute up to a phase
\bea{any}
V_{qc}(x)V_{q',c'}(x') = e^{iqq'\beta(x-x';c,c')}\cdot
V_{q',c'}(x')V_{qc}(x),\eea
where $\beta(x-x';c,c')$ is given by 
$$\beta(x-x';e,e') = -i[\phi(x,e),\phi(x',e')] = (ee')
(I_{-e'} I_eC_0)(x-x'),$$
smeared in $e$ and $e'$.  The commutator function $\beta(x-x';e,e')$ does not
suffer from the IR divergence because the Fourier transform of $C_0$
vanishes at $k=0$. It is a rather simple geometric quantity
that can be characterized in terms the intersection of the wedge $x+\RR_+e-x'-\RR_+e'$ with the
null cone \cite{R2,GRT}. In particular, whenever the (smeared)
strings $x_i+\bigcup_{e\in\supp(c_i)} \RR_+e$ are spacelike separated,
the phase is zero and the fields commute.\footnote{It was noted that
  braid group statistics does not exist \cite{Ha} in DHR sectors in
  four spacetime  dimensions. Indeed, the present commutativity up to a phase are
  not an instance of braid group statistics in the sense of DHR
  because a statistics operator cannot be defined. Namely, by lack of Lorentz
  covariance  of the infrared superselection sectors, one cannot rotate strings into spacelike separated positions.} 

Finally, the dressed Dirac field \eref{psiqc} is defined as
\bea{psiq-np}\psi_{qc}(x):= V_{qc}(x)\otimes \psi_0(x), \qquad \ol{\psi_{qc}}(x) :=
V_{qc}(x)^*\otimes\ol\psi_0(x),\eea
where $q$ is the unit of electric charge and the real string smearing $c\in
C_\RR^\infty(S^2)$ has unit weight $\int d\sigma(\vec e) c(\vec
e)=1$. The adjoint vertex operators are
$V_{qc}(x)^*=V_{-qc}(x)$.
This concludes the non-perturbative construction of 
the dressed Dirac field $\psi_{qc}(x)$.

Its correlation functions are products of free Dirac and vertex
operator correlations.  The power law decrease of the vertex operator correlations on top of
the canonical decrease of the free Dirac correlations, and details of the exponentiated
direction-dependent function $\wt H(x;c,c')$ are expected to become
important in the future scattering theory of  the dressed Dirac field,
cf.\ \sref{s:scatt}.

The commutation relations \eref{any}
modify those of the Dirac field, so that the dressed Dirac field
remains anti-local whenever the strings are spacelike separated.

One would like to know the Fourier transform of two-point function of
vertex operators, because it describes the dissolution of the
mass-shell in terms of the $c$-dependent
energy-momentum distribution of the
photon cloud created by the vertex operator. It is supported in
the interior of $V^+$ and has to be added to the
mass-shell energy-momentum of the free Dirac particle. The sum is the
spectrum of the infraparticle state generated by the dressed Dirac field.

This Fourier transform is impossible to compute for general
  $c$. However, because the two-point function drastically simplifies
  for the constant smearing function $c=c_0$, one can quantify
the ensuing dissolution of the mass-shell in this special case.

In \aref{a:vertex} we display the two-point function
  \bea{V0V0}
\Erw{V_{qc_0}(x_1)V_{-qc_0}(x_2)} =
\Bigg[\frac
{\Big(\frac{x^0-r-i\eps}{x^0+r-i\eps}\Big)^{\frac{x^0}{
      r}}}{- (x^2)_-}\Bigg]^ {\frac{q^2}{8\pi^2}} \qquad (x=x_1-x_2,
\; r = \vv x).
\eea
of the vertex
operator with constant smearing. It can be written as
$(x^0-i\eps)^{-\frac\alpha\pi}$ (where $\alpha=\frac{q^2}{4\pi}$ is
the fine structure constant) times a power
  series in $\frac{r^2}{(x^0)^2}$. This structure allows to extract
  quantitative details of its Fourier transform, and
  hence of its rotationally invariant energy-momentum distribution
  $\rho(\omega,\vec k)$ in the interior of $V^+$ \cite{R2}. By putting $r=0$, one concludes that the distribution $\rho(\omega)=\int
  d^3k\,\rho(\omega,\vec k)$ of energies 
  decays like $\omega^{\frac\alpha\pi-1}$. By applying powers of the Laplacian 
  before putting $r=0$, one can compute averages of powers of $\vv k^2$ at fixed
  energy $\omega$. E.g., the average of the invariant masses 
  $\omega^2-\vv k^2$ at given energy $\omega$ is found to be 
  $\frac\alpha\pi\cdot\omega^2 + O(\alpha^2)$ with variance
  $\frac49\frac\alpha\pi\cdot\omega^4 + O(\alpha^2)$. These data are
  roughly consistent with a power law behaviour $\rho(\omega,k)\sim (\omega^2-\vv
  k^2)^{\frac\alpha\pi-1}$ for the dissolution of the mass-shell at fixed energy $\omega$, see Fig.\ 1. 
\begin{figure}
  $$\includegraphics[width=50mm]{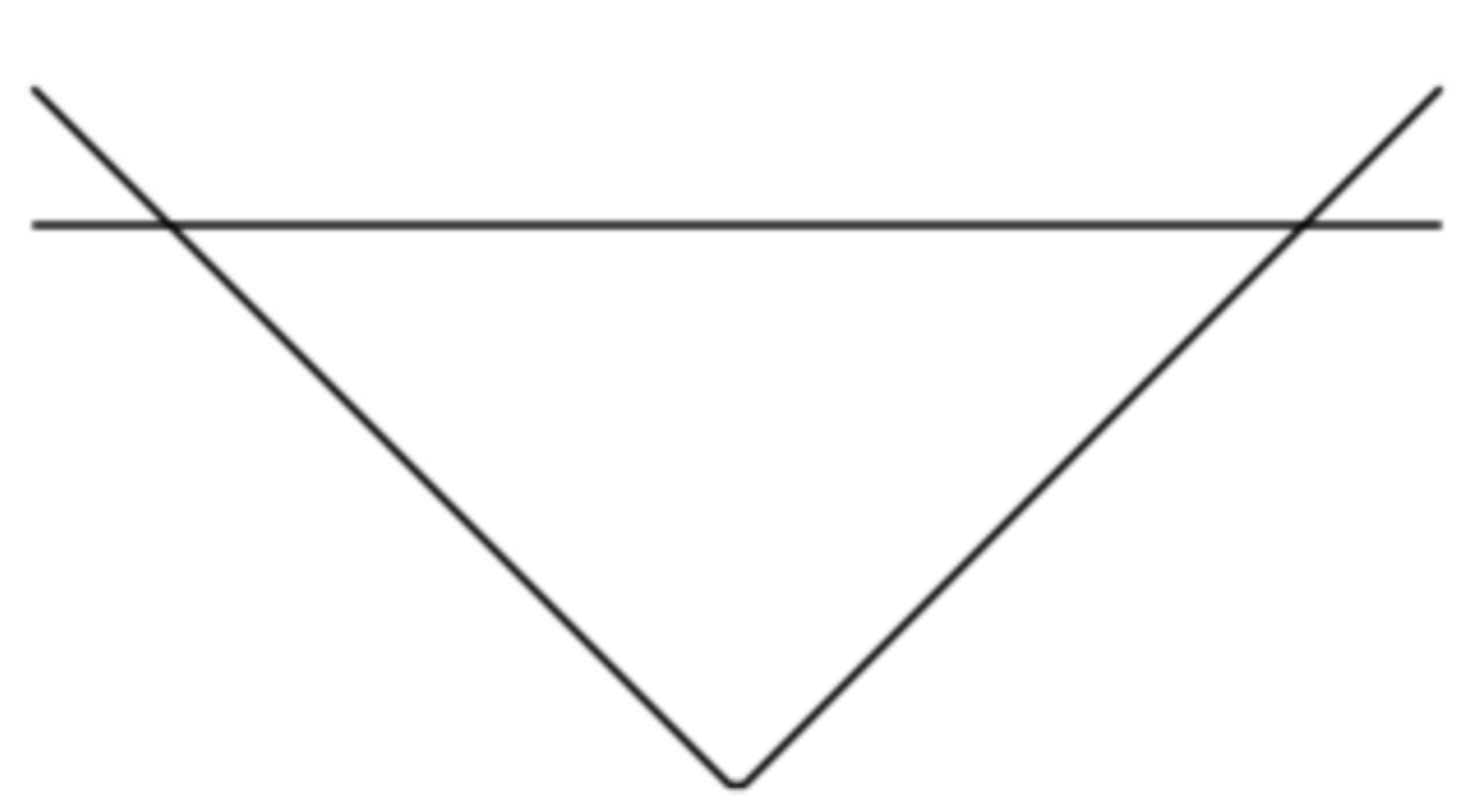} \qquad \includegraphics[width=70mm]{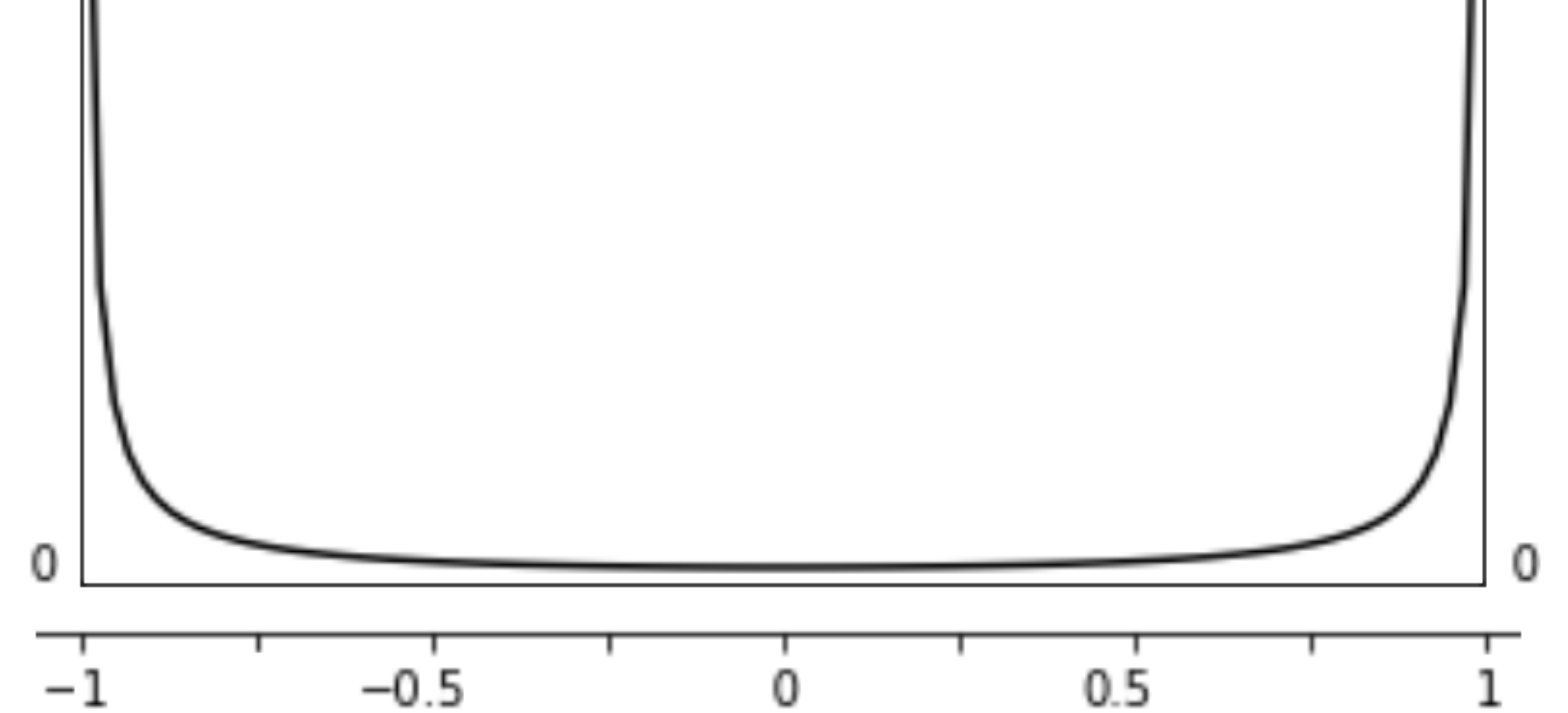} $$
\caption{\small Left: Cut at fixed energy through the momentum space forward
lightcone. Right: The Fourier transform $\rho$ of the two-point
function as a function of momentum (i.e., along the cut) seems to
exhibit an inverse power law peak at the maximum value $\frac{\vv
  k}\omega=1$. (Qualitative)}
\end{figure}

\subsection{Hilbert space and Lorentz invariance}
\label{s:Lor}

\subsubsection{Fixed Lorentz frame}

Until now, we were working in a fixed Lorentz frame, distinguished by
the time unit vector $u_0\in H^+_1$ and restriction of the support of $C(e)$ to $H_1\cap
  u_0^\perp$. We shall now exhibit the structure of the Hilbert space
  in the fixed Lorentz frame, and then restore Lorentz invariance. 

The massless limit \eref{corr4} of the correlation functions defines a state whose GNS
Hilbert space is of the highly nonseparable form
\bea{HGNS}
\HH_{u_0}^{\rm bos}=\bigoplus_{C\in C_\RR^\infty(S^2)}\HH^{\rm bos}_{u_0,C},
\eea
due to the superselection rule $C=0$ in \eref{corr4}. The operators
$\wickv{e^{i\phi(g\otimes c)}}$ in \eref{Wv}
take $\HH^{\rm bos}_{u_0,C}$ to $\HH^{\rm bos}_{u_0,C+\wh g(0)c}$. In particular, each sector
$\HH^{\rm bos}_{u_0,C}$ carries a representation of the IR-regular
Weyl subalgebra generated by
$W(g_0,c')=e^{i\phi(g_0,c')}$
with $\wh g_0(0)=0$, whose generators are $\partial_\mu
\phi(x,e)$. These representations are obtained from the vacuum
representation by automorphisms
$$W(g_0,c')\mapsto e^{[\phi(g\otimes c),\phi(g_0,c')]}\cdot W(g_0,c') \qquad
(\wh g(0)c=C),$$
where the commutator is an IR-finite imaginary number because  $\wh g_0(0)=0$.

The quantum dressed Dirac field \eref{psiq-np} is defined on the
Hilbert space
\bea{H0dressD} \HH_{u_0}^{\rm dress}= \bigoplus_{n\in \ZZ} \HH^{\rm bos}_{u_0,n}\otimes \HH^{\rm Dirac}_n,\eea
where $\HH^{\rm Dirac}_n$ is the subspace of $\HH^{\rm Dirac}$ of
Dirac charge $n$, and $\HH^{\rm bos}_{u_0,n}$ is the subspace of $\HH_{u_0}^{\rm
  bos}$ spanned by $\HH_{u_0,C}^{\rm
  bos}$ with $C$ of total weight $n$. The cyclic subspace
of $\psi_{qc}$ with a fixed $c$ will be some very small subspace (only
$C=nc$ occur) of $\HH_{u_0}^{\rm dress}\subset \HH_{u_0}^{\rm
  bos}\otimes \HH^{\rm Dirac}$. 

\subsubsection{Lorentz transformations}

Because of the restriction $e\in H_1\cap u_0^\perp$ on the string directions, $\HH^{\rm bos}_{u_0}$ is not Lorentz invariant.
But one may repeat the same construction in any other
Lorentz frame given by a timelike unit vector $u\in H_1^+$, with strings $e$
restricted to $H_1\cap u^\perp$. This gives rise to GNS Hilbert spaces
$\HH_{u}^{\rm dress}$. 
In the Krein space, all massive fields
$\wickmv{e^{i\phi(g\otimes c)}}$ with $c$ smooth smearing functions in $H_1\cap u^\perp$,
are simultaneously defined for all $u\in H_1^+$. For different $u\neq u'$, the inner
products $\erw {\wickmv{e^{i\phi(g\otimes
      c)}}\wickmv{e^{-i\phi(g'\otimes c')}}}$ 
where $c$ and $c'$ have equal total weight and $\wh g(0)=\wh{g}'(0)=q\neq
0$, vanish in the massless limit. The only exception is $c=c'=0$ which
corresponds to the vacuum representation in either frame. The proof of this remarkable
claim can be found in \aref{b:orth}.

Therefore, the charged sectors in \eref{H0dressD} of equal Dirac charge, but constructed in different
Lorentz frames are mutually orthogonal. Subspaces of different
electric charge
are orthogonal because of the conservation of
the Dirac charge. Therefore, $\HH^{\rm bos}_{u,n}\otimes \HH^{\rm
  Dirac}_n$ are mutually orthogonal for different $(u,n)$ with the
exception of the common vacuum subspace $\HH^{\rm bos}_{u,C=0}\otimes \HH^{\rm
  Dirac}_0\subset \HH^{\rm bos}_{u,0}\otimes \HH^{\rm Dirac}_0$. 
Let 
\bea{dressD-L} \HH^{\rm dress}:={\bigoplus_{u\in H_1^+}}^{\!\!*}\HH_{u}^{\rm dress}= {\bigoplus_{u\in H_1^+}}^{\!\!*}\Big(\bigoplus_{n\in \ZZ} \HH^{\rm bos}_{u,n}\otimes \HH^{\rm Dirac}_n\Big),\eea
where the
notation $\bigoplus_{u\in H_1^+}^{*}$ indicates the identification
of the vacuum subspaces ($C=0$) for all $u$. Lorentz transformations
act on $\HH^{\rm dress}$ by the Wigner
representation on the Dirac factor, and by 
$$U(\Lambda) \wickv{e^{iq\phi(g\otimes c)}}\Omega =
\wick{e^{iq\phi(g\circ\Lambda\inv,c\circ\Lambda\inv)}}_{v\circ\Lambda\inv}\Omega,$$ so that 
$U(\Lambda)\HH^{\rm dress}_u = \HH^{\rm dress}_{\Lambda u}$. (The
change of the regularization function $v(k)$ is a unitary equivalence.)

In \sref{s:expFu}, we shall exhibit the physical reason for the
breakdown of Lorentz covariance in individual sectors, cf.\ also \sref{s:intro}. Namely, we
shall compute the expectation values of asymptotic electromagnetic
fluxes, which are non-zero in sectors with photon clouds.  The asymptotic flux operators commute
with $\partial_\mu \phi$ and are therefore multiples of $\mathrm 1$ in
each irreducible representation $\HH^{\rm bos}_{u,C}$ and coincide with their expectation
values in states from $\HH^{\rm bos}_{u,C}$. But Lorentz transformations
transform the asymptotic flux operators in a nontrivial way, unless
they are zero. Hence
they cannot exist as unitary operators in a single irreducible
sector. This abstract argument for the broken Lorentz invariance in
irreducible representations of QED observables is long known
\cite{FMS}, but it is here for the
first time realized in an exactly solvable model.

\subsection{The dressed Maxwell-Dirac model}
\label{s:dressMD}

The Maxwell field cannot be expressed in terms of
the escort field $\phi(x,e)$, so it is not defined in $\HH^{\rm
  bos}$.
We need to add the free Maxwell field to the dressed Dirac model. 

\smallskip

\subsubsection{Positivity problem}
\label{s:pos}

The immediate problem is that (for $m>0$) $\phi(e)$ and $F_{\mu\nu}=\pa_\mu
A^K_\nu-\pa_\nu A^K_\mu$ are not defined on a common positive-definite
subspace of the Krein space $\KK$. But positivity is essential in
\sref{s:dressD} in order to
arrive at a GNS Hilbert space in the first place, and in order to
maintain the superselection structure which requires $d_{m,v}(C,C)$ to
diverge to $+\infty$ when $m\to0$. 

Let $u\in H_1^+$ be any future timelike unit vector. For
explicit formulas, we choose $u=u_0$, but the construction is the same
for any choice. 

For $e=\bpm0\\[-1mm]\vec e\epm\in H_1\cap u^\perp$, the escort field
$\phi(e)$, given by \eref{phi}, is
defined on the positive-definite subspace $\HH^u\subset \KK$, given by
\eref{eq:Hu}, that is generated by the 
spatial components of creation operators $a_i^*(k)$ of the Krein
potential $\vec A^K$. We want to exploit the positivity of this
subspace also for the other relevant fields in our construction.

This is achieved with the new potential and its field already introduced
in \sref{s:roadmap}
\bea{AuFu}A^u :=A^K + u I_u(\pa A^K) , \quad F^u:=\pa \wedge A^u= F^K
-(u\wedge \pa) I_u (\pa A^K) \eea
They are also defined\footnote{\label{fn:u}Timelike strings don't need a
smearing \cite{MdO} because the denominator $(uk)$ in \eref{a:I-mom} can never vanish except at
$k=0$ -- where the singularity is controlled by the infrared
regularization.} on the 
subspace $\HH^u$ characterized by \eref{eq:Hu} because $I_{-u}\pa_0=I_{-u}(u\pa)=1$ (see footnote \ref{fn:Ie}) and $u_\nu=\eta_{0\nu}$, hence
$$\erw{A^K_0 A^u_\nu} = -\eta_{0\nu}W_0(x-x')
-u_\nu (I_{-u}\pa_0W_0)(x-x') =0.$$
They differ only by the null field $(\pa A^K)$ from the Feynman gauge
fields. Hence, $F^u_{\mu\nu}$ (on $\HH^u$) has the
same two-point function as $F^K_{\mu\nu}$ (on the
Krein space), and the same is true for their
string-localized potentials $A^K_\mu(e)$ and $A^u_\mu(e)$. 
For the same reason, $F^u$ and $F^K$ are local fields relatively to
themselves and to each other. However, $A^u_\mu(x)$ and  $\phi(x,e)$ are
non-local relatively to $F^u_{\mu\nu}$ (because $x$ cannot be spacelike
from $y+\RR_+u$).

Apart from positivity, there is a physical reason for the definitions
\eref{AuFu} on $\HH^u$. The definition of $A^u$ replaces the
negative-definite zero component of $A^K$ by $I_u \Skp \nabla A$, so that all its
components involve only the three spacelike components of the Krein field $A^K$.
By using $I_u\partial_0=-1$ and 
$\pa_0^{-2}=\Delta\inv$,
one may write this as 
\bea{Au} A^{u}_0 = A^{K}_0 +I_u(\partial_0 A^{K0}+ \Skp\nabla {A^K}
= -\pa_0\Delta\inv\Skp\nabla {A^K},
\quad \vec A^u = \vec A^K.
\eea
This is gauge equivalent (by $\partial_\mu \Delta\inv\Skp\nabla {A^K}$) to the transverse Coulomb gauge potential
$$A^C_0=0, \quad \vec A^C = \vec A^K - \vec \nabla\Delta\inv \Skp\nabla {A^K}.$$
Consequently, $F^u$ is the physical Maxwell field \eref{eq:EB}
creating only the two physical photon states, embedded into the Krein space. 
As a consequence, the equations
\bea{dA0}\pa^\mu A^u_\mu=0, \qquad \pa^\mu F^u_{\mu\nu}=0,\eea
hold as operator identities. In contrast, the naively embedded Maxwell field
$F^K$ contains unphysical longitudinal degrees of freedom,
that are seen in its correlations with the unobservable Krein field $A^K$,
and $\pa^\mu F^K_{\mu\nu}=-\pa_\nu(\pa A^K)$ is
the fictitious current \eref{eq:fict} of
\cite{BLOT}. Accordingly, only $F^u$, when extended to the charged sectors, will
enjoy the physical transversality properties, see footnote
\ref{fn:trans} below.

\subsubsection{Extension to the physical Maxwell field}
\label{s:ext}

We shall now proceed to extend the construction of \sref{s:dressD} to
include also the physical Maxwell field $F^u$. By avoiding
the field $F^K$ altogether, which carries both physical and unphysical
degrees of freedom, our construction neatly separates the 
observable {\em field} $F^u$ from the unobservable escort field
responsible for the photon
clouds in charged {\em states} generated by the unobservable charged Dirac field. 

To extend the construction of \sref{s:dressD}, it suffices to consider the
``multi-component'' field
$$\Phi(g\otimes c\oplus f)=\phi(g\otimes c) + A^u(f)= \int dx\, \Big[g(x)\int d\sigma(\vec e) c(\vec e)\,
\phi(x,e) + f^\mu(x)\, A^u_\mu(x)\Big]$$
at mass $m$. Its two-point function
$$\erw{\phi(g\otimes c)\phi(g'\otimes c')}_m+\erw{\phi(g\otimes
  c)A^u(f')}_m+\erw{A^u(f)\phi(g'\otimes c')}_m+\erw{A^u(f)A^u(f')}_m
$$
is positive definite. Only the first term is logarithmically divergent in the limit $m\to0$,
and this term will be regularized as in \eref{wm}. Then the field
$$\wickv{e^{i\Phi(g\otimes c\oplus f)}}\;:= e^{-\frac12
  \wh g(0)^2d_{m,v}(c,c)}\cdot \wick{e^{i\Phi(g\otimes c\oplus f)}}$$
has a massless limit, whose correlation functions look exactly like \eref{corr4}
with the obvious additional factors
\bea{factors}e^{-\erw{\phi(g\otimes
    c)A^u(f')}-\erw{A^u(f)\phi(g'\otimes c')}-\erw{A^u(f)A^u(f')}}
\eea
As before, the factor $\delta_{\sum_i q_ic_i,0}$ in \eref{corr4} defines the superselection rule. 

Since $g$ and $f$ can be chosen independently,
one arrives at an algebra containing the fields $V_{qc}(x)$ and
$e^{iA^u(f)}$, represented in a larger GNS Hilbert space
\bea{H0dressM}\wt\HH_{u_0}^{\rm bos, M} = \bigoplus_{C}\wt \HH_{u_0,C},
\eea
where each $\wt\HH_{u_0,C}$ is a proper extension of $\HH^{\rm bos}_{u_0,C}$. 
Since the test functions $f^\mu$ of the potential do not contribute to the
superselection rule, one can freely vary w.r.t.\ $f^\mu(x)$. One obtains
the fields $A^u_\mu$, preserving each of the subspaces $\wt\HH_{u_0,C}$,
i.e., the latter are representation spaces of $A^u$, and consequently 
of their exterior derivatives $F^u$. Thus, each summand in \eref{H0dressM}
carries a representation of the fields $\partial_\mu\phi$ and $A_\mu^u$.

Correlations of vertex operators
with $A^u_\mu(x)$ are obtained by variation of the factors
\eref{factors} w.r.t.\ the test
functions $f^\mu(x)$. E.g.,
$$\erw{V_{qc}(y_1)\wick{e^{iA^u(f)}}V_{qc}^*(y_2)}=\erw{V_{qc}(y_1)V_{qc}^*(y_2)}\cdot
e^{-q\erw{\phi(y_1,c)A^u(f)}   + q\erw{A^u(f)\phi(y_2,c)}}$$
implies
\bea{3pt}
\erw{V_{qc}(y_1)A^u_\mu(x)V_{qc}^*(y_2)} =\erw{V_{qc}(y_1)V_{qc}^*(y_2)}\cdot
i q(\erw{\phi(y_1,c)A^u_\mu(x)}  -\erw{A^u_\mu(x)\phi(y_2,c)}).\qquad\quad
\eea
Correlations involving several $A^u$ are obtained
similarly. Correlations with $F^u(x)$ are exterior
derivatives of correlations with $A^u(x)$.

Now, the ``dressed'' Dirac field with string smearing functions in
$H_1\cap u_0^\perp$ is defined on the Hilbert space 
\bea{H0dressMD} \HH_{u_0}^{\rm dress,M}= \bigoplus_{n\in \ZZ} \wt\HH^{\rm bos}_{u_0,n}\otimes \HH^{\rm Dirac}_n,\eea
where $\HH^{\rm Dirac}_n$ is the subspace of $\HH^{\rm Dirac}$ of
Dirac charge $n$, and $\wt\HH^{\rm bos}_{u_0,n}$ is the subspace of $\wt\HH_{u_0}^{\rm
  bos}$ spanned by $\wt\HH_{u_0,C}^{\rm
  bos}$ with $C$ of total weight $n$. 

Lorentz invariance is restored in the same way as in \sref{s:Lor}, by
taking a direct sum of $\wt\HH^{\rm dress, MD}_u$ over all unit time 
vectors $u\in H_1^+$, $\supp\,c_u\subset H_1\cap u^\perp$, similar to \eref{dressD-L}.

{\em Remark:} The same construction can be done with $A^K_{\mu}$ rather
than $A^u_{\mu}$. This involves Weyl operators of $A^K(g)$ on an indefinite
space. Because the
superselection structure via the factor $\delta_{C,0}$ in
\eref{corr4} required only positive-definiteness on the escort part of
the multi-component correlations, the construction with $A^K$ gives rise to a variant of
\eref{H0dressM} with indefinite extensions of the
GNS Hilbert spaces $H^{\rm bos}_u$, where correlations like \eref{3pt},
involving $A^K$ rather than $A^u$, violate positivity. These are the arena where causal
perturbation theory can guarantee locality of the interacting observables,
whereas causal perturbation theory on \eref{H0dressM}
guarantees positivity, see Item 5 in \sref{s:roadmap}. Since the modification terms due to the
passage from $A^K$ to $A^u$ in \eref{AuFu} cancel each other order by order in
perturbation theory, cf. \sref{s:systcons}, the resulting perturbative
expansion will enjoy both
features.

\subsection{Expectation values of the Maxwell field}
\label{s:expFu}

We want to compute the expectation value
\bea{expFu}\erw{F^u_{\mu\nu}(x)}_{f,c}:=\Erw{\psi_{qc}(f)F^u_{\mu\nu}(x)\psi_{qc}(f)^*}\Big/\Erw{\psi_{qc}(f)\psi_{qc}(f)^*}
\eea
of the electromagnetic field strength in charged states
from $\HH^{\rm dress,M}$, where
$$\psi_{qc}(f) \equiv \int d^4y \, f(y) V_{qc}(y)\otimes \psi(y).$$
This requires to compute the bosonic factor $\Erw {V_{qc}(y_1)F^u_{\mu\nu}(x)
V_{qc}(y_2)^*}$, multiply with the fermionic factor
$\erw{\psi_0(y_1)\psi_0^*(y_2)}$, smear with $f(y_1)\ol{f(y_2)}$, and
divide by the normalization.

The bosonic factor can be computed from \eref{3pt}:
\bea{VFuV}
\erw{V_{qc}(y_1)F^u_{\mu\nu}(x)
V_{qc}^*(y_2)}=\hspace{90mm}\notag \\ = -iq \ERW{(e\wedge \partial_x)_{\mu\nu}
I_{-e}^x +(u\wedge \partial_x)_{\mu\nu} I_u^x(W_0(x-y_2)-W_0(y_1-x))}_c,\quad\eea
where $\Erw\cdot_c$ stands for the smearing of a function of $\vec e$
with $c(\vec e)$, see \eref{b:erwFu}.

Of particular interest is the asymptotic behaviour of
$\erw{F^u_{\mu\nu}(x)}_{f,c}$ when $x$ goes to infinity in various
directions. The behaviour of the electric field $\vec E(x)$ at
$x=(0,r\vec n)$, $r\to\infty$, in a given state is an abstract criterium for the infraparticle states \cite{Bu3}: whenever the expectation value decays like
$r^{-2}\vec a(\vec n)$ with $\vec a(\vec n)\neq 0$ (and the
fluctuations are not too wild), then  
the state cannot be an eigenvector of the mass operator
$M^2=P_\mu P^\mu$, and instead is an 
infraparticle state \cite{Bu3}. This abstract criterium must be
satisfied in QED, because the Gauss Law requires the $r^{-2}$ decay of the
electric flux. In the next subsection, the asymptotic flux will be
computed in states \eref{expFu}, and the
infraparticle criterium will be established.  Of course, the
dissolution of the mass-shell in the spectrum of  the state is manifest because of the contribution
of the dressing factor (photon cloud) to the momentum. The point is
that the state is created by an ``infrafield'' (the dressed Dirac
field).

The limiting behaviour in general spacelike and timelike directions
will be presented in \sref{s:spl-tl}.
The limiting behaviour in lightlike directions is of particular
interest in connection with the Infrared
Triangle \cite{Str}, as briefly outlined in \sref{s:triangle}.

\subsubsection{Purely spatial asymptotics}
\label{s:flux}

We call ``purely spatial'' the directions in the hyperplane
$(ux)=x^0=const$.

We claim that $\erw{V_{qc}(y_1)F^u_{\mu\nu}(x_r)
  V_{qc}(y_2)^*}$, and consequently also $\erw{F^u_{\mu\nu}(x_r)}_{f,c}$
decay like $r^{-2}$ in purely spatial directions $x_r = x_0+\bpm0\\[-1mm]r\vec n\epm$, with a
nontrivial limit\footnote{unless $c=c_0$ is the constant function in which
  case the limit is identically zero. Namely, 
  $\phi(c_0)$ is longitudinal while $F^u$ is transversal, cf.\
  \eref{phi0} and \ref{s:pos}, so that $V_{qc_0}$ commutes with
  $F^u$.}. The string-integrated distributions $I_{-e}W_0(z)$ and
$I_uW_0(z)$ appearing in \eref{VFuV} are given in \eref{fxe} and \eref{fxu}.
When $x_r-y_2$ resp.\ $y_1-x_r$ are inserted for $z$, then in the limit 
$r\to\infty$ the finite values $x_0-y_i$ may be
neglected in \eref{VFuV}. Namely, the behaviour of $((x_r-y)e)$ and $(x_r-y)^2$ is dominated by
$-r(\vec n\cdot \vec e)$ resp.\ $-r^2$ which are independent of $x_0-y$. This entails 
that in \eref{expFu}, the norm $\erw{\psi_{qc}(f)\psi_{qc}(f)^*}$ also
appears in the numerator, and the result is independent of the
smearing function $f$. Moreover, in \eref{VFuV} the difference of
two-point functions may be replaced by the commutator function. Thus \eref{expFu} simplifies to
\bea{erwcomm}
\lim_{r\to\infty}r^2\Erw{F^u_{\mu\nu}(x_r)}_{f,c}=
\lim_{r\to\infty}r^2\erw{V_{qc}(0)F^u_{\mu\nu}(x_r)V_{qc}^*(0)}
=\hspace{40mm}\notag \\ =
q \lim_{r\to\infty}r^2 \ERW{\Big[
(-e\wedge \partial_z)_{\mu\nu}
I_{-e}^z - (u\wedge \partial_z)_{\mu\nu}
I_u^z\Big]C_0(z)\big\vert_{z=(0,r\vec n)}}_c.
\eea
Now, at equal time, $C_0((0,\vec z))=0$ and $\pa_0C_0((0,\vec
z))=\delta(\vec z)$. 
Because $n^0=0$ and $e^0=0$, only the
electric field $E_i=F_{0i}$ has a non-zero asymptotic expectation
value\footnote{The magnetic expectation value will be non-zero when the
  asymptotic limit is taken in a boosted hyperplane $u'^\perp$ different from the
  hyperplane $u^\perp$ in which $c(e)$ is supported. }:
$$\lim_{r\to\infty}r^2\Erw{E_i(x_r)}_{f,c}=q \lim_{r\to\infty}r^2 \Big[
\int d\sigma(\vec e)\, c(\vec e)e_i\ioi ds\, \delta(r\vec n-s\vec e)
-\frac1\pi rn_i \ioi ds\, \delta'(s^2-\lambda^2)\Big] =$$\vskip-7mm
\bea{fluxd}=
-q\Big[ c(\vec n)-\frac1{4\pi}\Big]n^i.\eea
When the flux density is integrated over the ``sphere at infinity'', the
first term gives rise to the charge $-q$ (the charge of the
electron). But the first term is the contribution of $F^K_{\mu\nu}$ within
$F^u$, so by Stokes' theorem, its flux arises from the
fictitious charge density \eref{eq:fict}, commuted with the escort
field. The second term due to the modification \eref{AuFu} of the 
Maxwell field  cancels the fictitious total charge\footnote{\label{fn:ave}In fact, it
  subtracts the average over the sphere of the first term. This is not
  an accident. Namely, by \eref{a:c0} the average of
$\frac{e}{(ke)}$ equals $\frac u{(uk)} - \frac{k}{(uk)^2}$. The second
term is
annihilated by the exterior derivative, and $\frac{u}{(uk)}$ is the
subtracted term (with $e$ replaced by $u$).}.

The vanishing of the
total charge is a necessity in
the dressing model, because $\pa^\mu F^u_{\mu\nu}=0$ is an operator
identity. {\em It is physically expected because the model is only designed to implement the infrared features of QED
states, notably the non-vanishing asymptotic flux density, without the need to ``turn on'' the actual QED interaction with the Dirac
field, cf.\ \sref{s:roadmap} and \sref{s:ptII}.}

In \cite{MRS2}, we had done the same computation with $F^K_{\mu\nu}$
rather than $F^u_{\mu\nu}$, and found the same results without the
subtraction of the $u$-terms, and in particular the ``correct'' physical value $-q$
of the total charge. But the
evaluation of $F^K$ in dressed states is a formal
prescription, because $F^K$ is not defined on the dressed
Hilbert space, as explained in the remark at the end of \sref{s:ext}.
In particular, dressed expectation values of
polynomials in $F^K_{\mu\nu}$ violate positivity in the bulk. On the
other hand, the asymptotic flux operators
$\lim_{r\to\infty}r^2F^K_{\mu\nu}(x_r)$ commute with all
observables and are therefore multiples of $\mathbf 1$ in each sector,
hence coincide with their expectation values. By the asymptotic absence of
fluctuations (which decay like $r^{-4}$), positivity as a state on the commutative asymptotic algebra is
automatic. Therefore, the result in \cite{MRS2} is {\em justified as a
state on the asymptotic flux operators of $F^K_{\mu\nu}$, but not as a state on its full
algebra in the bulk.}

In contrast, the present computation is a state also on the bulk
algebra of the physical Maxwell field $F^u_{\mu\nu}$. Also its
asymptotic flux operators $\lim_{r\to\infty}r^2F^u_{\mu\nu}(x_r)$ commute with all
observables and are multiples of $\mathbf 1$ in each
sector. This classical behaviour of a genuine quantum model is a prerequisite for the existence
of a classical limit. It emerges not in some
limit sending $\hbar$ to zero, but because in the limit $r\to\infty$
commutators decay faster than expectation values due to long-range photons.

\subsubsection{Spacelike and timelike asymptotics}
\label{s:spl-tl}

By the same methods as in \sref{s:flux}, one can compute expectation
values of  asymptotic fields in other spacetime directions than
$x_r=(0,r\vec e)$. They give functions on the entire spacelike Penrose boundary $\frak
i^0$ of
Minkowski spacetime, whose behaviour adjacent  to the lightlike Penrose
boundaries $\frak I^\pm$ plays a central role in the discussion of
asymptotic symmetries in \cite{Str}. Our model has the benefit that
we can compute these functions in a genuine quantum context, rather
than formulating assumptions on their behaviour based on classical
considerations. 

We shall compute the expectation values of asymptotic fields
$$\lim_{\lambda\to\infty} \lambda^2 \Erw{F_{\mu\nu}(x_\lambda)}_{f,c}$$
where $x_\lambda=x_0+\lambda d$ approaches infinity in arbitrary lightlike and timelike 
directions $d$. To be
specific, we parameterize
$$d_w^\pm = \bpm \pm
1\\[-1mm]w\, \vec n\epm,$$
where $0\leq w<1$ (timelike), $w=1$ (lightlike),  or
$w>1$ (spacelike) is an inverse ``velocity parameter''. The purely spatial
limit corresponds to the limit $w\to\infty$.

The argument as in \sref{s:flux} to conclude that in
the limit, $y_i$ may be replaced by zero, and the expectation value equals
the commutator as in \eref{erwcomm}, applies also for the general
spacelike and timelike limits. Consequently, these limits will not depend
on the smearing function $f$ of the dressed Dirac field. But the
argument does not apply for the lightlike limit,
because the leading $\lambda^2$-term in $(x_\lambda-y)^2$ is absent,
and the dominant terms depends on $x_0-y$. The lightlike  case $w=1$,
which is of particular interest in connection with the Infrared
Triangle \cite{Str}, will be discussed separately in  \sref{s:ll}.

The detailed computations for the spacelike and timelike cases can be
found in \aref{b:spl-tl}. The result is 
\bea{Fu-spl-tl}\lim_{\lambda\to\infty}\lambda^2\Erw{F^u_{\mu\nu}(x_0+\lambda d^\pm_w)}_{f,c}=\frac{\pm q}{4\pi}\ERW{(e\wedge d^\pm_w)_{\mu\nu}\frac{\nu}{(1-w^2\sin^2\alpha)^\frac32} -
(u\wedge
d^\pm_w)_{\mu\nu}\frac{\nu'}{w^3}}_c,\qquad
\eea
where $\alpha=\angle(\vec n,\vec e)$ is the angle of $\vec e\in S^2$ relative
to $\vec n\in S^2$, and the integers $\nu$ and $\nu'$ are functions of
$w$ and $\alpha$, specified as follows.
\begin{itemize}\itemsep-1mm
\item
On $\frak i^+$ (future timelike), one has $\nu=1$ and $\nu'=0$.
\item
On $\frak i^-$ (past timelike), one has $\nu=1$ and $\nu'=2$.
\item On $\frak i^0$ (spacelike, $w>1$), one has
  $\nu=2\theta(\arcsin(w\inv)-\alpha)$ and $\nu'=1$.\end{itemize}
The ``polar cap'' $\alpha < \arcsin(w\inv)\in (0,\frac\pi2)$ shrinks to the point
$\vec e=\vec n$ in the limit $w\to\infty$ (purely spatial), in agreement with the
result \eref{fluxd} for the purely spatial case.\footnote{When
  taking the limit $w\to\infty$, it should be noted that the 
scaling parameter $\lambda$ differs from $r$ in \sref{s:flux} by the
factor $w$.} 

The first term in the electric component of \eref{Fu-spl-tl} is
parallel to $\vec e$, the second term is parallel to $\vec n$. Unlike
in \eref{fluxd}, here $\vec e$ and $\vec n$ are not necessarily
parallel. 
The smearing in $\vec e$ will smoothen the step function in $\alpha$
in the spacelike commutator. By choosing the
smearing function $c(\vec e)$, one can
``design'' asymptotic field expectation values on the spacelike Penrose infinity. 

Of special interest in connection with the Infrared Triangle  are the limits
$ w \searrow 1$ where $\frak i^0$ is adjacent to lightlike infinity $\frak
I^\pm$. Here, the ``polar cap'' becomes the hemisphere $\alpha <
\frac\pi2$, and the expectation values coincide with the limiting
values at $\frak I^\pm_\mp$, cf.\ \sref{s:ll}.

\subsubsection{Lightlike asymptotics}
\label{s:ll}
In order to make contact with the properties of electromagnetic fields
at lightlike infinity $\frak I^\pm$ in \cite{Str}, we study the
asymptotic behaviour of expectation values of the Maxwell field $F^u$ in lightlike
directions $x_\lambda = x_0+\lambda\ell^\pm$, where $\ell^\pm =\bpm
\pm 1\\[-1mm]\vec n\epm$. This asymptotics is more subtle than in the
spacelike and  timelike cases,
because the dominating term in $z_\lambda^2$ is
$2\lambda(z_0\ell^\pm)$. It is therefore not only $O(\lambda)$ rather than
$O(\lambda^2)$, but it also depends on the initial point $z_0=x_0-y_1$
resp.\ $x_0-y_2$, where $y_i$ are points in the support of the
smearing function $f$ of the dressed Dirac field, as in \eref{VFuVraw}. In
particular, the limiting behaviour will depend on $x_0$ and $y_i$.

The details of the computation can
be found in \aref{b:ll}. For very narrow smearing functions $f$ of the
dressed Dirac field around a point $y$, we find:
\bea{Fudr} \lim_{\lambda\to\infty}\lambda\cdot \Erw {F^u(x_\lambda)}_{f,c} =
\frac {q}{4\pi} \cdot \ERW{\frac{\ell \wedge e}{(\ell e)}- \ell\wedge u}_c\cdot\delta((\ell
x_0)-(\ell y)),
\quad\eea
where again $\Erw\cdot_c$ stands for the smearing of a function of $\vec e$
with $c(\vec e)$. The leading $O(\lambda\inv)$ contribution to the electric
field is\footnote{The expression $\frac{\vec e}{\skp ne}$ here and in \eref{Einfcc0}, \eref{Erinf} and \eref{erwQ} is ill-defined as a distribution for general
smearing functions $c$. (That it is well-defined for the constant
function, is rather accidental.) With the rather crude method of
  the asymptotic expansion in \aref{b:ll}, we  were not able to
  pinpoint its  precise distributional meaning. Presumably,
  $\frac{1}{\skp ne^p}$ should be interpreted as
  $\frac12(\frac{1}{\skp ne_+^p}+\frac{1}{\skp ne_-^p})$ such
  that \eref{Erinf} and \eref{erwQ} average to zero with $c=c_0$ --
  which they should do because \eref{Einf} does.}
\bea{Einf} \Erw{\vec E_{\infty}(U,\vec n)}_{f,c} = 
\frac {q}{4\pi} \cdot \ERW{\frac{\vec e}{\Skp ne}-\vec
  n}_c\cdot\delta(U-(\ell y)), \qquad (\ell y)= y^0-\Skp ny.
\eea
Notice the $\vec n$-dependent shift in the argument $(\ell y)=
y^0-\Skp ny$, so that the support
$U=(\ell y)$ of the asymptotic electric field along $\frak I^+$
depends on $\vec n$.
 \eref{Einf}
is transversal, in accord with the fact that the radial
component $\Skp nE$ decays faster than
$\lambda\inv$.\footnote{\label{fn:trans}The
subtraction of $\vec n$ in \eref{Einf} is the electric component of
$\frac{u\wedge\ell}{(\ell u)}$ in \eref{Fudr}, which would be absent
in expectation values of $F^K$.  The transversality of the
asymptotic electric field reflects the transversality of the physical
Maxwell tensor $F^u$, while $F^K$ would fail to be transversal. This is another reason, besides positivity, why $F^u$ should
  be adopted as the correct electromagnetic field outside the vacuum
  sector. }

For the constant function $c_0=\frac1{4\pi}$, one has $\Erw{\frac{\vec
    e}{\skp ne}}_{c_0}=\vec n$ (cf.\ \eref{a:c0}). Thus,
the second term $-\vec n$ in \eref{Einf} is the subtraction of the
spherical average of the first term, and one may as well write
\bea{Einfcc0} \Erw{\vec E_{\infty}(U,\vec n)}_{f,c} = 
\frac {q}{4\pi} \cdot \ERW{\frac{\vec e}{\Skp ne}}_{c-c_0}\cdot\delta(U-(\ell y)),
\eea
where $c-c_0$ has weight zero. In particular, for constant smearing
$c=c_0$, one has $\erw{\vec E_\infty}_{f,c_0}=0$. This physically
reflects the fact that the longitudinal field $\phi(c_0)$ cannot generate a transversal electric field, cf.\ \sref{s:dressMD}.

\eref{Einf} is odd in $\vec n$, while the asymptotic
bremsstrahlung field of an accelerated particle
has no such symmetry. Therefore, one should not attempt to choose the smearing function
$c$ so that the  asymptotic transversal electric field in the dressed state would ``mimic'' the classical asymptotic
transversal electric field of general configurations of accelerated charged
particles. Instead, the field configurations \eref{Einf} in the
charged states generated by the  dressed Dirac field are of a
  novel, genuinely
quantum nature.

The computation of the subleading radial contribution $\erw{E_{r,\infty}(U,\vec n)}_{f,c}$ in dressed states
is much harder than that of $\erw{\vec E_{\infty}(U,\vec n)}_{f,c}$. It can be
found in \aref{b:ll}. The result is
\bea{Erinf} \Erw{E_{r,\infty}(U,\vec n)}_{f,c} = \frac q{4\pi}\ERW{\frac{2\theta(\Skp ne)
    -\theta(U-(\ell y))}{\Skp  ne^2} + \theta((\ell y)-U)}_c
+ \Erw{\Skp y{E_\infty(U,\vec n)}}_{f,c},\qquad\eea
where $(\ell y)= y^0-\Skp ny$.
\eref{Einf} and \eref{Erinf} fulfill the asymptotic Gauss Law 
\eref{dErE}. The $U$-derivative of the term $\Skp y{E_\infty}$ in
\eref{Erinf} on the left-hand
side of \eref{dErE} accounts for the angular
derivative of the $\vec n$-dependent shift in \eref{Einf} on the right-hand
side.

The limiting value of \eref{Erinf} as $U\to -\infty$ coincides with the limiting value of the radial electric part of \eref{Fu-spl-tl}
  as  $w\searrow 1$, where $\frak i^0$ touches on $\frak I^+$ along the
  sphere $\frak I^+_-$:
\bea{erwQ}\lim_{U\to-\infty}\Erw{E_{r,\infty}(U,\vec n)}_{f,c} = \frac q{4\pi}\ERW{\frac{2\theta(\Skp ne)}{\Skp    ne^2} + 1}_c.\eea

  Along the same lines, one can compute the lightlike asymptotics
  of  expectation values of the magnetic field:
\bea{Binf} \Erw{\vec B_{\infty}(U,\vec n)}_{f,c} = 
\frac {q}{4\pi} \ERW{\frac{\vec n\times \vec e}{\Skp ne}}_c\cdot\delta(U-(\ell y)),
\quad \Erw{B_{r,\infty}(U,\vec n)}_{f,c} = \Erw{\Skp y{B_\infty(U,\vec n)}}_{f,c}.\qquad
\eea
The same computations
  can also be done on $\frak I^-$. We refrain
  from doing so, because the result is related to the result on $\frak I^+$
  by the anti-unitary time-reversal operator $T$ that exchanges $\frak I^+$ and $\frak I^-$. In particular, the
  fulfilment of the matching conditions \cite{Str}
  \bea{match} E_{r,\infty}\big\vert_{\frak I_-^+}= E_{r,\infty}\big\vert_{\frak
    I^-_+}, \quad B_{r,\infty}\big\vert_{\frak I^+_-}= -B_{r,\infty}\big\vert_{\frak
    I^-_+}\eea
has a fundamental origin: {\em it is a
manifestation of the time-reversal (or PCT) invariance of the Dirac and Maxwell
fields,} as follows. We have (in standard Dirac conventions)
$$T\psi(y)T^* = \gamma^1\gamma^3 \psi(Ty), \quad T A^K_\mu(x)T^* =
-(TA^K)_\mu(Tx).$$
The transformation law of $A^K_\mu$ along with the fact that
$I_u=-I_{-u}$ on all correlations of $A^K$ (see \eref{u-u}), hence
$TI_u(\pa A^K)(x)T^* = -I_{-u}(\pa A^K)(Tx) = I_u(\pa A^K)(Tx)$, and
trivially $u=-(Tu)$, implies also $T A^u_\mu(x)T^* =
-(TA^u)_\mu(Tx)$, hence 
$$T\vec E(x)T^* = \vec E(Tx), \quad T\vec B(x)T^* = -\vec B(Tx), \quad
T\phi(x,e)T^* = -\phi(Tx,e).$$
By the antilinearity of $T$, $TV_{qc}(x)T^* = V_{qc}(Tx)$.  Because the expectation values
of $E_{\infty}$ at $\frak I^+_-$ and $\frak I^-_+$ vanish, those of 
$E_{r,\infty}$ are independent of
the smearing function $f(y)$ of the Dirac field, see \eref{erwQ}. Thus,
the two expectation values match by
virtue of $T$ invariance. Likewise, those of $B$ must differ by a
sign. This behaviour is not changed by charge conjugation $C$, that
takes $\vec E\to-\vec E$, $\vec B\to-\vec B$, $\phi\to-\phi$, $V_q\to
V_{-q}$, nor by parity $P$, that takes $\vec E(x)\to-\vec E(Px)$, $\vec B(x)\to\vec B(Px)$, $\phi(x,e)\to\phi(Px,-e)$, $V_{qc}(x)\to
V_{qc\circ P}(Px)$. Both $C$ and $P$ preserve $\frak I^+_-$ and
$\frak I^-_+$, where the latter swaps $\vec n\to-\vec n$.

Now, the operators in \eref{match} commute with all observables
(\aref{b:ash}) and hence are multiples of $\mathbf 1$ in each
superselection sector. They can therefore be identified with their
expectation values such as \eref{erwQ}, and the matching condition for
the expectation values entails the  matching condition \eref{match} for the operators.

The present time-reversal (or PCT) argument pertains to the dressed
  model. It also applies to full QED, because the QED interaction is time-reversal (and CP)
  invariant.

This remarkable conservation
  law \eref{match} is discussed at length in \cite{HMPS,KPS,Str}.
The smeared fields $E_{r,\infty}(\vec n)$ and
  $B_{r,\infty}(\vec n)$ at $\frak  I^+_-$:
  \bea{Qeps} Q_\eps^+ = \int_{S^2} d\sigma(\vec n)\, \eps(\vec n)
  E_{r,\infty}(U,\vec n)\big\vert_{U=-\infty}, \quad \wt Q_{\wt
    \eps}^+ = \int_{S^2} d\sigma(\vec n)\, \wt \eps(\vec n)
  B_{r,\infty}(U,\vec n)\big\vert_{U=-\infty}
  \eea
are the electric and magnetic     generators of ``large gauge
    transformations'' \cite{Str} that locally transform the
    potential by a derivative. When they are  written with the help of
    the asymptotic Gauss Law \eref{dErE} and its magnetic analogue as integrals along $\frak I^+$, their
    commutators with the asymptotic potential on $\frak I^+$ can be
    worked out as in \aref{b:ash}, giving angular derivatives of
    $\eps(\vec n)$, i.e., gauge transformations. Being generators of gauge transformations,
    $Q^+_\eps$ and $\wt Q^+_\eps$ actually commute with all
    observables, in particular $\vec E_\infty$ and $\vec B_\infty$. This is consistent with the commutation of
    $E_{r,\infty}$ and $B_{r,\infty}$ with $\vec E_\infty$ and $\vec
    B_\infty$ on $\frak I^+$
    (see \aref{b:ash}). 

By the matching condition, $Q^+_\eps=Q^-_\eps$, the gauge
    transformation generated by them can be computed both on $\frak
    I^+$ and on $\frak I^-$, and is necessarily globally topologically nontrivial (``large''). This
    feature is interpreted as an ``infinite degeneracy of
    the vacuum'' \cite{Str}.

 The superselection structure of our model is ``dual'' to this
    infinite degeneracy in the sense that the large gauge
    transformations transform the escort field by a shift (which is
    possible because its localization reaches out to infinity), and hence
    the sector-creating charged fields by a complex
    phase. Conversely, the sectors assign expectation values
    (\eref{erwQ} smeared with $\eps(\vec n)$ for $Q^+_\eps$) to the generators. In full QED, first order
    corrections to the generators are expected to cancel this phase
    and render the interacting dressed Dirac field invariant.

\subsection{Perturbative dressing transformation}
\label{s:ptI}

We give a perturbative motivation for the non-perturbative
construction of \sref{s:dressMD}. Namely, when the free Dirac field is
perturbed with the dressing
density $L_{\rm dress}(c)=q\,\pa_\mu\phi(c)j^\mu$, the tree diagrams
can be seen to organize into the Wick-ordered exponential series of
$e^{iq\phi(c)}\cdot \psi_0$.

Causal perturbation theory based on Bogoliubov's formula
\cite{Bog} (see also \cite{MRS2} for a discussion of the
string-localized case)
gives the dressed field $\psi_0\big\vert_{L_{\rm dress}(c)}(x)$  as a power series in
integrals over retarded multiple commutators of $\psi_0(x)$ with $L_{\rm
  dress}(y_i,c)$, see \sref{s:roadmap}.

In the case at hand, because $L_{\rm dress}$ is a total
derivative, an integration by parts turns the retarded integrals for the tree diagrams in each order into $\frac1{n!}(iq\phi(x,c))^n\psi_0(x)$, summing up to
$e^{iq\phi(x,c)}\psi_0(x)$. E.g., in first order,
$$\psi^{(1)}(x) =i \int d^4y \, R[\psi_0(x),j^\mu(y)] \cdot\pa_\mu \phi(y,c) =
-i\int d^4y \, \phi(y,c) \pa^y_\mu R[\psi_0(x),j^\mu(y)] =i
\phi(x,c)\psi_0(x),$$
thanks to the Ward identity
\bea{ward}\pa^y_\mu R[\psi_0(x)j^\mu(y)]=-\delta(x-y)\psi_0(y).\eea
Loop diagrams have to be properly renormalized. This has been achieved
up to second order. Apart from UV renormalization, the logarithmical
IR divergence in the propagator of the escort field accounts for 
a multiplicative regularization of the exponential fields as in \eref{Wv}.

The characteristic trait of this perturbative construction is the
``collapse'' of retarded integrals to local expressions in $x$, as illustrated. Renormalizations have to be fixed
such that this structure is preserved order by order.   

By a theorem of Borchers \cite{Bor}, if two (sets of) fields belong to the same
Borchers class, then they have the same S-matrix. The Borchers class
of a set of fields is the class of all fields on the cyclic Hilbert space of
the given fields, that are relatively local w.r.t.\ the given fields,
i.e., they mutually commute at causal separation.
In the case at hand, the dressed field belongs to the
Borchers class (admitting relative string-locality) of the free
fields. Anticipating that the theorem can be extended to the string-localized case, 
it would imply absence of scattering also in the case at hand.
$S=1$ is the ``classically expected'' feature of a quantum theory
whose interaction density is a total derivative; it is, however, not automatic in the quantum
case, where it must be secured beyond tree level by appropriate renormalization 
conditions. From this perspective, the above renormalization condition 
(``preservation of the string-localized Borchers class'') is
equivalent (or at least not weaker) than the condition that $S=1$ in
each order. Technically, it amounts to requiring that all retarded
multiple commutators with $L$ appearing in the perturbative expansion
collapse (with the
help of integrations by parts) into $\delta$-functions.

One may as well subject the Maxwell field $F^u_{\mu\nu}$ to the
dressing transformation with the dressing density $L_{\rm
  dress}(c)$. It turns out that $F^u\vert_{L_{\rm dress}(c)}=F^u$ is
invariant under this transformation.
This is a consequence of the ``neutral'' Ward identity $\pa^y_\mu
R[j^\mu(y),j^\nu(y')]=0$. E.g., in first and second order,
$$(F^u)^{(1)}(x) = \int 
d^4y \, \pa^y_\mu R[F^u(x),\phi(y,c)]j^\mu(y)=-\int 
d^4y \, R[F^u(x),\phi(y,c)]\pa^y_\mu j^\mu(y)=0, $$
$$(F^u)^{(2)}(x) = -\int d^4y \int d^4 y'\,
R[F^u(x)\phi(y,c)]\pa^y_\mu R[j^\mu(y),R^\nu(y')]
\pa_\nu\phi(y',c)=0$$
by integration by parts, 
where we have used that $R[F^u(x),\pa_\mu\phi(y,e)] = \pa^y_\mu
R[F^u(x),\phi(y,e)]$ has no freedom of renormalization.

\section{Towards QED: Perturbation of the dressed Dirac field}
\label{s:ptII}

The dressing transformation is
only the first, ``kinematical'' step towards the full QED. It produces a free field of a new kind: the
dressed Dirac field, that captures essential infrared properties of the
actual interacting Dirac field of QED. In order to arrive at QED, the
dressed Dirac field has to be subjected to the interaction $L^u$ or $L^K$, which can only be done perturbatively. 

As pointed
out in \sref{s:roadmap} Item 5, there are two options: to perturb the
positive-definite model $\{\psi_{qc},F^u\}$ with the non-local
interaction density $L^u$, or to perturb the indefinite model $\{\psi_{qc},F^K\}$ with the local density $L^K$. Order by order in perturbation theory, the two options
give the same result for interacting correlation functions (see \sref{s:pt-gauss}). Because
the former option preserves positivity and has no contributions from
the fictitious current, and the
latter preserves locality of the observables, the resulting
formulation of QED enjoys both properties. This mechanism can
work because all $u$-dependent terms in the former model cancel each
other.

The discussion of the local Gauss Law in \sref{s:pt-gauss} will illustrate
these cancellations in an important instance. In \sref{s:systcons} we
present systematic considerations concerning the structure of the
perturbative expansion of the interacting infrafield.  The important
message will be that the dressing factor of the free infrafield 
remains untouched (and along with it the string-localization and the infrared features of the
field) in the perturbative expansion. But it causes additional
vertices connecting the dressing factor to QED vertices, thus contributing additional terms
with novel ``cloud propagators'' to the perturbative
expansion. We discuss the expected difficulties of the future scattering theory for
infrafields in \sref{s:scatt}. 

In \sref{s:joint}, we illustrate (by way
of a nontrivial example) how the diagrams with cloud propagators give rise to
an ``interference'' between the infrared
divergencies of QED and those of the free infrafield. Recall that the
latter determine the superselection structure of the Hilbert space of
the free dressed Dirac-Maxwell theory as in
\sref{s:dressMD}. If the observed pattern persists
in higher orders, then the dressing divergencies do not simply cancel
the QED divergencies. {\em Rather, the interplay of dressing and QED
interaction would ``deform'' the superselection structure of the
dressed model in a momentum-dependent way.}

\subsection{Local Gauss Law}
\label{s:pt-gauss}

We test the validity of the quantum Maxwell equation $\erw {\pa_\mu
  F^{\mu\nu}}= -q \erw{j^\nu}$, whose zero component is the local (=
differential) Gauss Law, in a charged state created by the interacting
charged field, in first order of
perturbation theory. In the standard QED approach $\{\psi_0,F^K\}\big\vert_{L^K}$,
there is a source term that can be attributed to the fictitious
current, which makes the total charge vanish. This is the failure of  the gobal Gauss
Law, that cannot be avoided in a local theory (\sref{s:clouds}). In contrast, in the 
dressed model
$\{\psi_{qc},F^K\}$ without interaction the global Gauss Law holds, i.e., the total
charge is the correct one, but the local charge density can still be
identified with the fictitious current density. The most important result is that, when the dressed model is perturbed
with the local QED interaction density $L^K$, the fictitious source term
is  cancelled and replaced by the
Dirac current. 

Subsequently, we turn to the perturbation of the manifestly
positive-definite model
$\{\psi_{qc},F^u\}$ with the non-local interaction density
$L^u$. Here, one result will be that $F^u$ without interaction is
sourcefree, as expected for a free theory in a Hilbert space where
there is no fictitious current. The main result will be that all non-local contributions
arising from $F^u-F^K$ and from $L^u-L^K$ cancel, thus
illustrating the equivalence of the two constructions \eref{eq:hyb} and the power of
the present approach, doing justice to positivity, locality and the
infrared structure of QED at the same time.

The result is that, in first order of perturbation theory, all terms
due to the non-local term $u_\mu I_u(\pa A^K)$ present both in
$F^u_{\mu\nu}$ and in the interaction density $q\,A^u_\mu j^\mu$
cancel each other, see Table 2 below. It follows that
$$\Erw{\psi_{qc}(y_1)F^K_{\mu\nu}(x)\psi_{qc}^*(y_2)}\big\vert_{qA^K  j}=\Erw{\psi_{qc}(y_1)F^u_{\mu\nu}(x)\psi_{qc}^*(y_2)}\big\vert_{qA^u
  j}=\Erw{\psi_0(y_1)F_{\mu\nu}(x)\psi_0^*(y_2)}\big\vert_{qA(c)  j},$$
where the second equality holds thanks to \sref{s:ptI}. The first
expression arises formally in the local setting on the GNS Krein
space, cf.\ the remark at the end of \sref{s:dressMD}. The expression
in the middle arises in the non-local setting on the extended GNS
Hilbert space \eref{H0dressMD}. The last expression is perturbatively defined in the
positive-definite string-localized setting on the Wigner Hilbert space. 
It was computed in \cite[Sect.\ 5]{MRS2}. 

Here, we start by computing the first expression
\bea{psiFKpsi}\Erw{\psi_{qc}(y_1)F^K_{\mu\nu}(x)\psi_{qc}^*(y_2)}\big\vert_{qA^K_\mu j^\mu}^{(1)}\eea
of interacting fields with the interaction $L^K=q\,A^K_\mu j^\mu$. 
In first order in $q$, it  consists of the
non-perturbative term as in \sref{s:expFu} (with only the
contributions from $F^K$)
$$X_0=-iq\big(\erw{F^K_{\mu\nu}(x)\phi(y_2,c)}-\erw{\phi(y_1,c)F^K_{\mu\nu}(x)}\big)
\cdot \erw{\psi_0(y_1)\psi_0^*(y_2)}$$
plus the perturbative terms
$$X_1+X_2+X_3=\Erw{\psi_0^{(1)}(y_1)F^K_{\mu\nu}(x)\psi_0^*(y_2)}+\Erw{\psi_0(y_1)F^K_{\mu\nu}(x)\psi_{0}^*{}^{(1)}(y_2)}+\Erw{\psi_{0}(y_1)F^K_{\mu\nu}{}^{(1)}(x)\psi_{0}^*(y_2)},$$
where in this order, one may replace the free infrafield
by the free Dirac field, and the field perturbations are given by \eref{eq:Phi1}.
$X_0$ is computed using
$$\erw{\phi(y_1,c)F^K(x)} = \pa^x\wedge \erw{\phi(y_1)A^K(x)} =
\Erw{e\wedge \pa^x  I_e^{y_1}}_cW_0(y_1-x)$$ etc.
$X_1+X_2$ and $X_3$ are identical with the expressions computed in
\cite{MRS2} {\em without} the string-dependent parts in that
computation. The latter arise when the dressing transformation is
implemented perturbatively. As expected from \sref{s:ptI}, they coincide
after an integration by parts and use of \eref{ward}, 
with the non-perturbative contributions $X_0$ in the present computation.

Table 1 shows the various contributions as follows.

\begin{table}
\begin{center} \begin{tabular}{c||c|c||c|c||l}
  &$\erw {F^K}$&$\pa\erw{F^K}$&interaction\cr
            \hline\hline                                      
$X_1+X_2$&retarded&$+\pa C_0\!\cdot\!\erw{\psi\psi^*}$&$A^K\,j$\cr
$X_3$&retarded&$-\erw{\psi j \psi^*}$&$A^K\,j$\cr
           \hline\hline$X_0$&$I_{-e}(\pa\wedge e)C_0\!\cdot\!\erw{\psi\psi^*}$&$-\pa C_0\!\cdot\!\erw{\psi\psi^*}$&non-pert\cr
    \hline\hline
$X_1+X_2$&$I_{-e}(\pa\wedge e)C_0\!\cdot\!\erw{\psi\psi^*}$&$-\pa C_0\!\cdot\!\erw{\psi\psi^*}$&$\pa\phi(e)\,j$\cr
$X_3$&0&0&$\pa\phi(e)\,j$\cr
           \hline\hline
         \end{tabular}
\end{center}         
\caption {\small Contributions and cancellations for Gauss' Law.}
\end{table}

For the sake of readability, the
presentation is quite schematic. In the headline, $\erw Q$ stands
for the charged expectation value of the quantity of interest $Q$, to which the
entries in the respective column contribute according to the indicated
interaction. $X_1+X_2$ resp.\ $X_3$ are the contributions due to the perturbations
of $\psi$, $\psi^*$ resp.\ $F$; $X_0$ is the non-perturbative
contribution due to the dressing. $\erw{\psi\psi^*}$ stands
for $\erw{\psi(y_1)\psi^*(y_2)}$, and $C_0$ stands symbolically for the
difference of massless two-point functions $i(W_0(x-y_2)-W_0(y_1-x))$
(which asymptotically becomes the commutator function).
Every entry has to be smeared with the test functions
$f(y_1)\ol{f(y_2)}$ and (where applicable) with $c(e)$. The overall
factor $q$ is suppressed.

This is what Table 1 tells us.
\begin{enumerate}\itemsep0mm
\item The first two rows give the perturbative contributions due to the
interaction $L^K$, as in standard QED.
The ``retarded'' expressions are the usual retarded
integrals in the standard Krein space QED in Feynman gauge. The sum of these two
rows is the result of standard perturbation theory. The source entry
in row 1 is the ``fictitious'' contribution from $\pa F^K= j_{\rm fict} = -\pa(\pa
A^K)$ (where $\erw{A^K_\mu(y)j^\nu_{\rm
    fict}(x)}=-\pa_\mu^y\pa_x^\nu W_0(y-x) $ and $\erw{j^\nu_{\rm
    fict}(x)A^K_\mu(y)}=-\pa_\mu^y\pa_x^\nu W_0(x-y) $ yield the displayed
result after an integration by parts and use
of \eref{ward}). As mentioned in footnote \ref{fn:fict}, its total charge cancels the total charge of the current term in
row 2. The presence of this term is the failure of the Gauss Law.
\item
The third row is the non-perturbative contribution of the dressing
factor, computed as in \eref{3pt}. This contribution alone (i.e., the dressed model without QED
interaction) yields quantitatively the correct expectation value $-q$ of the total
charge; 
but its origin is still the fictitious current. However, in QED it
cancels the source term
of fictitious provenience in row 1, so that only the Dirac source term
in row 2 remains, i.e., the Gauss Law holds for the interacting fields (in this order).
\item The last two rows
give the corresponding contributions when the dressing
transformation were done perturbatively. In fact, the last row is
identically zero
because of the conservation of the Dirac current. 
\item The equality of rows
3 and 4 illustrates the equivalence of the perturbative and
non-perturbative dressing transformation in this order, see
\sref{s:ptI}, as well as the equivalence (cf.\ \sref{s:roadmap} Item 5)
$\{\psi_0,F^K\}\big\vert_{L^K(c)}=\{\psi_{qc},F^K\}\big\vert_{L^K}$. 
\item The expectation
value of $F^K$ is the sum of a retarded term with source $\erw{\psi
j\psi^*}$ (row 2) and the two entries in
rows 1 and 3 that add up to a source-free vacuum solution of the
Maxwell equations.
\end{enumerate}
The main message is Item 2: {\em With the local interaction density $L^K$, the interacting Maxwell
  field evaluated in the state created by
the interacting infrafield satisfies the local Gauss Law.} 
In the perturbative approach, this was already one of the main 
messages of \cite{MRS2}: the string-localized total
interaction $q\, A(c)j$ (in Wigner space or in Krein space) yields the
correct Gauss Law without a fictitious source term compensating the
total charge.

We now compute the expectation value 
\bea{psiFupsi}\Erw{\psi_{qc}(y_1)F^u_{\mu\nu}(x)\psi_{qc}^*(y_2)}\big\vert_{qA^u_\mu j^\mu}^{(1)}\eea
of interacting fields with the interaction $L^u=q\,A^u_\mu
j^\mu$. Again, there is a non-perturbative term $X_0$ and three
perturbative terms $X_1$, $X_2$, $X_3$. Given the result for
\eref{psiFKpsi}, it is sufficient to consider only the additional contributions
due to $F^u-F^K= -I_u(u\wedge \pa)(\pa A^K)$, and concentrate
on their
cancellation. The contributions due to the modification of the
interaction $L^u-L^K=q\,I_u(\pa A^K)(uj)$
  neither contributes to $X_1+X_2$ because the null field $(\pa A^K)$ has vanishing
  correlation with $F^K$ and $F^u$, nor to $X_3$ because it has
  vanishing retarded commutator with $F^K$ and $F^u$.

The $u$-contributions in the non-perturbative term $X_0$ are computed using $$\erw{\phi(y_1,c)(\pa A^K)(x)}= -\Erw{(e\pa^x)I_e^{y_1}}_cW_0(y_1-x)=
\Erw{(e\pa^{y_1})I_e^{y_1}}_cW_0(x-y_1) = -W_0(y_1-x)$$ etc. The
$u$-contributions to $X_3$ vanish by current conservation. Those to
$X_1$ and $X_2$ are computed with integration by parts and using again \eref{ward}.
We collect the results in another table. 

\begin{table}
\begin{center}         
\begin{tabular}{c||c|c||c|c||l}
  &$\erw {F^K}$&$\erw{F^u-F^K}$&$\pa\erw{F^K}$& $\pa\erw{F^u-F^K}$&interaction\cr
           \hline\hline
$X_1+X_2$&$**$&$-I_{u}(\pa\wedge u)C_0\!\cdot\!\erw{\psi\psi^*}$&$**$&$-\pa
C_0\!\cdot\!\erw{\psi\psi^*}$&$A^Kj$ or  $A^uj$
    \cr
$X_3$&$**$&0&$**$&0&$A^Kj$ or  $A^uj$\cr
            \hline\hline                                      
$X_0$&$**$&$I_{u}(\pa\wedge u)C_0\!\cdot\!\erw{\psi\psi^*}$&$**$&$+\pa C_0\!\cdot\!\erw{\psi\psi^*}$&non-pert\cr
    \hline\hline
$X_1+X_2$&$**$&$I_{u}(\pa\wedge u)C_0\!\cdot\!\erw{\psi\psi^*}$&$**$&$+\pa C_0\!\cdot\!\erw{\psi\psi^*}$&$\pa\phi(e)\,j$\cr
$X_3$&0&0&0&0&$\pa\phi(e)\,j$\cr
\hline \hline
         \end{tabular}
\end{center}                  
\caption{\small Cancellations of non-local contributions. (Entries indicated by $**$ are
identical to Table 1.)}
\end{table}

This is what Table 2 tells us.

\begin{enumerate}\itemsep0mm
\item The total source term in the dressed
  model $\{\psi_{qc},F^u\}$ without interaction (the sum of the
  contributions in row 3) vanishes. 
\item The previous equivalence (Item 4 after Table 1) between the non-perturbative (row 3) and
  perturbative (rows 4 and 5)
  dressing transformations extends to $F^u$. 
\item   All additional non-local terms $I_u(\dots)$ either vanish or
  cancel each other one-by-one. Therefore, the
  present a priori non-local expansion is in fact local.
  \item Because the present construction is manifestly positive, while
    the additional terms cancel exactly, the previous model (Table 1)
    is also positive.
    \item Items 3 and 4 together illustrate the power of the equivalence (cf.\ \eref{eq:equ} and \eref{eq:hyb})
      $$\{\psi_{qc},F^K\}\big\vert_{L^K}=\{\psi_{qc},F^u\}\big\vert_{L^u}$$
 (one manifestly local, the other one manifestly positive).
\item 
The same cancellation also illustrates the equivalence (cf.\ \eref{eq:equ} and \eref{eq:hyb})
$$\{\psi_0,F\}\big\vert_{L(e)}=\{\psi_{qc},F^u\}\big\vert_{L^u}$$
(both manifestly positive). 
\end{enumerate}
In higher order, the systematic cancellation of $u$-dependent terms
(in particular those due to the nonlocal part $L^u-L^K$ of the interaction density)
is harder to see, but it must happen for the abstract reasons given in
\sref{s:roadmap}.

\subsection{Systematic considerations}
\label{s:systcons}

The perturbation theory
of the infrafield can be done as a power series expansion in the coupling constant
$q$ of $L^u$, while not at the same time expanding the
non-perturbative vertex operator
in the free infrafield $\psi_q$ as a power series in $q$. It will be
crucial that the perturbation leaves the vertex operators ``intact'' in the
interacting infrafield, so that one 
can expand their (finite) correlation functions \eref{VVV} (with
$q_i=\pm q$). One might distinguish two (in the beginning independent) couplings $q$
(for the dressing) and $q'$ (for perturbation theory with $L^u$.)
However, the requirement that the resulting S-matrix and observable fields of QED must be
independent of the string $e$ or the smearing function
$c(e)$ fixes $q'=q$ already in first order.

The infrafield $\psi_{qc}$ subjected to the interaction density
$L^u$ or $L^K$ will be a power series whose
coefficients are retarded multiple commutators of $\psi_{qc}$ with
several operators $L^u$ or $L^K$. The bosonic part contains only a single
vertex operator $V_{qc}(x)$ and several potentials $A^u(y_i)$. Such
terms  require a treatment beyond standard perturbation theory, as follows.

As explained in \sref{s:ext}, correlations involving one or
more potentials $A^u$ are well defined by variation of correlations
involving Weyl operators $e^{iA^u(f)}$. This implies that commutators and retarded commutators are of the form
$$[V_{qc}(x),A^u(y)] =V_{qc}(x)\cdot iq[\phi(x,c),A^u(y)]
\quad\hbox{and}\quad R[V_{qc}(x),A^u(y)] =V_{qc}(x)\cdot iqR[\phi(x,c),A^u(y)].$$
{\em Consequently, the vertex operators associated with the
charged fields remain ``intact'' in the expansion of the interacting
infrafield. But they contribute additional
string-localized ``cloud propagators'', with lines in Feynman diagrams
connecting vertex operators (``clouds'') with interaction
vertices (Fig.\ 2 in \sref{s:joint}).} These new propagators are of order 1 in $q$. They are obtained by string-integration
over ordinary propagators. ``Contractions of vertex
operators'' need not to be considered since they are already contained in the vertex operator correlations. 

There is a potentially important observation: Recall that for the
  construction of vertex operators in 
  \sref{s:dressMD}, we had to restrict smearing functions $c$ to be
  supported in $u^\perp$ for some $u\in H_1^+$. This was necessary for
two reasons: (i) to ensure
that the superselected correlation functions of vertex operators
satisfy positivity, and (ii) to make sure that the divergent exponent
$d_{m,v}(C,C)$ (see \aref{b:constr}) diverges to $+\infty$, so that
$e^{-\frac{q^2}2d_{m,v}(C,C)}\to \delta_{C,0}$. 

The observation is that the restriction on the
support of smearing functions can be relaxed when vertex operators are
tied to the Dirac field as in \eref{psiq-np} {\em and} the QED interaction
is added! As for (i), the positivity of the perturbative expansion of the
free infrafield $\psi_{qc}$ with the QED interaction is indirectly secured by the
equivalence (cf.\ \eref{eq:equ} and \eref{eq:hyb})
$$\{\psi_0,F\}\big\vert_{L(e)}=\{\psi_0,F^K\}\big\vert_{L^K(e)}=\{\psi_{qc},F^K\}\big\vert_{L^K}$$
for arbitrary smearing functions; as long as one keeps the strings
spacelike so as to maintain
sufficient causal separability. Thus, one may admit
$c_i$ supported on $H_1$ of unit total weight. As for (ii), charge conservation of the free Dirac 
field ensures, that in non-vanishing correlations of the dressed Dirac
field, 
$C=\sum_iq_ic_i$ ($q_i=+q$ or $-q$ for each field $\psi_{qc}$ or $\ol
\psi_{qc}$) has  the total weight $\sum_iq_i=0$. But
if $C$ has total weight zero, then
$(T_C(k)k)=\sum_i q_i (T_{c_i}(k)k)=0$ (see \eref{b:Uc}), hence for
$k^2=0$, $T_C(k)$ is spacelike or lightlike, and $$d_{m,v}(C,C)=-\int d\mu_m(k)\,v(k)(\ol
{T_C(k)}T_C(k))\geq 0$$ still diverges to $+\infty$ unless $T_C(k)$ is a multiple
of $k$ for all $k$ on the zero mass-shell. Thus, its exponential
converges to 1 if $C$ satisfies this condition, and zero
otherwise. This suffices for a finite result.
When $C$ may be supported on $H_1$, we do presently not know whether the
latter condition implies $C=0$, giving rise to the ``Kronecker delta''
$\delta_{C,0}$ as in \sref{s:dressD}; the kernel of the quadratic form
$d_{m,v}$ in the limit $m\to0$ could be larger.

\subsection{Scattering theory}
\label{s:scatt}

Scattering amplitudes of QED are infrared divergent.   Bloch and Nordsieck \cite{BN} had noticed that the divergence is cancelled by
  admitting real soft photons below the observation threshold accompanying the charged
 particles. Later, Weinberg \cite{Wsoft} recognized that the real part of
the singularities systematically comes with the characteristic
``soft photon'' factors
\bea{IRdiv} \mathrm{Re}(\alpha B)=\frac{q^2}2\lim_{m\to0}\int
  d\mu_m(k)\Big(\frac {p_{\rm in}}{(p_{\rm in}k)}-\frac {p_{\rm out}}{(p_{\rm
    out}k)}\Big)^2 =
  -\infty,
  \eea
  where $p$ are electron momenta and $k$ virtual soft photon
  momenta. $\alpha=\frac{q^2}{4\pi}$ is the fine structure constant. 
  The divergence is logarithmic. These factors sum up order by
order to yield exponentials of the form $e^{\alpha B}\to0$  with the consequence that all S-matrix elements in charged
    states vanish unless $p_{\rm out}=p_{\rm in}$. The physical
    reason is that in- and out-states lie in different superselection
    sectors.

Chung \cite{Chu} recognized that the same effect as in \cite{BN}
  to cancel the IR divergence \eref{IRdiv} is
 achieved by ``dressing'' the initial and final charged particle states with
momentum-dependent coherent real photon states of the form
  \bea{Chudiv}
  \exp \pm q \lim_{m\to0}\int d\mu_m(k)\,\Big(\sum_{\ell=1,2}\frac{(pe^\ell(k))}{(pk)}a^{\ell*}(k)-h.c.\Big)\ket0
\eea
  where $a^{\ell*}(k)$ create states of linear polarization $e^\ell(k)$,
  and $\pm q$ is the charge of the respective
  particle. (In the limit $m\to0$, these states again do not lie in
  the Fock space.) Faddeev and  Kulish \cite{FK} (see also \cite{Dy,Du3}) proposed to
  include a similar dressing factor into a redefinition of the
  S-matrix, stating that their prescription is equivalent
    to Chung's up to a certain gauge transformation to remove
    longitudinal photons.\footnote{Unfortunately, the gauge transformation 
    destroys the necessary condition \cite[Eq.\ (16)]{FK}, so there remains some
    doubt if this beautiful picture is entirely self-consistent.}

  In contrast, in our approach, the dressing factors of charged fields are
$x$-dependent and a priori independent of
  the momentum $p$ of the charged particles. But with $p=Mu$ ($M$ the
  electron mass) and $c_u(e)$ the constant  smearing function on
  $H_1\cap u^\perp$, i.e., in the rest frame of $p$, one has by \eref{a:c0} and \eref{b:Uc}
  \bea{ppk} \frac
  p{(pk)}= \frac u{(uk)} = T_{c_u}(k) + \frac
  k{(uk)^2}. \eea
 One might therefore expect that with the choice of
 $c=c_u$ adjusted to the momentum $p=Mu$ of the charged particles, one can
 achieve with dressed Dirac fields the same cancellation as in
 \cite{Chu,FK}. This is, however, not the case, as will be elaborated in
\sref{s:joint}. Instead of an exact cancellation of the IR divergence, one finds
an ``interference'' between the superselection rule $c=c'$ for the
free infrafield and the velocity superselection $p=p'$ of standard
QED into a novel joint superselection rule.

Another distinction (unrelated to scattering theory) is this: In Chung states, the expectation value of the total electric flux at
  infinity of the free (Wigner) Maxwell field $F$ is zero (in accord with
  \cite{BCRV}). This is the same finding as with the physical Maxwell
  field $F^u$
  in the positive version of our intermediate model. But while in our approach, the QED correction 
produces the correct value $\pm q$ (\sref{s:pt-gauss}), the QED correction in Chung states
also vanishes in $O(q)$ when $F$ is embedded into Krein space as $F^K$
or as $F^u$ (Coulomb gauge, cf.\ \sref{s:pos}). Thus these states do not resolve the problem with the fictitious charge,
cf.\ footnote \ref{fn:fict}. (This admittedly never was their purpose.)

Apart from these differences, there is the
marked distinction that the Chung resp.\ Faddeev-Kulish dressing
factors are attached ``by hand'' to the states resp.\ to the S-matrix,
while our dressing factors are ``dynamically'' attached to
the charged infrafield, and offer more flexibility concerning the
choice of $c$. This changes the large-time asymptotic behaviour of the
charged field, and we have not yet
succeeded to formulate the correct asymptotic-time limits that would
properly define the S-matrix.

The problems are manifold, starting from the fact that a suitable
  scattering theory for infraparticle fields does not yet exist. The LSZ
reduction formula is not applicable because it requires a sharp
mass-shell; the Haag-Ruelle
scattering theory \cite{Ha} is not applicable because by the absence of a free
equation of motion there are no immediate candidates
for ``asymptotic creation operators'' $a_t(f)$ that would create
time-independent one-particle states and time-convergent
many-particle states.

In non-relativistic QED (interaction of the quantized
  electromagnetic field  with nonrelativistic, quantum mechanical
  electrons), the infraparticle nature of the electron has been
  rigorously derived \cite{CFP2}, and scattering states including one
  electron and an arbitrary number of photons
  (Compton scattering) have been constructed \cite{CFP1}. The same has been
  done in the Nelson model, where the electromagnetic field is
  substituted by a scalar massless boson \cite{P1,P2,BDG}. In the Nelson
  model, also scattering states of one infraparticle (electron) and
  one massive Wigner particle (``atom'') have been constructed
  \cite{DP}.

  It is not obvious how to transfer these results to full
  perturbative QED, since they rely on a constructive Hamiltonian
  formulation. We emphasize that even in the non-relativistic setting,
  the scattering of two or more infraparticles is still an open
  issue.

Buchholz, Porrmann and Stein \cite{BPS} have put forward a (relativistic)
  collision theory in terms of ``particle weights'' that are
  functionals generalizing normalizable particle states when there are
  no asymptotic states with a sharp mass-shell. When the method is
  applicable, it produces cross-sections (rather than scattering
  amplitudes), and interference effects between different momenta
  cannot be addressed. 

A formulation due to
Herdegen \cite{He2} aims at restoring the possibility of
interference, by assuming the existence of charged fields with suitable asymptotic
decay properties of their commutators. The approach is more flexible
than LSZ and Haag-Ruelle scattering theory, but it requires a
spectral condition on the charged field that is not easy to establish
in our approach.

Buchholz and Roberts \cite{BR} have argued that states with
  uncontrolled incoming photons ``from the past'' could produce large
  fluctuations of the asymptotic electromagnetic field, so that the algebraic conclusion \cite{Bu3} of the absence of the sharp
  mass-shell may fail. The photons from the past cannot be seen by any
observations in a future lightcone. Indeed, the authors found that the
photon cloud superselection structure disappears upon restriction to
observables in future lightcones. Based on this idea, Alazzawi and
Dybalski \cite{AD} have constructed an outgoing M\o ller operator for
Compton scattering in 
Buchholz-Roberts representations; but because of the time-asymmetry of the
setting, there is no incoming M\o ller operator in the same
representation, and hence no S-matrix.

In all our attempts up to this point to apply either of the mentioned
strategies, the main problem is to 
  establish convergence of the relevant large-time limits, already in
  the free infrafield model. The
  failed attempts seem to indicate the structural feature that the (desired) S-matrix
  cannot arise as a product of two M\o ller operators that could be
separately defined; but can be constructed only in a simultaneous
limit of future and past asymptotic times.
Apart from the lack of the mass-shell a major
obstacle is the product structure of vertex operator correlations. It
seems to indicate that the space of
asymptotic ``free infraparticle states'' may not have the structure of a Fock space.

We mention here just a specific instance: A generalization of
Buchholz' scattering theory of massless waves  \cite{Bu1} can be
applied to the vertex operator  
fields of the two-dimensional model \sref{s:2D}, i.e., without the
spinor field \cite{DM}. This gives a finite result, despite the fact that the assumption of a zero
mass-shell is not fulfilled. The S-matrix turns out to be a complex phase
proportional to the product of charges. The same strategy applied to
the vertex operators in \sref{s:dressD} in four dimensions seems
promising at first sight: it also produces a complex phase 
(proportional to the product of charges and the bilinear form
$\erw{c,c'}$ in \eref{cc}) -- but the product structure of vertex
  operator correlations produces another factor arising from to the
  transcendental homogeneous function $H$ 
\eref{H}  \cite{GRT}. This factor being possibly of modulus $>1$,
jeopardizes the interpretation of the computed quantity 
as a scattering amplitude.

\subsection{Dynamical  superselection structure}
\label{s:joint}

In order to exhibit the cancellation of IR singularities in each order of
  perturbation theory, one has to think of the factor $\delta_{C,0}$
  in the correlator of vertex operators as the power series expansion of $e^{-\frac{1}2
    d_{m,v}(C,C)}$ (given by \eref{b:dmvH}) in the limit $m\to0$, as in
  \sref{s:dressD} and \aref{b:constr}, where
  $C=\sum_iq_ic_i$ with $q_i=\pm q$ is of order $q$. In
  perturbation theory, the IR-divergent term $e^{-\frac{1}2
  d_{m,v}(C,C)}$ will combine with IR-divergent contributions from the
  ordinary Feynman diagrams as in \eref{IRdiv} that have a similar form $e^{-\frac{1}2
  d_{m,v}(C_0,C_0)}$ (with $C_0$ specified below) and with further IR-divergent  contributions from 
  the cloud propagators. We shall present evidence of an (if
  confirmed in higher orders) exciting scenario: all these terms conspire to give rise to
  $$e^{-\frac{1}2
  d_{m,v}(C+C_0,C+C_0)},$$ where the diagrams with cloud propagators contribute
the mixed terms like $d_{m,v}(C,C_0)$. Thus, a novel interference effect seems to ``deform''
the superselected quantity from $C$ to $C+C_0$!

Two simple examples in low orders shall illustrate this
  observation. The two-point function of the interacting infrafield involves the two-point function $\erw{V_{qc}(x)V_{qc'}(x')} =
  \delta_{c,c'} e^{-q^2 w_v(x-x',c,c)}$, which is separately IR-finite. In second order, the usual QED mass correction of the Dirac
  field arises, which is IR-finite, hence in this case $C_0=0$. Two
  new diagrams with cloud propagators cancel each other thanks to $c=c'$.

The coupling of the interacting infrafield to an external
  potential $a_\mu(x)$ is a less trivial  instance. This is the
  simplest instance which allows a change of momentum, where the QED
  result is separately IR-divergent (and hence forbids
  scattering). For the sake of the illustration, we
  shall  tentatively appeal to the Gell-Mann--Low formula and apply
  the standard   LSZ prescription\footnote{We do not know
  whether the use of the GML formula is properly justified for
  infrafields. Duch shows that it is applicable for
  local fields also in the presence of IR divergences \cite{Du3}.  The standard LSZ
  formula can only be a proviso: it needs a modification in order to properly
  account for the absence of a sharp mass-shell. Yet, it turns out to be good
  enough to see how the IR-divergent result of QED is modified.}.

We shall compute in
  \bea{GML}\Erw{T \big[\psi_{qc'}(x') e^{iq \int d^4y \,
      j^\mu(y)(A_\mu^K+a_\mu)(y)}\ol \psi_{qc}(x)\big]}\eea
  the term linear in the external field $a_\mu$ in third order, 
  and concentrate on the IR-divergent terms when the LSZ  prescription is applied. 

We begin with the first order. Here, \eref{GML} equals
$$iq (-i)^2 \int d^4y \, S_F(x'-y)\gamma^\mu S_F(y-x)\cdot
a_\mu(y).$$
Inserting the Fourier representations, truncating the Dirac lines with the
  Klein-Gordon operators $(\square^{(\prime)}+M^2)$, and going
  on-shell (as in LSZ theory):
  $p^{(\prime)2}=M^2$, one finds the
  coefficient of $e^{-ip'x'}e^{ipx}$
  \bea{Gamma}\Gamma^{(1)}(p,p'):=-iq\,\wh a_\mu(p'-p) \cdot\frac{M+\sla
    p'}{(2\pi)^4} \gamma^\mu\frac{M+\sla p}{(2\pi)^4}.
  \eea
The two-point function of the vertex operators (without QED interaction) contribute (in the
massive approximation of \sref{s:dressD})  the factor
  $e^{-\frac{1}2d_{m,v}(C,C)}$ with $C=q(c-c')$. This factor gives in third
  order the IR-divergent term 
 $-\frac{1}2d_{m,v}(C,C)\cdot \Gamma^{(1)}(p,p')$.
The IR-divergent contribution from the QED diagrams, 
  given in  \cite[Eq.\ (8)]{Chu}, is of the form \eref{IRdiv} times
  $\Gamma^{(1)}(p,p')$.   With \eref{ppk}, it can be written as  $-\frac{1}2
  d_{m,v}(C_0,C_0)\cdot \Gamma^{(1)}(p,p')$, where $C_0=q(c_u-c_{u'})$.
  $c_u$ and  $c_{u'}$ are the constant smearing functions
  in the rest frames $u=p/M$ and $u'=p'/M$ of the in- and outgoing Dirac
  particle, see \eref{a:c0}. Notice that,
  unlike $C$, $C_0$ is momentum-dependent.

 In \aref{c:cloudv}, we compute the contributions from the four diagrams
 with cloud propagators. Two of them are depicted in Fig.\ 2,  the other two have the cloud vertex on the lower line.
 \begin{figure}
   $$\includegraphics[width=25mm]{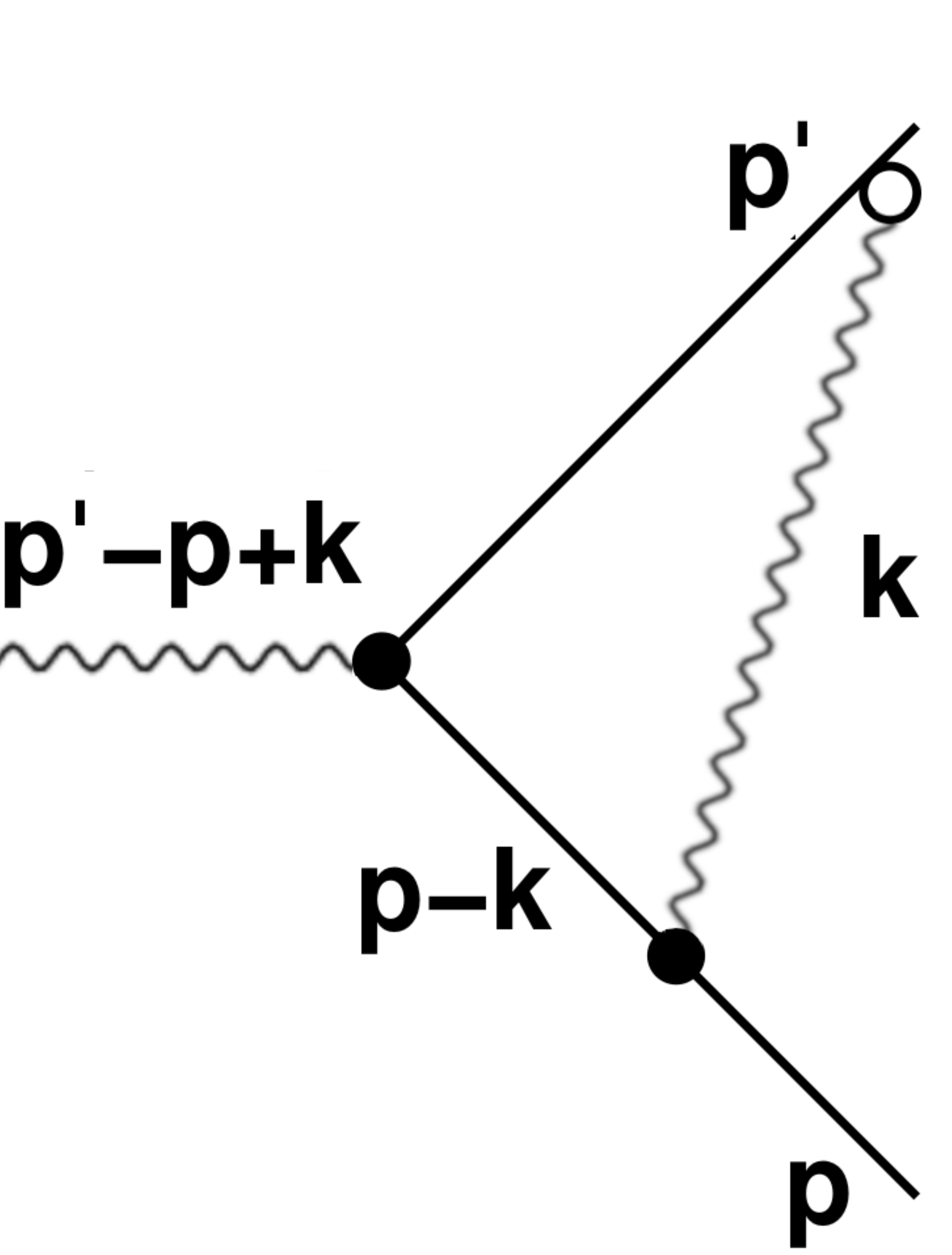} \qquad\qquad \includegraphics[width=25mm]{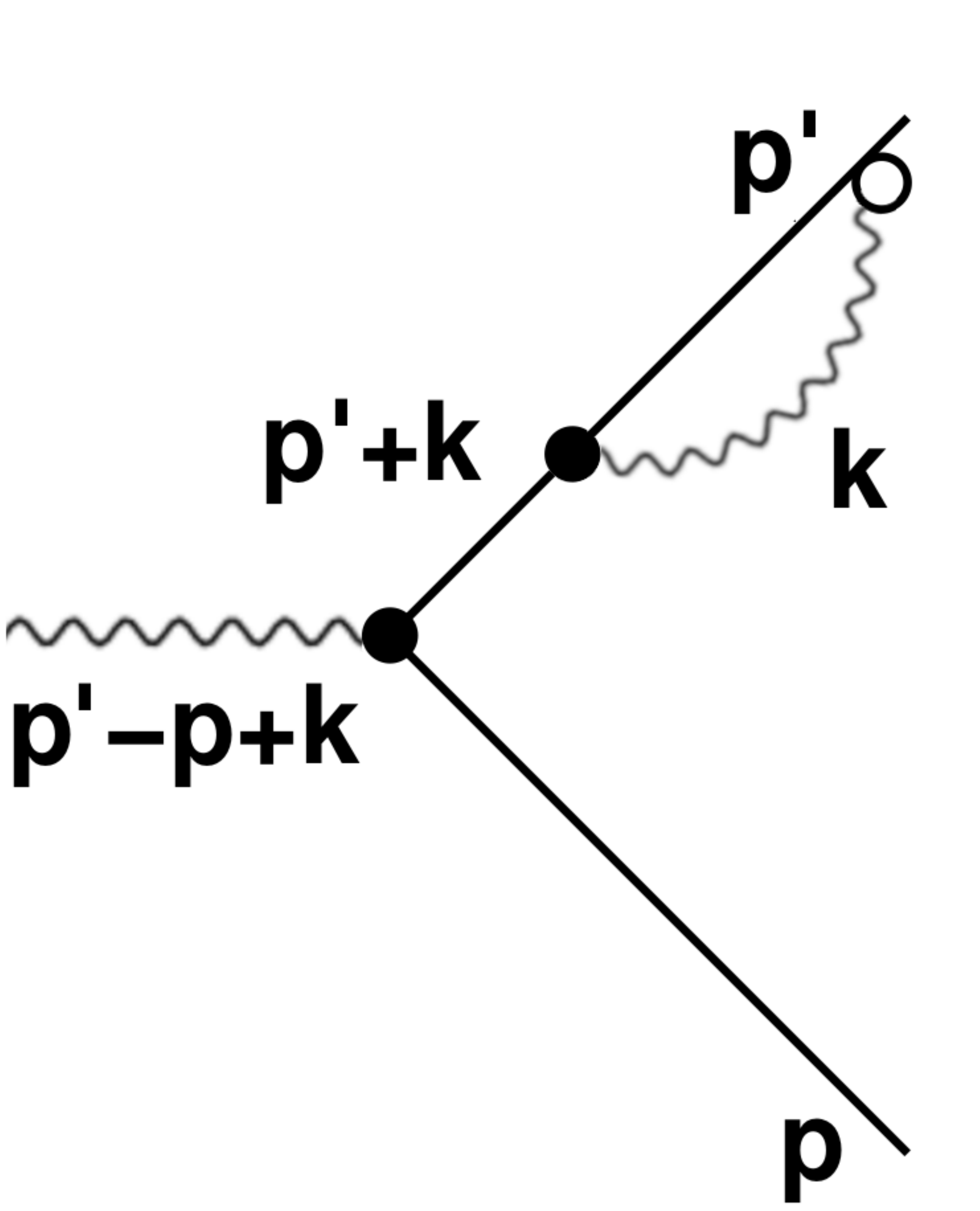} $$
\caption{\small Two diagrams with ``cloud propagators'' connecting
a vertex operator (``cloud'', open blob) with a QED vertex. Notice that the
open blob is attached to one Dirac line, and does not separate two
Dirac lines. The cloud propagator only adds the photon momentum to the outgoing infrafield
whose Fourier variable is $p'+k$}.
\end{figure}

 They contribute the divergent factor $-
  \frac12(d_{m,v}(C,C_0)+d_{m,v}(C_0,C))\cdot \Gamma^{(1)}(p,p')$. Thus, they provide the interference terms in
  $-\frac12 d_{m,v}(C+C_0,C+C_0)$ with
  \bea{Ctot} C+C_0= q(c-c' + c_u-c_{u'}).\eea
  We conjecture that this the onset of the perturbative expansion of the
  factor $e^{-\frac12
    d_{m,v}(C+C_0,C+C_0)}$. Because both $T_C(k)$ and $T_{C_0}(k)$ (see \eref{b:Uc})
  entering the definition of $d_{m,v}$  are orthogonal to
  $k$, the exponent is negative or $0$, and the exponential converges either to $0$ or to $1$.
At this point the observation at the end of
\sref{s:systcons} comes to bear: Recall that $c_{u^{(\prime)}}$ are supported in $H_1\cap
u^{(\prime)}{}^\perp$. When the smearing functions $c^{(\prime)}$
are suitably supported in $H_1$, the resulting dynamical selection rule
$$\lim_{m\to0} e^{-\frac12d_{m,v}(C+C_0,C+C_0)}\stackrel!=1\quad
  \Leftrightarrow\quad T_{C+C_0}(k)\stackrel!=\alpha(k)k \quad\hbox{for all $k$ with $k^2=0$}$$
can have nontrivial solutions
with $p'\neq p$.\footnote{$c^{(\prime)}=-c_{u^{(\prime)}}$ is not
a solution because $c^{(\prime)}$ and $c_{u^{(\prime)}}$ all must have unit
total weight. Solutions with unit total weight would be $c=c_{u'}$,
$c'=c_u$ or $c'=c+c_u-c_{u'}$, which all have $\alpha(k)=0$. We
conjecture, however, that solutions with $\alpha(k)\neq 0$ become important.}

The upshot is that the cloud contributions change the IR
  singularities of QED of the form $d_{m,v}(C_0,C_0)$, whose
absence requires $u=u'$ (see \aref{b:orth}) and hence the absence of scattering as in
\cite{Chu}, into a deformed ``dynamical'' superselection rule that does not require $p=p'$.  
Conversely, regarded as a perturbation of the free infrafield, the
QED interaction modifies the rigid momentum-independent superselection rule
$c=c'$ into a momentum-dependent one.

This mechanism of singularity cancellation and the coupling of
  the cloud-superselection to the momentum
  appears quite distinct  from the one proposed in \cite{BN,YFS} and \cite{Chu,FK}.

To be sure, the above scenario has been tested only in
  lowest nontrivial order. Moreover, an LSZ prescription for
  infrafields that might turn the ``suitable choice'' of smearing
  functions $c$, $c'$ (that would satisfy 
  the superselection rule) into an asymptotic
  automatism via some stationary phase mechanism, is still lacking.

\section{Emerging new paradigms for QFT and Outlook}
\label{s:paradigm}

There are ample new insights arising from string-localized QFT
  beyond the special case of QED.

Let us start our discussion with what we have learned about QED as a ``special instance''. We do not claim to construct a ``New QED'', and neither is this our
ambition. We rather reformulate off-shell QED in a more conceptual way, avoiding
unphysical features wherever possible. The final theory is expected to
be equivalent to ``ordinary'' QED plus off-shell charged fields. Yet, there are many differences.

Our approach distinguishes between observables and
states. Unobservable fields are needed to create charged
states from the vacuum -- simply because observables cannot change a
super\-selection sector. The extremely rich superselection structure of
QED (going well beyond electric charge conservation) is clearly addressed in
our approach.

The interaction density contains no unphysical quantum degrees of
freedom: It couples the Dirac current to a string-localized potential
$A(c)$ that is a functional of the Maxwell field strength $F$ -- in principle defined in
the physical Hilbert space of the Maxwell field with precisely two
polarization states, as obtained by second quantization of the
helicity-1 Wigner representation \cite{Wig,Wein}. However, it turns
out to be far more advantageous to embed the latter into the Krein
space and split $A(c)$ into the usual unphysical Krein potential
$A^K$ and a derivative of the ``escort field''. The latter ``dresses'' the free Dirac field,
while the former carries the honest QED interaction.

The embedding of the Maxwell field $F$ sharpens the view upon the Maxwell
field in Krein space. The usual Krein space Maxwell field $F^K= \pa
\wedge A^K$
involves unphysical photon degrees of freedom that are not visible in
its self-correlators but in correlators with the Krein potential. In
contrast, the embedded physical Wigner space Maxwell field $F$ appears
as $F^u$ in the Krein space, which is a local field relative to itself
but nonlocal relative to the unphysical Krein potential. Since
locality is an issue of commutation relations and not of mathematical
description and labelling, one may as well say that the Krein
potential is nonlocal with respect to the physical Maxwell field. In
this way, the picture makes more sense -- simply because $A^K$ is
unphysical.

It has been objected that a construction exclusively based on
observables (the current and the physical Maxwell field) cannot
construct QED with charged states \cite{BCRV}. The objection does not apply when the
Dirac field is included into the game, and the adiabatic
limit is taken. Namely,
the Dirac field couples to both the escort field and to the Krein
potential. The escort coupling alone ``transfers longitudinal photon degrees
of freedom'' to the charged field. This invalidates the
objection. The simultaneous coupling (our approach to QED) has better features than the
coupling to $A^K$ alone (which produces vanishing
scattering amplitudes due to infrared divergences). The latter are
cured by the dressing and intimately connected to the rich superselection
structure of charged states. In contrast to the prevailing cures of IR divergences, the
cure is included in the field that creates charged states, rather than
applied to the states. The highly interesting properties of the resulting
charged infrafield (including commutation relations \eref{any}) highlight the vanity of
attempts to construct off-shell QED within the Wightman axiomatic setting.

The most important message for general QFT is that quantum fields
of a new type have to enter the scene when their interactions are mediated by
fields of helicity 1 (or more).
In retrospect, it seems that the axiomatization of quantum fields
pioneered by  Wightman and G\aa rding must be questioned in the case
of quantum field theories with long-range interactions and infraparticles.
While it is very successful for interactions of fields of spin or
helicity below 1 ($\varphi^4$ or pion-nucleon interactions), it seems
to fall short when particles of spin 1 enter the stage -- in particular in the Standard
Model. It may be interesting to notice in this respect that the abstract analysis of
  \cite{BF} in the framework of Algebraic QFT can narrow down the
  localization of charges with a mass gap to spacelike cones, but
  the general argument can not be sharpened to compact localization.

The traditional dichotomy that observable fields must be local while
the natural candidates for charge-carrying
(charged-state creating) agents in a quantum-field
theoretical description may be anti-local, is shifted towards a new dichotomy between
point-localized and string-localized. To be sure, this does not refer
to the string-localized Wigner space
potential \eref{AFe} or the Krein space escort
field \eref{phi}: both appear only in the process of the {\em construction} of the
theory, where the former allows the formulation of the interaction on
a Hilbert space, and the latter serves as a 
catalyzer to transfer the string-localization of the former onto the
charged field.

The actual field content of the resulting QED is given by
the interacting Maxwell and infrafields. In contrast to the
potential and the escort field (no longer present in the final QED), that were just string-integrals over
local free fields, the
string-localization of the charged field is irrevocable. It reflects
physical features: photon clouds and the infraparticle nature of
charged particles.

The local observables of the final QED: the 
  interacting Maxwell field and current, are the same as in gauge
  theory. The difference is in the structure of the charged
  fields, whose string-localization is expected to
  cure the infrared vanishing of the S-matrix in the local approach.

The axiomatic perspective contemplates the field
  content of the final theory, not its making. In the orthodox
  approach, one would {\em only} consider observables. But with this attitude, one
  loses direct access to the agents that create charged states from the
  vacuum. This access was indirectly recovered in one of the great
  success stories of Algebraic QFT \cite{Ha}: the Doplicher-Roberts
  reconstruction of graded-local charged fields from the superselection structure
  of the observables. But the method works only for global symmetries,
  and its generalization to local ($=$ gauge)  symmetries remained an open
  problem. Our results do not solve this problem (reconstruction from
  the superselection structure as a general strategy), but it
  indicates with an exactly solvable model what a general stategy
  should be able to envisage. The answer radically departs from the
 axiomatization of electrically charged fields in the Wightman setting, and neither
 complies with the idea of a ``graded-local field net'' with
 anti-local Fermi fields.

With an eye on the Standard Model whose main theoretical challenges
are posed by massless and massive vector bosons, we make some comments
relating to
massive QED, Yang-Mills theory, and the Higgs model.

There are two main traditional setups to study QED with massive
photons: they describe the free
vector bosons by the Proca field, or by a gauge
field. The former is defined on a Wigner Hilbert space, but the Proca coupling to the current is power-counting
non-renormalizable. The latter allows a renormalizable coupling, but
is defined only on an indefinite Krein space.

In both variants, the idea of string-localization comes to rescue: the
string-localized Proca potential $A^P_\mu(e):=I_e F^P_{\mu\nu}e^\nu$ (possibly
smeared in $e$) has UV
dimension one and makes the coupling renormalizable \cite{MSY}. As in QED, the
interaction density differs from a local density by a total derivative
because $A^P_\mu(e)=A^P_\mu + \pa_\mu \phi^P(e)$, where $\phi^P(e) =
m^{-2}(\pa A^P(e))$ does not have a massless limit \cite{MRS1}. Similarly, the massive Krein potential can be
replaced by the potential $A^K_\mu(e)=I_e
F^K_{\mu\nu}e^\nu=A_\mu^K + \pa_\mu\phi^K(e)$ that can be defined on a positive-definite subspace
of the Krein space. The strategy of the present paper, dressing the
charged field via the total derivative part as  the dressing density, can be applied
in both approaches:
$$\psi_0\Big\vert_{L^P(e)} =
\Big(e^{iq\phi^P(e)}\psi_0\Big)\Big\vert_{L^P}, \qquad \phi^P(e)=\frac1{m^2}(\pa A^P(e)),$$
$$\psi_0\Big\vert_{L^K(e)} =
\Big(e^{iq\phi^K(e)}\psi_0\Big)\Big\vert_{L^K}, \qquad \phi^K(e)=
I_e(eA^K).$$
However, in contrast to the massless case, the dressings here do
not lead to a dissolution of the mass-shell. When S-matrix elements
are computed in the LSZ limit, the multiplicative contributions of the 
massive vertex operators  tend to 1 and may be ignored. The
dressed off-shell fields get ``undressed'' in the asymptotic time
limit of scattering theory. 

There is also a direct, point-localized connection between the
Proca and Krein approach (\cite{Du1}, implicitly appearing already in \cite[Sect.\ 3]{RR}):
$$A^P_\mu = A^K_\mu +
\pa_\mu\phi^{PK}, \qquad \phi^{PK} =
\frac1{m^2}(\pa A^K), \qquad \psi_0\Big\vert_{L^P} =
\Big(e^{iq\phi^{PK}}\psi_0\Big)\Big\vert_{L^K}, $$
which sheds an interesting new light on the ``unitary gauge'': $L^P$
is the non-renormalizable interaction with the Proca field, while the interaction $L^K$ is power-counting
renormalizable. But because $\phi^{PK}=\phi^K-\phi^P$ (which is
  $-1/m$ times the Stückelberg field) has
UV dimension $2$, unlike the massless escort field of dimension
zero in the present paper, the dressed field is a Jaffe field
\cite{Ja}. Jaffe fields are not polynomially bounded
  in momentum space, and can therefore only be
smeared with a restricted class of test functions of slow decay
in position space. This makes them of little use for local QFT.  Yet,
the dressed field is well-defined, and free of ambiguities. Therefore the right-hand
side is renormalizable in the sense of absence of infinitely many
undetermined constants, but with poor localization properties.

In general models involving string-localized fields, care must be taken
  that the quantum S-matrix is string-independent. In first order,
  this means that the string-dependence of the interaction density
  must be a total derivative, which clearly restricts the candidate
  interactions. More excitingly, string-independence as a
  renormalization condition in higher orders
  may lead to further constraints on coupling constants and to the necessity for
  additional couplings with fixed values of coupling constants. These are particularly interesting for
  Standard Model physics:

In the case of Yang-Mills theory, a most notable result
  \cite{GGM} states
  that every cubic self-coupling of string-localized massless vector bosons whose
  string-dependence is a total  derivative,
  must necessarily, in order to ensure string-independence of the
  quantum S-matrix at second order, come with coefficients that are the structure
  constants of a compact semisimple Lie algebra. The result and strategy of proof are
  quite similar to the analogous result in gauge theory \cite{AS} if
  BRST invariance is replaced by string-independence. Thus the
  usual paradigm of non-abelian gauge invariance as a starting point
  is reverted (in both settings) to become the consequence of a renormalization condition
in the service of fundamental physical principles. Moreover, the
symmetry is an off-shell property that may not be reflected in the
particle spectrum  (no physical quarks). An understanding of
  infrared features of non-abelian theories like confinement, although qualitatively
  very different, is 
  hardly conceivable without lessons from the abelian case.

String-localized vector potentials can also be used in order to
  couple massive vector bosons to spin-$\frac12$ matter 
without paying with ghosts for power-counting renormalizability. It
turns out that string-independence as a renormalization condition in
higher orders necessarily requires the presence of a scalar field
that must be coupled to the vector potential and to the matter fields like the
Higgs field, and that must have the self-coupling of the Higgs
field. This field does therefore not ``trigger a spontaneous symmetry
breakdown mechanism'' (note that gauge invariance is not a fundamental
principle in the string-localized approach whose breaking needs an explanation,
but it rather emerges as a renormalization condition). On the contrary 
it serves to maintain an invariance property. We shall discuss this in
a forthcoming publication \cite{MRS5}.

Besides spinor QED, one may also consider scalar QED. With the same
coupling $q\, \pa_\mu\phi(c)j^\mu$ where $j^\mu =
-i\chi^*\!\!\stackrel\lra{\pa^\mu}\!\! \chi$, one gets again a charged
infra field $$\chi_{qc}(x)= e^{iq \phi(x,c)}\chi_0(x)$$ creating
charged states with superselected smearing function $c$. Again, this
field belongs to the
free Borchers class (\sref{s:ptI}) and leads to a trivial S-matrix.
But in contrast to the spinor case, string-independence of the theory
requires an additional quartic interaction $-\frac{q^2}2A(c)^2\chi^*\chi$
that ``completes the square'' of the covariant derivatives.
Thus, gauge symmetry emerges as a renormalization condition.
Remarkably, this quartic term can as well be omitted when a different
convention (renormalization) for the derivative propagator
$\erw{T\pa\chi^*\pa\chi}$ is chosen. This scenario was first proven in the local
Krein space QFT setting \cite{DKS} -- showing that gauge symmetry of
a classical Lagrangian is dispensable when the consistency of the
theory can be achieved by a different renormalization condition.

Scalar QED also has an UV divergent ``box diagram'' (double photon
  exchange) that requires a counter term of the form $(\chi^*\chi)^2$ with an undetermined
coefficient. This UV divergence is not removed in the string-localized
approach -- simply because the UV behaviour of string-localized scalar
QED turns out to be the
same as that of Krein space scalar QED. This may be taken as another indication that string-localized quantum
field theory models do
not per se  describe any ``New Physics'' -- except that they may allow
otherwise ``power-counting forbidden'' interactions. 
On the other hand, the term $(\chi^*\chi)^2$ should not
  contribute at Penrose infinity because of its rapid fall-off. This may be an instance where the
  physics at Penrose infinity cannot capture all features of the
  bulk theory.

Work in progress by the present authors \cite{MRS4} studies the coupling of
massless helicity-2 ``gravitons'' to a matter stress-energy tensor. The general structure of
the theory largely parallels QED with one marked difference: the
dressing transformation acts not by an operator-valued phase but like
an operator-valued coordinate transformation (of a
scalar matter field for simplicity)
$$\chi_{qc}(x)=\chi_0(x-q\beta(x,c)),$$ where $\beta(c)$ is a vector-valued
string-localized helicity-2 escort field. As coordinate transformations are the natural
analogue of phase transformations with helicity 2, such a formula
is less surprising as it may appear at first sight.
In momentum space, it is again a phase
transformation by $e^{\pm iq (p\beta(x,c))}$ that can be defined as a
limit of Weyl operators.

In the case of the helicity-2 dressing transformation, the renormalization condition of
string-independence (hence triviality) of the S-matrix (or a
perturbative version of the Borchers class condition as indicated in
\sref{s:ptI}) seems to require an infinite number of higher-order couplings with
unique coefficients, similar to the quartic term $-\frac{q^2}2A(c)^2
\chi^*\chi$ in the scalar QED case; but in the
$h=2$ case these cannot be absorbed into a renormalization of
derivative propagators of the matter field. At least the lowest such terms
coincide with the expansion of the coordinate-invariant Lagrangian in general
curvilinear coordinates \cite{Br}. 
It thus seems that the classical symmetry (coordinate invariance) of the coupling of
matter to helicity-2 massless particles (``gravitons'') is again
determined by string-independence!

Also the other features of QED: superselection structure,
asymptotic fluxes and the infraparticle nature of the
dressed matter field, arise in close analogy to $h=1$. This presumably
also includes the behaviour at lightlike
infinity and the relation to the gravitational memory effect.  An
interesting difference is that the relevant asymptotic flux (the analogue of
the Gauss total charge) is not an integral over the (linearized) curvature at
infinity (which would decay too fast), but rather over a ``partial''
curvature with only one exterior
derivative of the tensor potential. The resulting charge
quantities can be interpreted as boosted versions of the (linearized)
Komar mass \cite{MRS4}.

To wrap this up: Gauge symmetry is no longer an a priori postulate. In
the standard view, it is a symmetry principle imposed on the unobservable
part of the theory, that constrains the form of the interaction and allows -- via BRST -- renormalized interactions mediated
by particles of spin or helicity $1$ with a unitary S-matrix. But 
in our view, the same structure of the interactions is rather a
consequence of
the need to implement the fundamental requirements of Hilbert space positivity
and renormalizability in the off-shell setting. This can be achieved for
spinor and scalar QED, Yang-Mills, the abelian (and perhaps also
non-abelian) Higgs model, 
by means of string-localized interaction densities in such a way that observables remain
string-independent and local, while charged fields become
string-localized in a way that ``looks like a gauge
transformation''. Ockham would opt for string-localizated charged fields
without conflict with Gauss' Law, rather than
for Krein with local charged fields in indefinite metric, or BRST
without charged fields.

We conclude with some speculative remarks.

Theorists are accustomed to employ classical ``external'' fields to describe a broad range of effects,
including anomalous magnetic moments, the Lamb shift, and 
the process of measurement which invokes  ``classical''
observations. But classical fields do not exist in nature: they are a
simplifying idealization while in reality, everything is quantum \cite{La}. One
may wonder why this idealization yields so stunningly exact results. 
We have seen in \sref{s:flux} that the asymptotic electric flux attached to the
charged field exhibits a classical behaviour in spacelike
directions. The infrafield thus carries along with
it classical features (the atomic radius is ``asymptotically large''  compared to the
Compton wave length of the electron and of the nucleus). So the question may be turned around into
asking whether, and how, the ``built-in'' infrared classicality may
lead to a new under\-standing of external field effects. It may
even help soothening, in the case of helicity 2, the disconcerting
``incompatibility'' of classical General Relativity with Quantum Theory.

In the case of self-coupled massless fields, there is no known
  way to separate off a total derivative part from the interaction
  density that could give rise to a dressing
  transformation. This is also true for self-coupled helicity-2 fields
  (``quantum gravity''). Therefore, the infrared features in these cases must
  make their appearance in a different manner. Confinement comes to
  one's mind which excludes charged fields altogether from the field
  content of the interacting theory. The answer must be
  quite different in the $h=2$ case.

\appendix

\section{String-localized correlation functions}
\label{a:sloc}

Correlation functions and commutators of string-localized fields are
string integrals over local correlation functions and commutators.
String integrals can be computed in momentum space by
the distributional prescription  
\bea{a:I-mom}I_e e^{\pm ikx} \equiv \ioi ds\, e^{\pm ik(x+se)}=\frac{\pm i}{(ke)\pm
  i\eps}\cdot e^{ikx} \equiv \frac{\pm i}{(ke)_\pm}\cdot e^{\pm
  ikx},\eea
where, the limit
$\eps\downarrow 0$ is understood in the sense of
distributions.
In view of \eref{a:I-mom}, one could therefore have departed from the expression
on the Krein Fock space
\bea{phia}\phi(x,e) = \int d\mu_0(k)\, \Big[-i\frac{e^{-ikx}}{(ke)_-} (a^*(k)e) + h.c.\ \Big]\eea
as a definition of the escort field; but only the position space
definition \eref{phi} properly exhibits
its localization properties, that are instrumental for perturbation theory (\sref{s:ptII}) and  
essential in scattering theory (\sref{s:scatt}).

An important special case arises when $\phi(e)$ is smeared over the
sphere $e\in H_1\cap u^\perp$ with the constant smearing function $c_u(e)=\frac 1{4\pi}$. Namely, one has \bea{a:c0}\frac1{4\pi}\int_{S^2}
    d\sigma(\vec e)\frac{\vec e}{\Skp ke_\mp} = \frac{\vec k}{\vv
      k^2}\qquad\Leftrightarrow\qquad \frac1{4\pi}\int_{H_1\cap u^\perp}
    d\sigma(e)\frac{e}{(ke)_\pm} = \frac{u}{(uk)} -
    \frac{k}{(uk)^2},\qquad\eea where the latter is the covariant
    formulation of the former for $u=u_0$ \cite{MRS1}. In position space,
    \eref{a:c0} can be written as $\frac1{4\pi}\int_{S^2}
    d\sigma(\vec e)\, e_\mu I_e  = I_u^2\pa_\mu +
    u_\mu I_u$. Inserting \eref{a:c0} into \eref{phia}, one obtains
    \bea{phi0}\phi(x,c_0)=\Delta\inv \Skp\nabla{A^K}(x).\eea

\subsection{String-integrations in position space}
\label{a:sloc-x}

For purposes in this paper, we need the string-integrations
in position space. The massless two-point function is 
\bea{W0}W_0(x_1-x_2) =\int d\mu_0(k) \, e^{-ikx}= -\frac{1}{(2\pi)^2}
\frac1{(x^2)_-},\eea
where $x=x_1-x_2$, and $\frac{1}{(x^2)_-}$ is the distributional
limit $\lim_{\eps\downarrow0} \frac 1{(x-i\eps u)^2}$, $u\in H_1^+$
arbitrary. Form this, one gets the two-point function of the Feynman gauge potential
\bea{eq:AKAK}\erw{A_\mu(x)A_\nu(x')} = -\eta_{\mu\nu} \cdot W_0(x-x')=
-\eta_{\mu\nu} \int d\mu_0(k) \,e^{-ik(x-x')}
\eea
and the two-point function of the escort field
\bea{phiphi}\erw{\phi(x,e)\phi(x',e')} = -(ee')\cdot (I_{-e'}I_e W_0)(x-x') =
-(ee')\int d\mu_0(k)
\frac{e^{-ik(x-x')}}{((ke)_-)((ke')_+)}.\qquad\eea
The latter is logarithmically divergent at $k=0$. Its IR regularization is defined (in
\sref{s:dressD}) as a distribution in $x-x'$ and in $e,e'$, see below. The
smearing function in $e$ will be called $c$.

String integrated correlation functions in position space were computed in \cite{GRT} for spacelike
$e$. For one string integration, one gets (with the change of integration variable 
$s'=s+\frac{(xe)}{e^2} = s-(xe)$)
$$f(x,e) := (2\pi)^2 (I_eW_0)(x) = -\ioi
\frac{ds}{(x+se)^2-i\eps(x+se)^0}=\int_{-(xe)}^\infty
\frac{ds'}{s'^2-((xe)^2+x^2-i\eps x^0)}.$$
This is an elementary integral, giving
\bea{fxe}f(x,e) =
\frac{\log\frac{-(xe)+\sigma_-}{-(xe)-\sigma_-}}{2\sigma_-}, \qquad (\sigma_-=\sqrt{(xe)^2+x^2-i\eps x^0}).
\eea
Here and below, the log's of fractions are always understood as differences of log's,
with phases determined by $i\eps$. This means that it is not allowed
to cancel negative factors in the fractions. Note that the branch cut
of the square root is not a singularity of \eref{fxe} and \eref{fxu}.

The behaviour for timelike strings $u$ is different because the $i\eps$ appears in
different places. With the substitution $s'=s+\frac{(xu)}{u^2} =
s+(xu)$, one finds 
\bea{fxu}
f(x,u)  := (2\pi)^2 (I_uW_0)(x) = -\frac{\log\frac{(xu)-i\eps+\rho}{(xu)-i\eps-\rho}}{2\rho},
\qquad (\rho=\sqrt{(xu)^2
  - x^2}).
\eea
For $u=u_0$ the standard future timelike unit vector, $(xu)=x^0$ and $\rho = \vert\vec x\vert$.

Incidentally, one sees by extending the lower integration limit to
$-\infty$, that
\bea{u-u}
f(x,e)+f(x,-e) = -2\pi i\cdot \frac{\sign(x^0)}{2\sigma_-},
\quad\hbox{while}\quad f(x,u)+f(x,-u)=0.\eea

For the string-localized commutator, one may use the representation
$C_0(z)=\frac1{2\pi}\sign(z^0)\delta(z^2)$ of
the massless commutator function:
$$2\pi (I_aC_0)(z) = \ioi ds \, \sign((z+sa)^0)\delta((z+sa)^2).$$
When the sign of $z+sa$ is constant, either because $a=e$ with $e^0=0$,
or because $a=u$ with $z^0>0$, then the integral essentially counts the
number $\nu$ of {\em positive} zeroes of the polynomial
$(z+sa)^2= a^2s^2+2s(za)+z^2$: 
\bea{Idelta-nu}\ioi ds\, \delta(a^2s^2+2s(za)+z^2) =
\theta((az)^2-a^2z^2)\cdot\frac{\nu}{2\sqrt{(az)^2-a^2z^2}}. \eea
In the purely spatial case $a=e$ with $e^0=0$ and $z^0=0$, $\vec
z=r\vec n$, the equal-time
commutator $C_0(\vec z)=0$ and $\pa_0C_0(\vec x)=\delta(\vec x)$ is
more convenient:
\bea{IedC}(I_e\pa_0C_0)(\vec z) = \ioi ds\, \delta(r\vec n+s\vec e) =\frac
{\delta(r-s)}{r^2} \cdot \delta_{\vec e,\vec n},\eea
where $\delta_{\vec e,\vec n}$ is the $\delta$ function on $S^2$
w.r.t.\ the normalized invariant measure $d\sigma(\vec e)$.

The double string-integrated two-point function appearing in \eref{phiphi} is regularized
by
\bea{IvWdef}(I_{-e'}I_e W_0)_v(x):=\lim_{\eps\downarrow0}\lim_{\eps'\downarrow0}\int d\mu_0(k)
\frac{e^{-ikx}-v(k)}{((ke)-i\eps)((ke')+i\eps')}\eea
as explained in \sref{s:dressD}. Multiplied
by $-(ee')$, smeared
in $e$, $e'$ and exponentiated, it appears in
correlation functions of dressed Dirac fields. In scattering theory, 
the causal ``separation of wave
packets'' of the dressed Dirac field at late and early times is controlled by its asymptotic
properties, when the smeared strings are spacelike separated.

\eref{IvWdef} with two purely spatial 
strings ($e^0=e'^0=0$) has been computed in
\cite{GRT}. 
The result is symmetric in $e_1=e\lra e_2=-e'$, and can be written as
 \bea{IIDelta}
 (2\pi)^2 (I_{e_2}I_{e_1}W_0)_v(x) = -\frac12 f(e_1,e_2)
 \log\big(-\mu_v^2\cdot(x^2)_-\big)+\frac{H(x;e_1,e_2)}{(e_1e_2)},\eea
 where for $\gamma=\angle(\vec e_1,\vec
 e_2)\in [0,\pi]$ (i.e., $\cos\gamma=\Skp{e_1}{e_2}=-(e_1e_2)$)
 \bea{fe1e2} f(e_1,e_2)=\frac\gamma{\sin\gamma}\eea
is the same function as \eref{fxe} with $e_1$, $e_2$ substituted for
$x$, $e$, and 
$H(x;e_1,e_2)$ is defined for purely spatial $e_i$ as follows. The
dimensional factor $\mu_v^2$ depends on the regulator function $v(k)$ in \sref{s:dressD} and may be a
function of $\theta =
\angle(\vec e,\vec e \,')=\pi-\gamma$.

Because the operation $I_e$ is homogeneous
 of degree $-1$ in the string $e$, it is advantageous to relax the
 normalization $e^2=-1$, while keeping $e\in u_0^\perp$
 spacelike. Then $H$ is homogeneous of degree zero, separately in all three variables.
 Let
$$\det{}_{x,e_1,e_2}=x^2e_1^2e_2^2 - x^2(e_1e_2)^2 - e_1^2(xe_2)^2 -
e_2^2(xe_1)^2 + 2 (xe_1)(xe_2)(e_1e_2)$$
be the Gram determinant of the vectors $x$, $e_1$, $e_2$, which in singular expressions is always
understood distributionally as the boundary value from the forward
tube $x-i\eps u_0$. For $\{i,j\}=\{1,2\}$, let $\Lambda_i=(xe_i)(e_ie_j)-e_i^2(xe_j)$ be the cofactors
(signed subdeterminants) of the entries $(xe_j)$ of the Gram matrix. Then it holds
$$\Lambda_i^2 - \det{}_{e_1,e_2}\det{}_{x,e_i} = - e_i^2\cdot
\det{}_{x,e_1,e_2}.$$
Therefore, one can define the homogeneous variables $\zeta_1$,
$\zeta_2$ by
$$\pm e^{\pm \zeta_1} = \frac{\Lambda_1\pm
  \sqrt{\det_{e_1,e_2}\det_{x,e_1}}}{\sqrt{e_1^2\det_{x,e_1,e_2}}}, \qquad
\pm e^{\pm \zeta_2} = \frac{\Lambda_2\pm
  \sqrt{\det_{e_1,e_2}\det_{x,e_1}}}{\sqrt{e_2^2\det_{x,e_1,e_2}}}.$$
They are real if also $x$ is purely spatial, because in this case
all diagonal cofactors are $\geq0$ and $\det_{x,e_1,e_2}\leq0$. Otherwise, they are defined distributionally as boundary
values from the forward tube. Finally, it is convenient to define
$\gamma=\angle(\vec e_1,\vec
e_2)$ as before, so that $\det_{e_1,e_2}=e_1^2e_2^2\sin^2\gamma$, and $D=\frac{\det_{x,e_1,e_2}}{x^2e_1^2e_2^2}$. In these
variables \cite{GRT},
\bea{H}H(x;e_1,e_2) = \frac1{\sin\gamma}\left[\gamma\log\big(\frac{\sin^2\gamma}{\sqrt D}\big)+
\frac{\pi}{2}(\zeta_1+\zeta_2)-\frac
i{4}\left\{\Li_2\Big(e^{i\gamma}e^{\zeta_1}e^{\zeta_2}\Big)\ba{c}+(e^{\zeta_1}\lra
  -e^{-\zeta_1})\\ +(e^{\zeta_2}\lra -e^{-\zeta_2})\\-(e^{i\gamma}\lra
  e^{-i\gamma})\ea\right\}\right]\!\!. \quad\qquad
\eea
The branch cuts of the dilog functions can be seen to secure that the limit $\gamma\to0$ ($e_1$ and $e_2$ parallel) is regular, while the limit $\gamma\to\pi$ ($e_1$ and $e_2$
antiparallel) is singular (as expected because $I_{-e}I_e$ is not
defined). The singularity is 
$\frac{O(\log(\pi-\gamma))}{\pi-\gamma}$ and hence integrable in the
string directions w.r.t.\ the invariant measure on $S^2$.

 For the two-point functions of the
 string-localized potential $A_\mu(e)$, only the derivative of the
 distribution \eref{IIDelta} is needed. The derivative needs no infrared regularization, and takes a much simpler and highly symmetric
 form \cite{GRT}:
 \bea{dIIDelta}
(2\pi)^2 (I_{e_2}I_{e_1}\partial^x_\mu W_0)(x)=-\frac { \big[f(e_1,e_2)\pa^x_\mu +
f(x,e_2)\pa^{e_1}_\mu + f(x,e_1)\pa^{e_2}_\mu\big]\det{}_{x,e_1,e_2}}{2\det{}_{x,e_1,e_2}}.\eea
Remarkably, despite the fact that $\det_{x,e_1,e_2}$ can in Lorentz
metric vanish in far
more configurations than $x$ being linearly dependent of $e_1$ and
$e_2$ \cite{GRT}, thanks to cancellations of singularities in numerator and
denominator, the singular support of this distribution after smearing
in $e_i$, is just $x^2=0$, exactly as for the point-local two-point
function $W_0$.

\subsection{Vertex operator correlations}
\label{a:vertex}

From \eref{IIDelta} we conclude the regularized two-point function of the escort
  field in \sref{s:dressD}
\bea{a:phiphi}
w_v(x,e,e') = \erw{\phi(x_1,e)\phi(x_2,e')}_v = \frac{(ee')}{8\pi^2}
\wt f(e,e')
\log\big(-\mu_v^2\cdot(x^2)_-\big)+\frac{\wt H(x;e,e')}{4\pi^2},\qquad\quad\eea
where $x=x_1-x_2$, $\wt f(e,e') :=f(e,-e')$ and $\wt H(x;e,e'):=
H(x;e,-e')$ (both symmetric under $e\lra -e'$). Because positive and negative powers of $-(x^2)_-$ are
well-defined 
distributions, and the homogeneous distribution $\wt H$ can be
exponentiated, one may exponentiate \eref{a:phiphi} without smearing
in $x$. When 
$\wickv{e^{i\phi(g\otimes c)}}$
with $g=q\delta_x$ are inserted into \eref{corr4}, the factors become 
\bea{power}e^{-q_iq_jw_v(x_i-x_j; c_i,c_j)} = e^{\frac{q_iq_j}{8\pi^2}\lambda_v(c_i,c_j)}\cdot
\Big(\frac{-1}{(x_i-x_j)^2_-}\Big)^{-\frac{q_iq_j}{8\pi^2}\erw{c_i,c_j}}\cdot
e^{-\frac{q_iq_j}{4\pi^2}\wt H(x_i-x_j;c_i,c_j)},\eea
where the quadratic form
\bea{cc}\erw{c,c'}:= \int d\sigma(\vec e)\,c(\vec e)\int d\sigma(\vec
e\,')c'(\vec e\,')\,\skp e{e\,'}\wt f(e,e') \qquad \big(\wt f(e,e') =\frac{\pi-\theta_{\vec e,\vec
    e\,'}}{\sin(\theta_{\vec e,\vec e\,'})}\big)\eea
determines the power law
  behaviour. In contrast, the quadratic form  
$$\lambda_v(c_i,c_j):=\int d\sigma(\vec e)\,c(\vec e)\int d\sigma(\vec
e\,')c'(\vec e\,')\,\skp e{e\,'}\wt f(e,e')\log\mu_v(e,-e')$$
just sets a scale and can be eliminated altogether: because it 
is symmetric, $\sum_iq_ic_i=0$ implies
$\sum_{i<j} \frac{q_iq_j}{8\pi^2}\lambda_v(c_i,c_j) =
-\sum_i\frac{q_i^2}{16\pi^2}\lambda_v(c_i,c_i)$, and by conveniently
defining
\bea{a:V} V_{qc}(x):= e^{\frac{q^2}{16\pi^2}\lambda_v(c,c)}\cdot
\wickv{e^{iq\phi(x,c)}},
\eea
one arrives at \eref{VVV} which is independent of the regulator function $v(k)$.

When smeared with the constant function $c_0(e_i)=\frac1{4\pi}$,
\eref{a:phiphi} simplifies drastically, see \cite{GRT}: one has $\erw
{c_0,c_0}=1$, and  
\bea{c:H0}\wt H(x;c_0,c_0) =\frac {x^0}{2r}
\log\frac{(x^0-i\eps)+r}{(x^0-i\eps)-r} \qquad (r=\vv x),\eea
hence 
\bea{c:V0V0}
\Erw{V_{qc_0}(x_1)V_{-qc_0}(x_2)} =
\Bigg[\frac
{\Big(\frac{x^0-r-i\eps}{x^0+r-i\eps}\Big)^{\frac{x^0}{
      r}}}{- (x^2)_-}\Bigg]^ {\frac{q^2}{8\pi^2}} \qquad (x=x_1-x_2).
\eea
Among the smearing functions of unit weight, $c_0$ is a stationary point for
$\erw{c,c}$. It is presumably a minimum: that is, the power law decay
for general $c\neq c_0$ is faster than for $c_0$.

\subsection{Photon cloud superselection structure}
\label{b:constr}

We present the details of the non-perturbative construction of the
four-dimensional dressing transformation, starting from \eref{phiphi}
and \eref{wm}.

Smearing the escort field with real test functions $g(x)$ and $c(e)$
we get from \eref{wm}
$$\erw{\phi(g\otimes c)\phi(g'\otimes c')} = w_{m,v}(g\otimes c;g'\otimes c') + \ol{\wh
  g(0)}\wh g'(0) \cdot d_{m,v}(c,c').$$
Now, we define 
\bea{E-reg}
V_c(g)\equiv \wickmv{e^{i\phi(g\otimes c)}} = e^{-\vert \wh g(0)\vert^2 d_{m,v}(c,c)}\cdot
\wick{e^{i\phi(g\otimes c)}}
\eea
and compute the massive correlation functions of operators
$\wickmv{e^{i\phi(g_i\otimes c_i)}}$. When all commutator contributions from
the Weyl formula and renormalization factors are collected, the
correlations exhibit an overall factor
$$e^{-d_{m,v}(C,C)}, \qquad
\hbox{where} \quad C(\vec e)=\sum_i \wh g_i(0)c_i(\vec e).$$
$d_{m,v}(C,C)$ can be written as 
\bea{b:dmvS}
d_{m,v}(C,C) =\int d\mu_m(k) \, v(k) \vert \vec T_C(k)\vert ^2,\eea
where 
\bea{b:Tc}\vec T_C(k) := \int d\sigma(\vec e)\,C(\vec e) \frac{\vec
  e}{(ke)+i\eps}\equiv\ERW{\frac{\vec e}{(ke)+i\eps}}_C.\eea
In the massless limit, \eref{b:dmvS} diverges logarithmically at $k=0$ unless $\vec T_C(k)$
vanishes for all $k$ on the zero mass-shell. This is possible only if $C(\vec e)=0$
\cite{MRS2}.\footnote{For more detailed properties of the function
  $T_C$, see \aref{b:orth}, in particular the inversion formula \eref{b:cU}.}
Therefore, the prefactor $e^{-\frac12d_{m,v}(C,C)}$ converges to the
superselection rule
\bea{sss}\delta_{C,0}=\left\{\ba{cl} 1 & \hbox{if}\,\, C=0\\ 0 &
  \hbox{if}\,\, C\neq0\ea\right.\eea
The resulting finite massless correlation functions among
$\wickv{e^{i\phi(g\otimes c)}}=\lim_{m\to0}\wickmv{e^{i\phi(g\otimes c)}}$ are \eref{corr4}.

\subsection{Orthogonality of Lorentz transformed sectors}
\label{b:orth}
Let $u\neq u'$ forward unit vectors, and $c$, $c'$ smooth real functions on $H_1\cap u^\perp$, 
$H_1\cap u'^\perp$, respectively, of equal total weight. We claim in \sref{s:Lor} that states of the form
$$\ket{f,c}=\wickv{e^{i\phi(f,c)}}\Omega, \quad
\ket{f',c'}=\wickv{e^{i\phi(f',c')}}\Omega,  \qquad (\wh f(0)=\wh{f'}(0)=q\neq0)$$
are mutually orthogonal, unless $c=c'=0$. This happens because the two-point
function $\erw{\phi(f,c)\phi(f',c')}$ diverges to $+\infty$.

The argument is as follows. For a momentum four-vector $k$ on the
closed forward lightcone, consider 
$\frac{e}{(ke)+i\eps} = -i \ioi ds\, e^{is(ke)}$ as a distribution on $H_1$. For a smooth real function $c$, supported
on $H_1\cap u^\perp$, define (generalizing \eref{b:Tc} to general
forward unit vectors)
\bea{b:Uc}T_c(k) := 
  \int_{H^1\cap u^\perp}
  d\sigma_u(e)\,\frac{c(e)e}{(ke)+i\eps} \equiv \ERW{\frac{e}{(ke)+i\eps}}_c\eea
where  $d\sigma_u(e)$ is the invariant measure on the sphere $H_1\cap u^\perp$. Because
$\frac1{(ke)+i\eps}$ is rotationally invariant and 
$c(e)$ is smooth, $T_c(k)$ is a smooth function. By definition, 
\bea{b:conds}(uT_c(k))=0 \qquad \hbox{and} \qquad \ol{T_c(k)} + T_c(P_uk)=0,\eea
where $P_uk = 2(uk)u-k$ is the parity reflection in the
frame $u$. The second property is a reality condition reflecting the
fact that $c$ is a real function. Moreover, 
the function $T_c(k)$ is homogeneous in $k$ of degree $-1$, and
$(T_c(k)k)=\erw 1_c$ is the total weight of $c$. 

The smearing function $c(e)$ can be recovered from the restriction of
$T_c(k)$ to any mass-shell:
\bea{b:cU} c(e) e = ir^2\int d\mu_m(k)\cdot 2(uk)\,e^{-ir(ke)} T_c(k),\eea
where $r>0$ is arbitrary. We prove \eref{b:cU}
(without
loss of generality in the standard frame $u=u_0$, where $2(u_0k)\,d\mu_m(k)= \frac{d^3k}{(2\pi)^3}$). Let accordingly $e=\bpm 0\\ \vec
e\epm$, and let $\Erw\cdot_c$ denote the smearing with $c(e')$: 
$$ir^2\int d\mu_0(k)\cdot 2(uk)\,e^{-ir(ke)} T_c(k)
=\frac{r^2}{(2\pi)^3}\int d^3k\, \ERW{e \ioi ds \, e^{ir\skp ek-is\skp
    {e\,'}k}}_c =$$
$$=r^2 \ERW{e \ioi ds\, \delta(r\vec e-s\vec e\,')}_c=\ERW{e  \ioi
ds\, \delta(r-s)\delta_{\vec e,\vec e\,'}}_c=\ERW{e
\,\delta_{\vec e,\vec e\,'}}_c=c(e)e.$$
After these preparations, we turn to the issue at hand. Let $c$, $c'$ and $f$, $f'$ as
specified above.  
The divergent part of the massive two-point
function $\erw{\phi(f,c)\phi(f',c')}$ is
\bea{b:dmvH}
  q^2 d_{m,v}(c-c',c-c') = -q^2\int d\mu_m(k) \,
  v(k)\vert T_c(k)-
  T'_{c'}(k)\vert^2.
  \eea
If $c$ and $c'$ have equal weight, then $T_c(k)-T'_{c'}(k)$ is
orthogonal to $k$, and consequently, in the massless limit where
$k^2=0$, it is either spacelike or a multiple of
  $k$. Thus, the integral \eref{b:dmvH} diverges to $+\infty$ unless
  $T_c(k)-T'_{c'}(k)$ is a multiple of $k$ for all $k$ on the zero
  mass-shell. We claim that this is impossible if 
  $c$ and $c'$ are  both real.

 Let $T_c(k)-T'_{c'}(k)=\alpha(k)\cdot k$ for all $k$ on the
 mass-shell. Then, because   $T_{c}(k)\in u^\perp$ and  
 $T'_{c'}(k)\in u'^\perp$, the coefficient is uniquely fixed:
 \bea{b:alpha}\alpha(k) = \frac{(u'T_{c}(k))}{(u'k)} = -
 \frac{(uT'_{c'}(k))}{(uk)}.\eea
 Now assume that the reality condition (the second in
 \eref{b:conds}) holds for both $T_c$ and $T'_{c'}$. Then 
$$\frac{(u'T_{c}(k))}{(u'k)} = - \frac{(uT'_{c'}(k))}{(uk)} =
\frac{\ol{(uT'_{c'}(P_{u'}k))}}{(uk)} =-
\frac{\ol{(u'T_{c}(P_{u'}k))}(uP_{u'}k)}{(u'P_{u'}k)(uk)} =
\frac{(u'T_{c}(P_uP_{u'}k))(uP_{u'}k)}{(u'P_{u'}k)(uk)},$$
where also \eref{b:alpha} was used twice.
Because $(u'P_{u'}k)=(u'k)$ and $(uP_{u'}k)=(uP_uP_{u'}k)$, this is
$$f(k)=f(P_uP_{u'}k), \quad\hbox{where}\quad f(k):= (uk)(u'T_c(k)). $$
When $u'=\Lambda u$ with a boost $\Lambda$ of rapidity $\tau>0$, then
one has $P_uP_{u'}=\Lambda^{-2}$. Because $k$ is arbitrary, this implies that the homogeneous
function $f(k)$ must be $\Lambda^2$-periodic:
$$f(\Lambda^{2}k)=f(k).$$
For the boost $\Lambda$ there is a (unique
up to a positive factor) massless
four-momentum $k_0$ such that $\Lambda k_0=e^\tau k_0$. Then, for
every $k$, the sequence $e^{-2n\tau}\Lambda^{2n}k$ converges to a
multiple of $k_0$. It follows by continuity and homogeneity
that $f(k)=\lim_{n\to\infty}
f(e^{-2n\tau}\Lambda^{2n}k)=f(k_0)=:f_0$. In particular, $f(k)$ must be
constant, hence for all $k$
$$(u'T_c(k))=\frac {f_0}{(uk)}.$$
We now insert this into the inversion formula \eref{b:cU} (without
loss of generality for
$u=u_0$), contracted with $u'$. The integral is a
multiple of the massless two-point function $W(r e)$, hence a
multiple of $r^{-2}$. It then follows that $c(e)(u'e)$ must be
constant. Because $(u'e)=0$ for  $e\in u^\perp\cap u'^\perp$, the
constant is zero, and $c(e)$ must be supported on the circle $H_1\cap u^\perp\cap
u'^\perp$. Since $c$ is smooth on $H_1\cap u^\perp$, this is
impossible unless $c=0$.

\section{Asymptotics}

\subsection{Dressed expectation values}
\label{b:erwFu}
The expectation value of the electromagnetic field in a dressed Dirac
state requires the computation of the bosonic factor $\Erw {V_{qc}(y_1)F^u_{\mu\nu}(x)
V_{qc}(y_2)^*}$. This is obtained by varying the expectation value of
$\wick{e^{i F^u_{\mu\nu}(f^{\mu\nu})}}$ w.r.t.\ $f^{\mu\nu}$, as
explained in \sref{s:dressMD}. It yields
\bea{VFuVraw}\Erw {V_{qc}(y_1)F^u_{\mu\nu}(x)
V_{qc}(y_2)^*} = -iq\big(\erw{F^u_{\mu\nu}(x)\phi(y_2,c)}-\erw{\phi(y_1,c) F^u_{\mu\nu}(x)}\big)\cdot\Erw {V_q(y_1,c)
V_q(y_2,c)^*}.\quad\qquad\eea
The expectation values in the first factor on the right-hand side are those in the
Krein vacuum, and can easily be computed in terms of the massless
two-point function $W_0$:
\bea{expFphi}\erw{F^u_{\mu\nu}(x)\phi(y,e)} = \erw{ F_{\mu\nu}(x)\phi(y,e)}
- (u_\mu\pa_\nu-u_\nu\pa_\mu)
  I_u \erw{ (\pa A^K)(x)\phi(y,e)}=\notag \\
=(e\wedge \partial_x)_{\mu\nu} I_{-e}^xW_0(x-y)
+ (u\wedge \partial_x)_{\mu\nu} I_u^xW_0(x-y) ,\eea
where we have used $\erw{(\pa A)(x)\phi(y)}=-W_0(x-y)$ and
$I^y_e=I^x_{-e}$. Notice that $\pa^\mu (e\wedge \partial)_{\mu\nu}
I_eW_0(x-y) = -\pa_\nu W_0(x-y)$ for every $e$, so that the
total divergence of  \eref{expFphi} is zero, in accord with the
cancellation of the fictitious current in $F^u_{\mu\nu}$, cf.\
\sref{s:dressMD}.

Thus, because $c(\vec e)$ has total weight 1,
\bea{b:VFuV}
\erw{V_q(y_1,c)F^u_{\mu\nu}(x)
V_q(y_2,c)^*}=\hspace{90mm}\notag \\ = -iq \ERW{(e\wedge \partial_x)_{\mu\nu}
I_{-e}^x +(u\wedge \partial_x)_{\mu\nu} I_u^x(W_0(x-y_2)-W_0(y_1-x))}_c,\qquad\eea
where $\Erw\cdot_c$ stands for the smearing of a function of $\vec e$
with $c(\vec e)$.

\subsection{Timelike and spacelike asymptotics}
\label{b:spl-tl}
In \sref{s:spl-tl}, we want to compute the asymptotic behaviour of \eref{VFuV} (in the
standard frame $u=u_0$) in the spacelike and timelike directions 
$x^\pm_\lambda=x_0+\lambda d_w^\pm$, where 
$d_w^\pm = \bpm \pm 1\\[-1mm]w\vec n\epm$, $w\neq 1$.

For $w\neq 1$, we
are again allowed to ignore $x_0$ and $y_i$ in \eref{VFuV} in the limits
$\lambda\to\infty$, and replace both $x_\lambda-y_i=x_0+\lambda d_w^\pm -y_i$ by $z_\lambda =\lambda
\cdot d_w^\pm$. Then, in \eref{VFuVraw}, the differences of
two-point functions may be replaced by the commutator: 
\bea{Atphi}-iq[A^u_\mu(x_\lambda),\phi(y,e)]
=q(e_\mu I_{-e}+u_\mu I_u)C_0(z_\lambda) = \frac{\pm q}{2\pi}
\big(e_\mu I_{-e}+u_\mu I_u\big)\delta(z_\lambda^2) + O(\lambda^{-3}),\qquad\eea
We have used that $C_0(z_\lambda) =
\frac1{2\pi}\delta(z_\lambda^2)\sign(z_\lambda^0)=\frac{\pm 1}{2\pi}\delta(z_\lambda^2)$ for $z_\lambda = \lambda
d^\pm_w$. Hence
\bea{Ftphi}-iq[F^u_{\mu\nu}(x_\lambda),\phi(y,e)]
= \frac{\pm q}{\pi}
\Big((z_\lambda\wedge e)_{\mu\nu}I_{-e}+(z_\lambda\wedge u)_{\mu\nu}I_u\Big)\delta'(z_\lambda^2) + O(\lambda^{-3}).\eea
The integrals are
$$I_{-e}\delta'(z_\lambda^2) = \frac{\nu}{4\sqrt{(z_\lambda
    e)^2+z_\lambda^2}^3}, \qquad I_{u}\delta'(z_\lambda^2) =
\frac{\nu'}{4\sqrt{(z_\lambda u)^2-z_\lambda^2}^3},$$
where $\nu$ and $\nu'$ are the numbers of {\em positive} ones among
the zeroes $s_\pm$ and $s'_\pm$ of the
polynomials $(z_\lambda-se)^2$ and $(z_\lambda+s'u)^2$, respectively, cf. \eref{Idelta-nu}.
$\nu$ and $\nu'$ are functions of  $d^\pm_w$ and
$e$, but independent of $\lambda$. One has $s_\pm= \lambda\cdot (-(d^\pm_we)\pm \sqrt{(d^\pm_we)^2+d^\pm_w{}^2})$ and $s'_\pm
= \lambda\cdot(- (d^\pm_w)^0\pm \sqrt{1-d^\pm_w{}^2})$. Since the zeroes may be positive
or negative or complex, several cases have to be distinguished.
\begin{itemize}\itemsep-1mm
\item
On $\frak i^+$ (future timelike,
$0<d^\pm_w{}^2\leq 1$), one has $\nu=1$ and $\nu'=0$.
\item
On $\frak i^-$ (past timelike), one has $\nu=1$ and $\nu'=2$.
\item On $\frak i^0$ (spacelike, $w<1$,
$d^\pm_w{}^2<0$), one has $\nu'=1$. $\nu$ depends on the angle $\alpha=\angle(\vec n,\vec e)$   of $\vec e\in S^2$ relative to $\vec n\in S^2$. One has $s_\pm = \lambda\cdot (w\cos\alpha\pm
\sqrt{1-w^2\sin^2\alpha})$, so that $\nu=2$ on the ``polar
cap'' $\alpha < \arcsin (w\inv)<\frac\pi2$, and $\nu=0$ otherwise. 
\end{itemize}
This yields finally, with $\nu$ and $\nu'$ as specified,
\bea{b:Fu-spl-tl}\lim_{\lambda\to\infty}\lambda^2\Erw{F^u_{\mu\nu}(x_0+\lambda d^\pm_w)}_{f,c}=\frac{\pm q}{4\pi}\ERW{(e\wedge d^\pm_w)_{\mu\nu}\frac{\nu}{\sqrt{1-w^2\sin^2\alpha}^3} -
(u\wedge
d^\pm_w)_{\mu\nu}\frac{\nu'}{w^3}}_c. \qquad
\eea

\subsection{Local Gauss Law at lightlike infinity}
\label{b:maxw}
We want to derive the local (= differential) Gauss Law on
$\frak I^+$. We take it source-free, although hypothetical charged
sources present on $\frak I^+$ can easily be added (see
\cite{BG}). The present computation is classical, and in the quantum
case the fields should be replaced by expectation values in suitable states.

We write the local Gauss Law in the bulk in Penrose coordinates
$(V,U,\vec n)$: 
$$V=x^0+\vv x, \qquad U = x^0-\vv x, \qquad \vec n = \frac{\vec x}{\vv
  x}.$$
The asymptotic fields arise from the expansion in $V\inv$:
\bea{asfield}
\vec E(V,U,\vec n) = \frac2V \vec E_1(U,\vec n) + \frac4{V^2} \vec E_2(U,\vec n) +
O(V^{-3})\eea
With 
\bea{asderiv}\pa_i =
n^i\,\pa_V-n^i\,\pa_U + \frac 2{V-U}\pa^\perp_{n,i},
\qquad\hbox{where}\quad \pa^\perp_{n,i}\equiv \pa_{n,i}-n_i\Skp n{\nabla_n},\eea
we get for large $V$ 
$$
0=\vnabla\cdot \vec E(V,U,\vec n)= \pa_V\Skp nE -\pa_U \Skp nE +
\frac 2{V-U}\Skp{\nabla^\perp_n}E =$$
$$= -2V^{-2}\Skp n{E_1} -2V^{-1}\pa_U\Skp n{E_1}-4V^{-2}\pa_U\Skp
n{E_2} + 4V^{-2}\Skp{\nabla^\perp_n}{E_1} + O(V^{-3}).$$
At leading order, this requires $\pa_U\Skp n{E_1}=0$. For physical
radiation fields, one imposes transversality: $\Skp
n{E_1}=0$. Then the next-to-leading order gives the asymptotic Gauss
Law at $V=\infty$:
\bea{asgauss}\pa_U \Skp
n{E_2} =\Skp{\nabla^\perp_n}{E_1}.\eea
This is \eref{dErE}, where $\vec E_\infty \equiv \vec E_1$ and
$E_{r,\infty}=\Skp n{E_2}$. Exchanging $U$ and $V$, one obtains the Gauss Law on $\frak I^-$.
One may as well include sources  in the bulk, but the charge density of massive
charged sources would vanish on $\frak I^\pm$. For massless sources,
see \cite{BG}.

\subsection{Commutation relations at lightlike infinity}
\label{b:ash}

Ashtekar \cite{Ash} has derived the commutation relations of the electromagnetic
field at lightlike infinity from the symplectic form on the
``characteristic data'' on $\frak I^\pm$, regarded as a surface of
constant $V$ as $V\to\infty$. We show here how one arrives at the same result by the corresponding limit of the commutation
relations in the bulk.

The massless two-point function in Penrose coordinates (\aref{b:maxw})
is
$$W_0(x-x') = \frac1{2\pi^2}\cdot \frac1{(V-U)(V-U')(1-\Skp
  n{n\,'}) -2(V-V')(U-U') + i\eps(V-V'+U-U')}.$$
Consequently, the commutator function at $V'=V$ in the limit $V\to\infty$ is:
$$\lim_{V\to\infty} V^2C_0(x-x')\big\vert_{V'=V}
=\frac i{2\pi^2} \cdot \Big(\frac 1{1-\Skp
  n{n\,'}+i\eps(U-U')}-(U\lra U')\Big)=
\sign(U-U')\cdot\delta(\vec n,\vec n\,'),$$
where $\delta(\vec n,\vec n\,')$ is the delta function w.r.t.\ the
invariant measure on the sphere.
Then $A^\mu_1(U,\vec n):=\lim_{V\to\infty} \frac 12V\cdot A^\mu(V,U,\vec n)$ have commutation
relations
\bea{AAas} i[A^\mu_1(U,\vec n),A^\nu_1(U',\vec n')] = -\frac14
\eta^{\mu\nu} \sign(U-U')\cdot\delta(\vec n,\vec n\,').\eea
The commutator of  $E_\infty^i = F_1^{i0}$, the leading $O(\frac 2V)$ contribution to the
electric field as in \aref{b:maxw}, is computed with
$\pa_0=\pa_V+\pa_U$ and \eref{asderiv}: 
\bea{EEas} i[E_\infty^i(U,\vec n),E_\infty^j(U',\vec n')] =
\frac12\delta'(U-U')\cdot(n^in'^j-\delta_{ij})\delta(\vec n,\vec n\,'),\eea
which is compatible with transversality $\Skp n{E_\infty}=0$. Using the identity
$$\nabla'^\perp_j(n^in'^j-\delta_{ij}) \delta(\vec n,\vec n') =
(2n^i-\nabla'^\perp_i)\delta(\vec n,\vec n') = \nabla'^\perp_i(1-\Skp
n{n\,'})\delta(\vec n,\vec n') =0,$$
one finds that the leading $O((\frac 2V)^2)$ contribution
$E_{r,\infty}=\Skp n{E_2}$ of the radial 
component $\Skp n{E} = -(\pa_V+\pa_U)\Skp nA
-(\pa_V-\pa_U)A_0$  
commutes with $\vec E_\infty$ and with itself. This is consistent with
\eref{Qeps} being the generators of large gauge transformations 
\cite{Str}, which leave the Maxwell tensor invariant, hence must
commute with $E_{r,\infty}(U',\vec n')$ and among each other. 

The same computations repeated at spacelike infinity would yield
zero because of the more rapid decay of Maxwell correlation functions
(and hence commutators) in spacelike than in lightlike
directions. This means that the algebra of asymptotic Maxwell fields
becomes classical (and decoherence occurs intrinsically) at $\frak
i^0$, while it remains (at least partially) non-commutative on $\frak I^\pm$.

\subsection{Lightlike asymptotics}
\label{b:ll}
We want to compute the asymptotic behaviour of \eref{VFuV} (in the
standard frame $u=u_0$) in the lightlike direction 
$x^\pm_\lambda=x_0+\lambda\ell^\pm$, where 
$\ell^\pm = \bpm \pm 1\\[-1mm]\vec n\epm$. The trajectories $x^\pm_\lambda$
reach the Penrose lightlike infinity $\frak I^\pm$ at the points $(U,\vec
n)$ with $U=(\ell^+ x_0)$, and $(V,\vec n)$ with $V=-(\ell^- x_0)$, respectively. 

In the lightlike limit, the dependence of \eref{VFuV} on the ``initial points''
$x_0-y_i$ ($i=1,2$) does not drop out, giving functions of $U-(\ell^+ y_i)$ on $\frak I^+$
resp.\ $V+(\ell^-y_i)$ on $\frak I^-$. Therefore, one should compute  the
differences of two-point functions in \eref{VFuV} rather than just
commutators. We first investigate the expexted deviation, by
inspection of \eref{fxe} and \eref{fxu}.

The differences in \eref{VFuV} would require to consider expressions like
$$\log\frac{(z_1e)+\sigma_{1,-}}{(z_2e)+\sigma_{2,+}}
-\log\frac{(z_1e)-\sigma_{1,-}}{(z_2e)-\sigma_{2,+}}\quad\hbox{and}\quad
\log\frac{z_1^0-i\eps+\rho_1}{z_2^0+i\eps+\rho_2}-\log\frac{z_1^0-i\eps-\rho_1}{z_2^0+i\eps-\rho_2}, $$
evaluated at $z_i=x_0+\lambda\ell^\pm -y_i$, in the limit
$\lambda\to\infty$. In this limit, the numerators and 
denominators of either one of the $\log$'s become small, so that
the behaviour is dictated by the $i\eps$ prescriptions. In particular, for
$y_1=y_2$, the $\log$'s would just be multiples of $2\pi i$ times step
functions. The corrections for $y_1\neq y_2$ can in principle be
worked out from these expressions. We want to spare that labour, by
just noting that a narrow smearing in $y$ would essentially smoothen the step functions.

Keeping this in mind, we shall from now on assume a narrow smearing
and set $y_1=y_2$, so that we
can compute the complex phases by returning to the
commutators. However, in contrast to \sref{b:spl-tl}, we have to keep
the dependence on $z_0=x_0-y$ in $z^\pm_\lambda=x_0+\lambda \ell^\pm-y$.

For $x_\lambda=x_0+\lambda\ell^+$, the Penrose coordinate is $V_\lambda=x^0_\lambda+ 
\vv{x_\lambda}\approx 2\lambda$. Therefore, in order to compute the leading order
$\vec E_1$ in \eref{asfield}, one may as well expand in $\lambda\inv$
(which simplifies the computation). The difference will become
effective only in the computation of $\vec E_2$. Similar for $x_\lambda=x_0+\lambda\ell^-$.

Using the massless commutator in
$C_0(z)=\frac1{2\pi}\sign(z^0)\delta(z^2)$ and
$\sign((z^\pm_\lambda)^0)=\pm 1$ for sufficiently large $\lambda$,
$$f(z^\pm_\lambda,-e)-f(-z^\pm_\lambda,e) =
-i(2\pi)^2 I_{-e} C_0(z^\pm_\lambda) =
\mp 2\pi i \ioi ds \,\delta((z^\pm_\lambda-se)^2).$$
 As in \aref{b:spl-tl}, this is
 $$f(z^\pm_\lambda,-e)-f(-z^\pm_\lambda,e) =\mp 2\pi i\cdot
 \frac{\nu^\pm}{2\sqrt{(z^\pm_\lambda e)^2+ z^\pm_\lambda{}^2}},$$
 where $\nu^\pm$ is the number of positive zeroes of
$(z^\pm_\lambda-se)^2=-s^2-2s(z^\pm_\lambda e)+z^\pm_\lambda{}^2$. For
sufficiently large $\lambda$, $\nu^\pm=2\theta(-(z^\pm_\lambda e)) +\sign((z^\pm_\lambda e))\theta(z^\pm_\lambda{}^2)$. Thus,
\bea{fze} f(z^\pm_\lambda,-e)-f(-z^\pm_\lambda,e) =\mp i\pi \cdot
 \frac{-2\theta(-(z^\pm_\lambda e))+ \theta(z^\pm_\lambda{}^2)}{(z^\pm_\lambda
   e)\sqrt{1+w^\pm_\lambda}}, \qquad \big(w^\pm_\lambda\equiv\frac{z^\pm_\lambda{}^2}{(z^\pm_\lambda e)^2}\big) .\eea
Similarly, for sufficiently large $\lambda$,
\bea{fzu}f(z^\pm_\lambda,u)-f(-z^\pm_\lambda,-u) =  \mp i\pi
\cdot\frac{2\theta(-(z^\pm_\lambda)^0)+\theta(-z^\pm_\lambda{}^2)}{(z^\pm_\lambda)^0\sqrt{1-v^\pm_\lambda}},
\qquad \big(v^\pm_\lambda \equiv \frac{z_\lambda^2}{(z_\lambda^0)^2}\big).\eea

For the asymptotic treatment in leading order, notice that $(z^\pm_\lambda e)\approx
\lambda(\ell^\pm e)$ and
$(z^\pm_\lambda)^0\approx\pm\lambda$ and $z^\pm_\lambda{}^2\approx
2(\ell^\pm z_0)$ are $O(\lambda)$, hence $w^\pm_\lambda$ and
$v^\pm_\lambda$ are $O(\lambda\inv)$. Thus, in the limit, 
\bea{fze-as} f(z^\pm_\lambda,-e)-f(-z^\pm_\lambda,e)\approx
 \frac{\pm i\pi}\lambda \cdot
 \frac{2\theta(-(\ell^\pm e))- \theta((\ell^\pm z_0))}{(\ell^\pm e)} +
 O(\lambda^{-2}).\eea
Similarly
\bea{fzu-as}f(z^\pm_\lambda,u)-f(-z^\pm_\lambda,-u)\approx \frac{\mp
  i\pi}\lambda \cdot\big(2\theta(\mp 1)+\theta(-(\ell^\pm z_0))\big) + O(\lambda^{-2}).\eea
We present the argument for $\frak I^+$, $\ell\equiv \ell^+$,
$x_\lambda=x_0+\lambda\ell$, $z_\lambda=x_\lambda - y$. 
From \eref{fze-as} and \eref{fzu-as}, one can read off the leading
behaviour (order $\lambda\inv$) of the expectation values of the
electromagnetic field
\bea{Finf}
\Erw{F^u_{\infty}(U,\vec n)}_{f,c} := \lim_{\lambda\to\infty}\lambda\cdot
\Erw{F^u(x_\lambda)}_{f,c} =
\frac{q}{4\pi}\ERW{\frac{(\ell \wedge e)}{(\ell e)} - (\ell \wedge
  u)}_c\delta (U-(\ell y)),\eea
where  $U
= (\ell x_0)$ and $\vec n$ are the corresponding coordinates on $\frak
I^+$. The electric field is accordingly
\bea{b:Einf}\Erw{\vec E_{\infty}(U,\vec n)}_{f,c} = \frac {q}{4\pi} \cdot
\ERW{\frac{\vec e}{(\vec n\cdot \vec e)}-\vec
  n}_c\cdot\delta(U-(\ell y)) \quad\hbox{where}\quad (\ell y)=y^0-\Skp ny.\quad
\eea
This is \eref{Einf}.
 \eref{b:Einf} is obviously transverse (orthogonal to $\vec n$), as expected for a
radiation field, see footnote \ref{fn:trans}.

The radial electric field decays like $\lambda^{-2}$. Its asymptotic
expectation value $\erw{E_{r,\infty}(U,\vec n)}_{f,c}\equiv \erw{\Skp n{E_2(U,\vec n)}}_{f,c}$ according
to \eref{asfield} is computed as
$$\Erw{E_{r,\infty}(U,\vec n)}_{f,c}=
\lim_{\lambda\to\infty} \frac{V_\lambda^2}4 \cdot \Erw{(\vec
  n_\lambda\cdot \vec E(V_\lambda,U_\lambda,\vec n_\lambda))}_{f,c}
\qquad \hbox{where}\quad \vec n_\lambda =\vec n + \lambda\inv (\vec
x-\Skp nx \vec n).$$
To compute it, we have to study the subleading orders of \eref{fze} and \eref{fzu}.

It is convenient to write $\Skp{n_\lambda}{E}$ as
$(u\wedge\ell_\lambda).F^u$, where  $(a\wedge b).(c\wedge d) \equiv
(ac)(bd)-(ad)(bc)$, and 
$\ell_\lambda = \bpm 1\\[-1mm]\vec
n_\lambda\epm$. This yields
$$\Erw{E_{r,\infty}(U,\vec n)}_{f,c}=\frac{-iq}{4\pi^2}\lim_{\lambda\to\infty} \frac{V_\lambda^2}4 \cdot (u\wedge
\ell_\lambda).\ERW{(e\wedge\pa)
\big(f(z_\lambda,-e)-f(-z_\lambda,e)\big) + (u\wedge\pa)
\big(f(z_\lambda,u)-f(-z_\lambda,-u)\big)}_c.$$
Inserting \eref{fze} and \eref{fzu},
$$\Erw{E_{r,\infty}(U,\vec n)}_{f,c}=\frac{q}{4\pi}\lim_{\lambda\to\infty} \frac{V_\lambda^2}4 \cdot (u\wedge
\ell_\lambda).\ERW{(e\wedge\pa)
\big( \frac{2\theta(-(z_\lambda e))- \theta(z_\lambda{}^2)}{(z_\lambda
   e)\sqrt{1+\frac{z_\lambda{}^2}{(z_\lambda e)^2}}}\big) + (u\wedge\pa)
\big(\frac{-\theta(-z_\lambda{}^2)}{z_\lambda^0\sqrt{1-\frac{z_\lambda^2}{z_\lambda^0{}^2}}}\big)}_c,$$
where $\frac{V_\lambda}2=\lambda + \frac12(x_0^0+\Skp n{x_0})$.
The derivatives ``see'' only $z_\lambda^2$. Acting on the
denominators of the two terms inside $\Erw{\dots}$, they are separately
$O(\lambda^{-2})$. Their contribution to $\erw{E_{r,\infty}}_{f,c}$ is 
$$\frac q{4\pi}\ERW{\frac{2\theta(\Skp ne) -\theta(U-(y^0-\Skp ny))}{\Skp
    ne^2} + \theta((y^0-\Skp ny)-U)}_c.$$
The contributions from the derivatives acting on the numerators are
proportional to $\delta(z_\lambda^2)$ which is $O(\lambda\inv)$. The
coefficients are $O(1)$, but the two terms in $\Erw{\dots}_c$ cancel
each other. The subleading terms contribute
$$\frac q{4\pi}\ERW{\frac{\Skp ey}{\Skp ne}-\Skp
  ny}_c\delta(U-(y^0-\Skp ny)) = \Erw{\Skp y{E_\infty(U,\vec n)}}_{f,c}.$$
Thus,
\bea{b:Erinf} \Erw{E_{r,\infty}(U,\vec n)}_{f,c} = \frac q{4\pi}\ERW{\frac{2\theta(\Skp ne)
    -\theta(U-(\ell y))}{\Skp
    ne^2} + \theta((\ell y)-U)}_c + \Erw{\Skp y{E_\infty(U,\vec n)}}_{f,c},\qquad\eea
where $(\ell y)= y^0-\Skp ny$. This is \eref{Erinf}.

\section{Perturbation theory in the presence of the infrafield}
\label{c:ptII}

\subsection{Cloud propagator contributions}
\label{c:cloudv}

For the argument in \sref{s:joint}, we still have to compute the
contributions to \eref{GML} due to the four
diagrams with cloud propagators, two of which 
are depicted in Fig.\ 2. We compute the first depicted diagram. It equals
$$2\cdot \frac{(iq)^2}2 \Erw{T\underbracket{\!\!\psi_0(x') j}\underbracket{{\!}^\mu(y_1) j}\underbracket{{\!}^\nu(y_2)\ol
  \psi_0\!\!}\,\,(x)}\cdot
a_\mu(y_1) \cdot iq \Erw{T\phi(x',c')A^K_\nu(y_2)} ,$$
where the brackets stand for
$\erw{T\psi(\cdot)\ol\psi(\cdot)}=-iS_F(\cdot)$ and $S_F$ is the
Dirac Feynman propagator. The last factor is the first-order contribution to 
$\erw{TV_{qc'}(x')A^K_\nu(y_2)}$, i.e., the 
cloud propagator $-q \Erw{e'_\nu I_{e'}G_{0,F}(x'-y_2)}_{c'}$
where $G_{0,F}$ is the massless scalar Feynman propagator. Inserting
the Fourier representations, we get
\bea{x}&&(iq)^2\cdot \int d^4y_1\,d^4y_2\, a_\mu(y_1) \int
\frac{d^4p'}{(2\pi)^4}\,\frac{d^4q}{(2\pi)^4}\,\frac{d^4 p}{(2\pi)^4} \,
e^{-ip'(x'-y_1)}e^{-iq(y_1-y_2)}e^{-ip(y_2-x)}\cdot \notag \\
&&\cdot  (-i)^3 \frac {M+\sla
  p'}{M^2-p'^2-i\eps}\gamma^\mu \frac {M+\sla q}{M^2-q^2-i\eps}\gamma^\nu
\frac {M+\sla p}{M^2-p^2-i\eps} \cdot iq \int \frac{d^4k}{(2\pi)^4}
\, \ERW{\frac{e'_\nu}{(ke')_-}}_{c'}\frac
{e^{-ik(x'-y_2)}}{m^2-k^2-i\eps}.\notag \eea
The $y$-integrations yield $(2\pi)^4\delta(q-p+k)\cdot \wh
a_\mu(p'-q)$. The remaining exponential factors 
$e^{-i(p'+k)x'}e^{ipx}$ become $e^{-ip'x'}e^{ipx}$ after a change of
the integration variable $p'$. Now, we are interested in the singular behaviour near $k=0$ when $p$ and
$p'$ are on-shell. Truncating with
$(\square^{(\prime)}+M^2)$ according to the LSZ prescription, cancels the denominator of the last
Dirac propagator. The denominator of the first Dirac propagator
(with the cloud vertex attached) is cancelled only up to $O(k)$. The (unknown) infraparticle truncation, that properly accounts for the
absence of a sharp mass-shell,  should justify to let $k\to0$ in this
term before
going on-shell with $p'$. Then we get the coefficient of
$e^{-ip'x'}e^{ipx}$:
$$q^3\cdot \wh a_\mu(p'-p) \int
\frac{d^4k}{(2\pi)^4} \cdot 
\frac{M+\sla p'-\sla k}{(2\pi)^4} \gamma^\mu \frac{M+\sla p-\sla
  k}{M^2-(p-k)^2-i\eps}\gamma^\nu \frac{M+\sla p}{(2\pi)^4} \cdot
\ERW{\frac{e'_\nu}{(ke')-i\eps}}_{c'}\cdot \frac {1}{m^2-k^2-i\eps}.
$$
With $\gamma^\nu(M+\sla p)$ $= 2p^\nu +(M-\sla
p)\gamma^\nu$, the first factor under the integral becomes
$\approx \frac{M+\sla p'}{(2\pi)^4} \gamma^\mu
\frac{2p^\nu}{2(pk)}\frac{M+\sla p}{(2\pi)^4}$ plus finite terms
$O(k^0)$. This allows to factor out the first-order vertex amplitude
$\Gamma^{(1)}(p,p')$ of \eref{Gamma}. The divergent coefficient has
the real part 
$$\mathrm{Re}\Big[q^2 \int
\frac{d^4k}{(2\pi)^4} \, \frac{p^\nu}{(pk)_-}
\ERW{\frac{e'_\nu}{(ke')_-}}_{c'}\cdot \frac {i}{m^2-k^2-i\eps}\Big] =-\frac  {q^2}2 \mathrm{Re}\Big[\int
\frac{d^4k}{(2\pi)^3} \, \frac{p^\nu}{(pk)}
\ERW{\frac{e'_\nu}{(ke')_-}}_{c'}\cdot \delta(k^2-m^2)\Big].$$
(For this equality, we have used the symmetry of the propagator under $k\to-k$).
As we think of this as the onset of an exponential whose phase does
not matter, we ignore the imaginary part. Furthermore, we may replace
$\frac{d^4k}{(2\pi)^3} \, \delta(k^2-m^2)$ by $2d\mu_m(k)$.

Proceeding similarly with the other three diagrams, we obtain the
total coefficient 
$$q^2 \mathrm{Re}\Big[\int d\mu_m(k) \, \Big(\frac{p^\nu}{(pk)}-\frac{p'^\nu}{(p'k)}\Big)\Big(
\ERW{\frac{e_\nu}{(ke)_-}}_{c}-\ERW{\frac{e'_\nu}{(ke')_-}}_{c'}\Big)\Big]=
-\frac12(d_{m,v}(C,C_0)+d_{m,v}(C_0,C)) + \hbox{finite}
$$
with $C=q(c-c')$ and $C_0=q(c_u-c_{u'})$, $u=p/M$, $u'=p'/M$. This is the interference
term between $- \frac12d_{m,v}(C,C)$ and $-\frac12d_{m,v}(C_0,C_0)$, anticipated in \sref{s:joint}.

\bigskip

{\bf Acknowledgments:} We thank W. Dybalski for his interest and
critical comments. JM has received financial support by the Brazilian research agency CNPq, and was partially supported by the Emmy Noether grant DY107/2-2 of the DFG.

\end{document}